%% file: oneloopI.tex
\input harvmacMv2.tex

\input amssym
\input epsf
\input localpaper.defs


\title Towards the n-point one-loop superstring amplitude I:

\title Pure spinors and superfield kinematics

\author
Carlos R. Mafra\email{\dagger}{c.r.mafra@soton.ac.uk}$^{\dagger}$ and
Oliver Schlotterer\email{\ddagger}{olivers@aei.mpg.de}$^{\ddagger,\ast}$

\address
$^\dagger$Mathematical Sciences and STAG Research Centre, University of Southampton,
Highfield, Southampton, SO17 1BJ, UK

\address
$^\ddagger$Max--Planck--Institut f\"ur Gravitationsphysik
Albert--Einstein--Institut, 14476 Potsdam, Germany

\address
$^\ast$Perimeter Institute for Theoretical Physics, Waterloo, ON N2L 2Y5, Canada

\abstract
This is the first installment of a series of three papers in which we describe a
method to determine higher-point correlation functions in one-loop
open-superstring amplitudes from first principles. In this first part, we
exploit the synergy between the cohomological features of pure-spinor superspace
and the pure-spinor zero-mode integration rules of the one-loop amplitude
prescription. This leads to the study of a rich variety of multiparticle
superfields which are local, have covariant BRST variations, and are compatible
with the particularities of the pure-spinor amplitude prescription. Several objects
related to these superfields, such as their non-local counterparts and the
so-called BRST pseudo-invariants, are thoroughly reviewed and put into new
light. Their properties will turn out to be mysteriously connected to products
of one-loop worldsheet functions in packages dubbed ``generalized elliptic
integrands'', whose prominence will be seen in the later parts of this series of
papers.

\Date {December 2018}


\lref\wipI{
C.R.~Mafra and O.~Schlotterer,
``Towards the n-point one-loop superstring amplitude I:
Pure spinors and superfield kinematics'', [arXiv:1812.10969 [hep-th]]\semi
C.R.~Mafra and O.~Schlotterer,
``Towards the n-point one-loop superstring amplitude II:
Worldsheet functions and their duality to kinematics'', [arXiv:1812.10970 [hep-th]]\semi
C.R.~Mafra and O.~Schlotterer,
``Towards the n-point one-loop superstring amplitude III:
 One-loop correlators and their double-copy structure'', [arXiv:1812.10971 [hep-th]]
}

\lref\PScohomology{
  N.~Berkovits,
  ``Cohomology in the pure spinor formalism for the superstring,''
JHEP {\bf 0009}, 046 (2000).
[hep-th/0006003].
\semi
N.~Berkovits and O.~Chandia,
  ``Lorentz invariance of the pure spinor BRST cohomology for the superstring,''
Phys.\ Lett.\ B {\bf 514}, 394 (2001).
[hep-th/0105149].
}

\lref\michos{
I. C. Michos, ``On twin and anti-twin words in the
support of the free Lie algebra.'' Journal of Algebraic Combinatorics 36.3 (2012) 355-388.
}

\lref\GreenED{
  M.~B.~Green and J.~H.~Schwarz,
  ``Infinity Cancellations in SO(32) Superstring Theory,''
Phys.\ Lett.\  {\bf 151B}, 21 (1985)..
}

\lref\PSS{
	C.R.~Mafra,
  	``PSS: A FORM Program to Evaluate Pure Spinor Superspace Expressions,''
	[arXiv:1007.4999 [hep-th]].
}
\lref\FORM{
	J.A.M.~Vermaseren,
	``New features of FORM,''
	arXiv:math-ph/0010025.
\semi
	M.~Tentyukov and J.A.M.~Vermaseren,
	``The multithreaded version of FORM,''
	arXiv:hep-ph/0702279.
}

\lref\AntoniadisVW{
  I.~Antoniadis, C.~Bachas, C.~Fabre, H.~Partouche and T.~R.~Taylor,
  ``Aspects of type I - type II - heterotic triality in four-dimensions,''
Nucl.\ Phys.\ B {\bf 489}, 160 (1997).
[hep-th/9608012].
}

\lref\oldMomKer{
	Z.~Bern, L.~J.~Dixon, M.~Perelstein and J.~S.~Rozowsky,
	``Multileg one loop gravity amplitudes from gauge theory,''
	Nucl.\ Phys.\ B {\bf 546}, 423 (1999).
	[hep-th/9811140].
}
\lref\MomKer{
	N.~E.~J.~Bjerrum-Bohr, P.~H.~Damgaard, T.~Sondergaard and P.~Vanhove,
	``The Momentum Kernel of Gauge and Gravity Theories,''
	JHEP {\bf 1101}, 001 (2011).
	[arXiv:1010.3933 [hep-th]].
}

\lref\KK{
	R.~Kleiss, H.~Kuijf,
  	``Multi - Gluon Cross-sections And Five Jet Production At Hadron Colliders,''
	Nucl.\ Phys.\  {\bf B312}, 616 (1989).
}
\lref\BGpaper{
	F.A.~Berends, W.T.~Giele,
  	``Recursive Calculations for Processes with n Gluons,''
	Nucl.\ Phys.\  {\bf B306}, 759 (1988).
}
\lref\BGSym{
	F.A.~Berends and W.T.~Giele,
	``Multiple Soft Gluon Radiation in Parton Processes,''
	Nucl.\ Phys.\ B {\bf 313}, 595 (1989).
}

\lref\Gauge{
	S.~Lee, C.R.~Mafra and O.~Schlotterer,
  	``Non-linear gauge transformations in $D=10$ SYM theory and the BCJ duality,''
	JHEP {\bf 1603}, 090 (2016).
	[arXiv:1510.08843 [hep-th]].
}

\lref\BernQJ{
  Z.~Bern, J.~J.~M.~Carrasco and H.~Johansson,
  ``New Relations for Gauge-Theory Amplitudes,''
Phys.\ Rev.\ D {\bf 78}, 085011 (2008).
[arXiv:0805.3993 [hep-ph]].
}

\lref\OchirovXBA{
  A.~Ochirov and P.~Tourkine,
  ``BCJ duality and double copy in the closed string sector,''
JHEP {\bf 1405}, 136 (2014).
[arXiv:1312.1326 [hep-th]].
}

\lref\verlindes{
	E.P.~Verlinde and H.L.~Verlinde,
	``Chiral Bosonization, Determinants and the String Partition Function,''
	Nucl.\ Phys.\ B {\bf 288}, 357 (1987).
}

\lref\threeloop{
	H.~Gomez and C.R.~Mafra,
  	``The closed-string 3-loop amplitude and S-duality,''
	JHEP {\bf 1310}, 217 (2013).
	[arXiv:1308.6567 [hep-th]].
}

\lref\Ree{
	R. Ree, ``Lie elements and an algebra associated with shuffles'',
	Ann. Math. {\bf 62}, No. 2 (1958), 210--220.
}
\lref\BGschocker{
	M. Schocker,
	``Lie elements and Knuth relations,'' Canad. J. Math. {\bf 56} (2004), 871-882.
	[math/0209327].
}

\lref\lothaire{
	Lothaire, M., ``Combinatorics on Words'',
	(Cambridge Mathematical Library), Cambridge University Press (1997).
}

\lref\NLSM{
	J.~J.~M.~Carrasco, C.R.~Mafra and O.~Schlotterer,
  	``Abelian Z-theory: NLSM amplitudes and $\alpha$'-corrections from the open string,''
	JHEP {\bf 1706}, 093 (2017).
	[arXiv:1608.02569 [hep-th]].
}

\lref\FTlimit{
	C.R.~Mafra,
  	``Berends-Giele recursion for double-color-ordered amplitudes,''
	JHEP {\bf 1607}, 080 (2016).
	[arXiv:1603.09731 [hep-th]].
}

\lref\DPellis{
	F.~Cachazo, S.~He and E.Y.~Yuan,
	``Scattering of Massless Particles: Scalars, Gluons and Gravitons,''
	JHEP {\bf 1407}, 033 (2014).
	[arXiv:1309.0885 [hep-th]].
}

\lref\LiE{
	M.A.A. van Leeuwen, A.M. Cohen and B. Lisser, ``LiE, A Package for Lie Group Computations'',
	Computer Algebra Nederland, Amsterdam, ISBN 90-74116-02-7, 1992
}

\lref\Richards{
	D.~M.~Richards,
	``The One-Loop Five-Graviton Amplitude and the Effective Action,''
	JHEP {\bf 0810}, 042 (2008).
	[arXiv:0807.2421 [hep-th]].
}

\lref\fiveptNMPS{
	C.R.~Mafra and C.~Stahn,
  	``The One-loop Open Superstring Massless Five-point Amplitude with the Non-Minimal Pure Spinor Formalism,''
	JHEP {\bf 0903}, 126 (2009).
	[arXiv:0902.1539 [hep-th]].
}

\lref\SiegelYI{
	W.~Siegel,
  	``Superfields in Higher Dimensional Space-time,''
	Phys.\ Lett.\ B {\bf 80}, 220 (1979).
}

\lref\towardsoneloop{
	C.R.~Mafra and O.~Schlotterer,
  	``Towards one-loop SYM amplitudes from the pure spinor BRST cohomology,''
	Fortsch.\ Phys.\  {\bf 63}, no. 2, 105 (2015).
	[arXiv:1410.0668 [hep-th]].
}

\lref\reutenauer{
	C.~Reutenauer,
	``Free Lie Algebras'', London Mathematical Society Monographs, 1993.
}

\lref\fourptpaper{
	C.R.~Mafra,
  	``Four-point one-loop amplitude computation in the pure spinor formalism,''
	JHEP {\bf 0601}, 075 (2006).
	[hep-th/0512052].
}
\lref\mafraids{
	C.R.~Mafra,
  	``Pure Spinor Superspace Identities for Massless Four-point Kinematic Factors,''
	JHEP {\bf 0804}, 093 (2008).
	[arXiv:0801.0580 [hep-th]].
}

\lref\MafraKJ{
  C.R.~Mafra, O.~Schlotterer and S.~Stieberger,
  ``Explicit BCJ Numerators from Pure Spinors,''
JHEP {\bf 1107}, 092 (2011).
[arXiv:1104.5224 [hep-th]].
}

\lref\MafraVCA{
  C.R.~Mafra and O.~Schlotterer,
  ``Berends-Giele recursions and the BCJ duality in superspace and components,''
JHEP {\bf 1603}, 097 (2016).
[arXiv:1510.08846 [hep-th]].
}

\lref\MafraNWR{
	C.R.~Mafra and O.~Schlotterer,
	``One-loop superstring six-point amplitudes and anomalies in pure spinor superspace,''
	JHEP {\bf 1604}, 148 (2016).
	[arXiv:1603.04790 [hep-th]].
}

\lref\BergWUX{
  M.~Berg, I.~Buchberger and O.~Schlotterer,
  ``From maximal to minimal supersymmetry in string loop amplitudes,''
JHEP {\bf 1704}, 163 (2017).
[arXiv:1603.05262 [hep-th]].
}

\lref\GregoriHI{
  A.~Gregori, E.~Kiritsis, C.~Kounnas, N.~A.~Obers, P.~M.~Petropoulos and B.~Pioline,
  ``R**2 corrections and nonperturbative dualities of N=4 string ground states,''
Nucl.\ Phys.\ B {\bf 510}, 423 (1998).
[hep-th/9708062].
}

\lref\BroedelTTA{
  J.~Broedel, O.~Schlotterer and S.~Stieberger,
  ``Polylogarithms, Multiple Zeta Values and Superstring Amplitudes,''
Fortsch.\ Phys.\  {\bf 61}, 812 (2013).
[arXiv:1304.7267 [hep-th]].
}

\lref\EOMbbs{
	C.R.~Mafra and O.~Schlotterer,
  	``Multiparticle SYM equations of motion and pure spinor BRST blocks,''
	JHEP {\bf 1407}, 153 (2014).
	[arXiv:1404.4986 [hep-th]].
}

\lref\oneloopbb{
	C.R.~Mafra and O.~Schlotterer,
  	``The Structure of n-Point One-Loop Open Superstring Amplitudes,''
	JHEP {\bf 1408}, 099 (2014).
	[arXiv:1203.6215 [hep-th]].
}

\lref\partI{
	C.R.~Mafra and O.~Schlotterer,
  	``Cohomology foundations of one-loop amplitudes in pure spinor superspace,''
	[arXiv:1408.3605 [hep-th]].
}

\lref\fivetree{
	C.R.~Mafra,
	``Simplifying the Tree-level Superstring Massless Five-point Amplitude,''
	JHEP {\bf 1001}, 007 (2010).
	[arXiv:0909.5206 [hep-th]].
}
\lref\nptFT{
	C.R.~Mafra, O.~Schlotterer, S.~Stieberger and D.~Tsimpis,
	``A recursive method for SYM n-point tree amplitudes,''
	Phys.\ Rev.\ D {\bf 83}, 126012 (2011).
	[arXiv:1012.3981 [hep-th]].
}
\lref\nptString{
	C.R.~Mafra, O.~Schlotterer and S.~Stieberger,
  	``Complete N-Point Superstring Disk Amplitude I. Pure Spinor Computation,''
	Nucl.\ Phys.\ B {\bf 873}, 419 (2013).
	[arXiv:1106.2645 [hep-th]].
}

\lref\wittentwistor{
	E.~Witten,
	``Twistor-Like Transform In Ten-Dimensions,''
	Nucl.\ Phys.\  B {\bf 266}, 245 (1986).
}
\lref\psf{
 	N.~Berkovits,
	``Super-Poincare covariant quantization of the superstring,''
	JHEP {\bf 0004}, 018 (2000)
	[arXiv:hep-th/0001035].
}
\lref\MPS{
	N.~Berkovits,
	``Multiloop amplitudes and vanishing theorems using the pure spinor formalism for the superstring,''
	JHEP {\bf 0409}, 047 (2004).
	[hep-th/0406055].
}

\lref\MafraIOJ{
	C.R.~Mafra and O.~Schlotterer,
  	``Double-Copy Structure of One-Loop Open-String Amplitudes,''
	Phys.\ Rev.\ Lett.\  {\bf 121}, no. 1, 011601 (2018).
	[arXiv:1711.09104 [hep-th]].
}

\lref\oneloopMichael{
	M.~B.~Green, C.R.~Mafra and O.~Schlotterer,
  	``Multiparticle one-loop amplitudes and S-duality in closed superstring theory,''
	JHEP {\bf 1310}, 188 (2013).
	[arXiv:1307.3534 [hep-th]].
}

\lref\DHokerPDL{
	E.~D'Hoker and D.~H.~Phong,
  	``The Geometry of String Perturbation Theory,''
	Rev.\ Mod.\ Phys.\  {\bf 60}, 917 (1988).
}

\lref\xerox{
	E.~D'Hoker and D.~H.~Phong,
  	``Conformal Scalar Fields and Chiral Splitting on Superriemann Surfaces,''
	Commun.\ Math.\ Phys.\  {\bf 125}, 469 (1989).
}

\lref\GreenMN{
	M.~B.~Green, J.~H.~Schwarz and E.~Witten,
	``Superstring Theory. Vol. 2: Loop Amplitudes, Anomalies And Phenomenology,''
	Cambridge  University Press (1987).
}

\lref\AnomalyGreen{
	M.B.~Green and J.H.~Schwarz,
  	``The Hexagon Gauge Anomaly in Type I Superstring Theory,''
  	Nucl.\ Phys.\ B {\bf 255} (1985) 93.
\semi
	M.B.~Green and J.H.~Schwarz,
  	``Anomaly Cancellation in Supersymmetric D=10 Gauge Theory and Superstring Theory,''
  	Phys.\ Lett.\ B {\bf 149} (1984) 117.
}

\lref\BernUE{
  Z.~Bern, J.~J.~M.~Carrasco and H.~Johansson,
  ``Perturbative Quantum Gravity as a Double Copy of Gauge Theory,''
Phys.\ Rev.\ Lett.\  {\bf 105}, 061602 (2010).
[arXiv:1004.0476 [hep-th]].
}

\lref\BernYXU{
  Z.~Bern, J.~J.~Carrasco, W.~M.~Chen, H.~Johansson and R.~Roiban,
  ``Gravity Amplitudes as Generalized Double Copies of Gauge-Theory Amplitudes,''
Phys.\ Rev.\ Lett.\  {\bf 118}, no. 18, 181602 (2017).
[arXiv:1701.02519 [hep-th]].
}

\lref\PolchinskiTU{
  	J.~Polchinski and Y.~Cai,
  	``Consistency of Open Superstring Theories,''
	Nucl.\ Phys.\ B {\bf 296}, 91 (1988).
}

\lref\PolchinskiRQ{
  J.~Polchinski,
  ``String theory. Vol. 1: An introduction to the bosonic string,'' Cambridge University Press (2007).
}

\lref\expPSS{
	N.~Berkovits,
  	``Explaining Pure Spinor Superspace,''
  	[hep-th/0612021].
}

\lref\KawaiXQ{
	H.~Kawai, D.~C.~Lewellen and S.~H.~H.~Tye,
  	``A Relation Between Tree Amplitudes of Closed and Open Strings,''
	Nucl.\ Phys.\ B {\bf 269}, 1 (1986).
}

\lref\NMPS{
  	N.~Berkovits,
  	``Pure spinor formalism as an N=2 topological string,''
	JHEP {\bf 0510}, 089 (2005).
	[hep-th/0509120].
}
\lref\SiegelXJ{
	W.~Siegel,
  	``Classical Superstring Mechanics,''
	Nucl.\ Phys.\ B {\bf 263}, 93 (1986).
}

\lref\anomalypaper{
	N.~Berkovits and C.R.~Mafra,
	``Some Superstring Amplitude Computations with the Non-Minimal Pure Spinor Formalism,''
	JHEP {\bf 0611}, 079 (2006).
	[hep-th/0607187].
}
\lref\StiebergerHQ{
  S.~Stieberger,
  ``Open \& Closed vs. Pure Open String Disk Amplitudes,''
[arXiv:0907.2211 [hep-th]].
}
\lref\BjerrumBohrRD{
  N.~E.~J.~Bjerrum-Bohr, P.~H.~Damgaard and P.~Vanhove,
  ``Minimal Basis for Gauge Theory Amplitudes,''
Phys.\ Rev.\ Lett.\  {\bf 103}, 161602 (2009).
[arXiv:0907.1425 [hep-th]].
}

\lref\twoloop{
  N.~Berkovits,
  ``Super-Poincare covariant two-loop superstring amplitudes,''
JHEP {\bf 0601}, 005 (2006).
[hep-th/0503197].
\semi
  N.~Berkovits and C.R.~Mafra,
  ``Equivalence of two-loop superstring amplitudes in the pure spinor and RNS formalisms,''
Phys.\ Rev.\ Lett.\  {\bf 96}, 011602 (2006).
[hep-th/0509234].
\semi
  H.~Gomez and C.R.~Mafra,
  ``The Overall Coefficient of the Two-loop Superstring Amplitude Using Pure Spinors,''
JHEP {\bf 1005}, 017 (2010).
[arXiv:1003.0678 [hep-th]].
}

\lref\vallilo{
  N.~Berkovits and B.~C.~Vallilo,
  ``Consistency of superPoincare covariant superstring tree amplitudes,''
JHEP {\bf 0007}, 015 (2000).
[hep-th/0004171].
}

\input labelII.defs
\input labelIII.defs

\listtoc
\writetoc
\filbreak

\newsec{Introduction}

This is the first part of a series of papers \wipI\
towards the derivation of $n$-point one-loop correlators of open- and 
closed-superstring states using the pure-spinor formalism \refs{\psf,\MPS}.
When we refer to section and equation numbers from the papers II and III,
these numbers will be prefixed by the roman numerals II and III accordingly.

A variety of recent developments revealed hidden simplicity and unexpectedly
rich structures in scattering amplitudes of string theories. Many of these
findings can be attributed to the boost in computational reach due to the
manifestly supersymmetric pure-spinor formalism
\refs{\psf\MPS\vallilo\twoloop\PScohomology{--}\NMPS}. For instance,
pure-spinor methods enabled the first three-loop computation in the low-energy
limit of the four-point closed-string amplitude \threeloop\ and gave rise to a
strikingly compact form of $n$-point tree-level amplitudes
\refs{\nptString,\nptFT}.

The pure-spinor computation at tree level paved the way for string-theory
realizations and extensions of recent unifying relations among field-theory
amplitudes. For instance, using a local representation of the massless $n$-point disk
correlation function, manifestly local numerators satisfying the
color-kinematics duality \BernQJ\ in gauge-theory amplitudes were
systematically constructed \MafraKJ\foot{Also see
\refs{\BjerrumBohrRD,\StiebergerHQ} for a string-theory derivation of the
resulting Bern--Carrasco--Johansson relations among color-ordered gauge-theory
amplitudes at tree level.}. Locality refers to the absence of kinematic poles
in the superspace kinematic factors of the correlator: All the propagators in
the color-kinematics dual gauge-theory amplitudes stem from the field-theory
limit of the moduli-space integrals over disk worldsheets. {\it The first main
goal of this series of papers is to generalize these results to loop level and to
construct local correlators on a genus-one surface}.

Moreover, the $n$-point disk amplitudes \nptString\ were later on found in
\BroedelTTA\ to share the structure of the Kawai--Lewellen--Tye (KLT) formula
\KawaiXQ\ for supergravity trees.  In relating the open superstring to
supergravity, one copy of the color-ordered gauge-theory trees in the KLT
formula are mapped to so-called Parke--Taylor integrands $(z_{12} z_{23}\ldots
z_{n1})^{-1}$ with $z_{ij}=z_i{-}z_j$. This mapping rests on the fact that disk
integrals of Parke--Taylor type share the Bern--Carrasco--Johansson (BCJ)
relations \BernQJ\ of gauge-theory tree amplitudes \BroedelTTA. We will refer to
these phenomena as a double-copy structure of disk amplitudes and a duality
between kinematics and worldsheet functions. {\it The second main goal of this
series of papers is to find a one-loop incarnation of the duality between kinematics and
worldsheet functions that results in a double-copy structure of open-superstring
amplitudes \MafraIOJ}.

While tree-level correlators are completely determined by their singularity
structure encoded in the OPEs of the vertex operators, the quest for genus-one
correlators is guided by additional constraints: The homology cycles of the
genus-one surface translate into a notion of double-periodicity in its complex
coordinate. As we will see in part II, double-periodicity does not hold term-by-term in
the genus-one correlators. Instead, the monodromies of individual terms cancel
in similar patterns as the BRST variations of kinematic factors in pure-spinor
superspace\foot{BRST-invariance of a kinematic factor in pure-spinor
superspace implies its components to be both gauge invariant and
supersymmetric \psf.}. This will not only be a crucial guiding principle in
constructing local representations of genus-one correlators in part III but also furnish a
key incarnation of the duality between kinematics and worldsheet functions.

Apart from the parallels in their BRST- and monodromy variations, the
kinematic building blocks and worldsheet functions in this work resonate in
their symmetry properties under exchange of external legs. In a local form of
the correlators, kinematic building blocks exhibit Lie-symmetries in several
groups of labels, which translate into kinematic Jacobi relations in the
tree-level subdiagrams of the field-theory limit \refs{\MafraKJ, \EOMbbs,
\towardsoneloop}. The worldsheet functions of part II in turn are designed to vanish
under shuffle products in several groups of labels which is reminiscent of the
Kleiss--Kuijf relations of gauge-theory tree amplitudes \KK. Combinations of
Lie- and shuffle symmetric objects are tailor-made to realize the permutation
invariance of the correlators. At the same time, this symmetry structure is
well known in the mathematics literature from a theorem by Ree \Ree\ in the
context of {\it Lie polynomials} \reutenauer. Therefore we say that the local
one-loop correlators of part III have a {\it Lie-polynomial} structure\foot{Note that this
same Lie-polynomial structure is already present in the calculation of the
tree-level correlator in \nptString, but it remained unnoticed until now.}.

We will also explore manifestly BRST-invariant but non-local representations
of the correlators in part III, where the kinematic building blocks are dressed with the
tree-level propagators. The resulting supersymmetric Berends--Giele currents
\refs{\EOMbbs, \Gauge} and their BRST-invariant combinations \partI\ realize
shuffle symmetries on the kinematic side. Similarly, monodromy invariance of
the correlators can be manifested by organizing the worldsheet functions into
so-called {\it generalized elliptic integrands} (GEIs), see \MafraIOJ\ and part II. In contrast
to conventional elliptic functions, GEIs may involve loop momenta of the
chiral-splitting formalism \refs{\verlindes,\DHokerPDL,\xerox} that transform
as well when punctures are taken along the homology cycle and cancel the
monodromies of Jacobi theta functions.

In part III, we will present expressions for $(n\leq 7)$-point correlators in terms of
BRST-invariant superfields and GEIs such that both kinds of invariances are
manifest. Given that BRST-invariant superfields and GEIs are shown in part II to obey the
same kinds of relations, the role of kinematics and worldsheet functions can
be freely interchanged. This generalizes the $(n{-}3)!$-term representations
of tree-level correlators \nptString, where gauge-theory trees and
Parke--Taylor factors enter on symmetric footing \BroedelTTA. In analogy to
these tree-level results, the one-loop amplitudes computed from such
correlators are said exhibit a double-copy structure \MafraIOJ. In the same
way as the double-copy representations of gravitational loop integrands
\refs{\BernUE, \BernYXU} hinges on the color-kinematics duality in gauge
theories, the duality between kinematics and worldsheet function reveals a
double-copy structure in one-loop open-superstring amplitudes.

A brief executive summary for the combined parts \wipI\
of this series is as follows. In the first sections we will have
self-contained discussions to set up preliminary notions regarding the pure-spinor 
formalism (section~\basicsec), local superfields (sections~\SYMsec\ and
\LocalBBsec), non-local superfields (section~\BRSTsec) and one-loop 
worldsheet functions (section~\ZEsec). Several
important relations and interplays among these first sections are pointed out
and thoroughly illustrated too. For instance, section~\dualitysec\ discusses
several parallels and dualities between the non-local kinematic building
blocks of section~\BRSTsec\ and worldsheet functions that are built from the
constituents in section~\ZEsec. Then, in section~\secexper, after a brief
discussion pointing out the Lie-polynomial structure of the local $n$-point
{\it tree-level} correlators, we will argue that also the local $n$-point {\it
one-loop} correlators of the open superstring have a Lie-polynomial form.
Namely,
\eqn\Kintro{
\cK_n(\ell) = \sum_{r=0}^{n-4}{1\over r!}\Big(
V_{A_1}T^{m_1 \ldots m_r}_{A_2, \ldots, A_{r+4}}\cZ^{m_1 \ldots m_r}_{A_1, \ldots, A_{r+4}}
+ \big[12 \ldots n|A_1, \ldots,A_{r+4}\big]\Big) + \hbox{ corrections}\, ,
}
where $V_A$ and $T^{m_1 \ldots m_r}_{A_1, \ldots,A_{r+3}}$ are local kinematic
building blocks satisfying Lie symmetries while $\cZ^{m_1 \ldots m_r}_{A_1,
\ldots,A_{r+4}}$ are functions on the genus-one worldsheet satisfying shuffle
symmetries (the unconventional notation for the permutations is explained in
detail in the appendix~\stirlingapp).

The need for ``+ corrections'' at $n\geq 7$ points will be elaborated in
detail invoking e.g., locality, BRST invariance, single-valuedness and several
other related technical aspects introduced in the first sections.  In
section~\explsec, a multitude of representations for the correlators with
$n=4,5,6,7$ is presented. The $n=8$ correlator is proposed and, while it
satisfies many non-trivial constraints, it fails to be BRST invariant by terms
proportional to the holomorphic Eisenstein series ${\rm G}_4$. Unfortunately,
this points to a certain weakness of our method since any Eisenstein
series is a monodromy-invariant function with no dependence on the worldsheet
punctures. (We leave it as an open challenge for future work to determine 
the kinematic coefficients of ${\rm G}_k$ in $(n\geq 8)$-point correlators.) Further
representations for the correlators are presented in section~\loopintsec,
which is concerned with the explicit integration of the loop momentum in both
open- and closed-string one-loop amplitudes. Several rather technical
discussions are left to the appendices. 

Finally, one should not be overwhelmed
by the total number of pages of this series; the wide areas of both
mathematics and physics that it touches lead to several relationships and
beautiful connections. The final results for the correlators are in fact quite
compact.

\newnewsec\basicsec Basic formalism

In this section we will review certain aspects of the one-loop amplitude
prescription in the minimal\foot{See \NMPS\ for the one-loop amplitude
prescription in the non-minimal pure-spinor formalism.} pure-spinor formalism
\MPS. In the later sections, this prescription will be used as a basis to
formulate a general approach to assemble integrands for $n$-point
open-superstring amplitudes at one loop from standard constraints such as
single-valuedness and BRST invariance, among others.

\subsubsec Conventions

Throughout this work, we will use the shorthands
\eqn\KNKNG{
z_{ij} \equiv z_i - z_j,
}
for the worldsheet positions and
\eqn\multmom{
 k^m_\emptyset\equiv0\,,\quad \quad k^m_{12\ldots p} \equiv k_1^m +k_2^m 
 + \cdots + k_p^m\,,\quad \quad
s_{12\ldots p} \equiv {1\over 2} k^2_{12\ldots p} = \sum_{i<j}^p s_{ij}
}
for multiparticle momenta and Mandelstam invariants, where $s_{ij}=(k_i\cdot
k_j)$. Vector and spinor indices of the ten-dimensional Lorentz group are
denoted by $m,n,\ldots=0,1,\ldots,9$ and $\alpha,\beta,\ldots=1,2,\ldots,16$,
respectively.

Our convention for (anti)symmetrizing $r$ vector indices does {\it not}
include a factor of ${1\over r!}$ and it always generates unit coefficients
for each inequivalent term, even in the presence of symmetric tensors; 
for instance $A^{(m}B^{n)} \equiv A^m B^n + A^n B^m$ as well as
\eqn\controversy{
\d^{(mn}k^{p)}\equiv \d^{mn}k^p + \d^{mp}k^n + \d^{np}k^m\,.
}

\newsubsec\purespinorsec The pure-spinor amplitude prescription

The prescription to compute $n$-point one-loop amplitudes
for open superstrings is \MPS\
\eqn\onepresc{
{\cal A}_n =
\sum_{\rm top} C_{\rm top} \int_{D_{\rm top}}\!\!\!\!
d\tau \,\langle\!\langle (\mu, b)\,\cZ\,
V_1(z_1)\prod_{j=2}^n \int\, dz_j\, U_j(z_j)\rangle\!\rangle\,,
}
where the Beltrami differential $\mu$ and the modulus $\tau$ encode the
topological information of the genus-one surface. The sum runs over all
open-string worldsheet topologies at one loop: the planar and non-planar
cylinder as well as the M\"obius strip. For each topology the integration
domain $D_{\rm top}$ for the modulus $t$ and the color factors $C_{\rm top}$
have to be adjusted and the region of integration over the $z_j$ variables
must reflect the ordering of the vertex operators insertions on its boundaries
\GreenMN, see section~\seconeonezero\ below. Using translation invariance of
the path integral, the position $z_1$ can be fixed to $z_1=0$, but it is
customary to carry it unfixed in the formulas.

Moreover, the prescription \onepresc\ uses picture-changing operators\foot{The
ingredients of ${\cal Z}=  Z_J \prod_{P=2}^{10} Z_{B_P} \prod_{I=1}^{11} Y_{C_I}$
are explained in \MPS.} collectively denoted by
${\cal Z}$ and detailed in \MPS\ as well as a composite $b$-ghost of schematic form
\eqnn\bghost
$$\eqalignno{
b & = (\Pi d + N\p\t + J\p\t)\,d\,\d(N) + (w\p\l + J\p N + N\p J + N\p N)\d(N) &\bghost \cr
&\quad{}+ (N\Pi + J\Pi + \p\Pi + d^2)(\Pi\d(N) + d^2\d'(N))\cr
&\quad{}+ (Nd + Jd)(\p\t\d(N) + d\Pi\d'(N) + d^3\d''(N))\cr
&\quad{}+ (N^2 + JN + J^2)(d\p\t\d'(N) + \Pi^2\d'(N) + \Pi d^2 \d''(N) + d^4 \d'''(N)) \ ,
}$$
where the worldsheet fields on the right-hand side will be introduced below.
The complicated expression \bghost\ poses difficulties in a direct evaluation
of \onepresc\ at multiplicities above four, especially when OPE contractions
involving the $b$-ghost $b(z_0)$ are considered. For the five-point
correlator, these contributions were shown to be total worldsheet derivatives
with respect to $z_0$ and therefore could be dropped in the integrated
amplitude \fiveptNMPS. However, starting at six-points, the OPE contributions
involving the $b$-ghost may introduce non-trivial functions of the punctures
into the correlator although the dependence on $z_0$ ultimately drops
out\foot{An explicit example of a related cancellation can be found in
appendix B of \MafraNWR\ in the parity-odd sector of one-loop amplitudes in
the RNS formalism.}. The expressions for one-loop correlators to be proposed
in the later sections provide evidence that such OPE contractions can be
reduced to zero-mode contributions.

When all the external states are massless, the unintegrated vertex operator
$V_i(z_i)$ of conformal weight zero is given by\foot{For ease of notation, the
dependence on $z_i$ via $\l^\alpha(z) ,x^m(z), \theta^\alpha(z)$ as well as
$\p\t^\a(z),\Pi^m(z)$, $d_\a(z),N^{mn}(z)$ is left implicit on the right-hand
side of \Vvertex, \integratedU\ and later equations.}
\eqn\Vvertex{
V_i(z_i) = \l^\alpha A^i_\alpha(x,\t) \ ,
}
while the integrated vertices $U_i(z_i)$ are
\eqn\integratedU{
U_i(z_i) = \p\t^\a  A^i_\alpha(x,\t) + \Pi^m A^i_m(x,\t)
+ d_\a  W_i^\a(x,\t) + \half N^{mn} F^i_{mn}(x,\t)\,.
}
The vectorial and spinorial polarizations of the $i^{\rm th}$ gluon and gluino
as well as their light-like momenta $k_i$ enter through the ten-dimensional
super-Yang--Mills superfields $[A^i_\a,A^i_m,W_i^\a,F_i^{mn}]$ to be
reviewed in section \SYMsec.

The bosonic pure spinor $\l^\alpha$ in \Vvertex\ has conformal weight zero and obeys
\eqn\puresp{
 (\l \gamma^m \l)=0 \ ,
}
where $\gamma^m_{\alpha \beta} = \gamma^m_{\beta \alpha}$ denote the $16\times
16$ Pauli matrices of $SO(1,9)$ subject to the Clifford algebra
$\gamma^{(m}_{\alpha \beta} \gamma^{n) \beta \gamma} = \delta^{mn} 
\delta^\gamma_\alpha$. Note that the symmetry properties of antisymmetrized
Pauli matrices are $\gamma^{mnp}_{\alpha \beta} = -\gamma^{mnp}_{\beta
\alpha}$ and $\gamma^{mnpqr}_{\alpha \beta} = \gamma^{mnpqr}_{\beta \alpha}$
for odd rank as well as $\gamma^{mn}{}_{\alpha}{}^{ \beta} = -\gamma^{mn
\,\beta}{}_{ \alpha}$ and $\gamma^{mnpq}{}_{\alpha }{}^{\beta} = \gamma^{mnpq
\,\beta}{}_{ \alpha}$ for even rank.

Integrated vertices \integratedU\ involve the Lorentz-current $N^{mn}$ of the
pure-spinor ghost, and $\Pi^m = \partial x^m+ {1\over 2}(\theta \gamma^m
\partial \theta)$ as well as $d_\alpha$ are supersymmetric combinations of the
matter variables in the pure-spinor worldsheet action \psf.

\newsubsubsec\seconeonezero Open-string integration domains

The vertex-operator locations or {\it punctures} need to be integrated over
the torus or over the boundary components of the cylinder and the M\"obius
strip. For open strings, the integration over the boundaries has to match the
cyclic orderings of the accompanying color factors: Each external state
carries color degrees of freedom encoded in Lie-algebra\foot{For the type-I
superstring, the gauge group has to be chosen as $SO(32)$ in order to
guarantee cancellation of infinities \GreenED\ and gauge anomalies \AnomalyGreen.}
generators $t^a$, and the color dependence of the amplitude enters via traces.

Given a torus of modular parameter $\tau$ in the parameterization of figure
\torusparam, one can obtain the open-string worldsheets via suitable involutions
\refs{\AntoniadisVW, \PolchinskiRQ}, resulting in a purely imaginary modular
parameter $\tau$ in case of the cylinder. The cylinder boundaries ${\cal
B}_{1,2}$ will be taken to be the $B$-cycle through the origin and the point
$z={1\over 2}$, respectively,
\eqn\basicsA{
{\cal B}_{1} = \{ z=i\nu , \ 0\leq \nu \leq |\tau| \} 
\ , \ \ \ \ \ \ 
{\cal B}_{2} = \{ z={1\over 2}+i\nu , \ 0\leq \nu \leq |\tau| \} \, ,
}
i.e. they are separated by half the $A$-cycle. After using translation
invariance to fix one puncture as $z_1=0$, the integration domain associated
with the single trace ${\rm tr}(t^1 t^2 \ldots t^n)$ is characterized by
$0<\Im(z_2)<\Im(z_3)<\ldots <\Im(z_n)<\Im(\tau)$, and similar choices can be
made for the M\"obius strip \refs{\AntoniadisVW, \PolchinskiRQ}. Double traces
${\rm tr}(t^1 t^2 \ldots t^j) {\rm tr}(t^{j+1} \ldots t^n)$ are exclusive to the
cylinder topology, and one may define analogous domains where the two cyclic
orderings are implemented on ${\cal B}_{1,2}$ in \basicsA.

Additionally, the modular parameters need to be integrated, e.g.\ over $\tau \in
i \Bbb R_+$ in case of the cylinder. As indicated in \onepresc, the
integrations over the different topologies along with their color traces
are denoted by the subscript $_{\rm top}$.

\subsubsec Functional integration and OPEs

The worldsheet fields $[\p\t^\a,\Pi^m,d_\a,N^{mn}]$ in the integrated vertices
\integratedU\ have conformal weight $+1$ and can be integrated after separating
off the zero modes. Using $d_\a(z)$ as an example, in a genus-$g$ surface one
gets
\eqn\zerom{
d_\a(z) = \sum_{I=1}^g d^I_\a \om_I(z) + \hat d_\a(z) \ ,
}
where $\om_I(z)$ are $g$ holomorphic one-forms normalized as
$\oint_{A_I}dz \, \om_J(z) = \d_{IJ}$ when integrated around the
$A$-cycles. The non-zero modes $\hat d_\a(z)$ in turn are characterized
by $\oint_{A_I}dz \, \hat d_\a(z) = 0$. In addition, when
the holomorphic one-forms are integrated around the $B$-cycles one
gets the period matrix $\Omega_{IJ}=\oint_{B_I}dz \, \om_J(z)$.
Note that $\om_1(z)=1$ at genus one, and the parameterization of the
torus in \torusparam\ involves the period matrix $\oint_{B}dz = \tau$ with
$\tau\in \Bbb C$ and $\Im \tau >0$.


\ifig\torusparam{Parameterization
of the torus through the lattice $\Bbb C/(\Bbb Z {+} \tau \Bbb Z)$
with an identification
of points $z$ with their translates $z{+}1$ and $z{+}\tau$ along the $A$- and $B$-cycle.}
{\epsfxsize=0.70\hsize\epsfbox{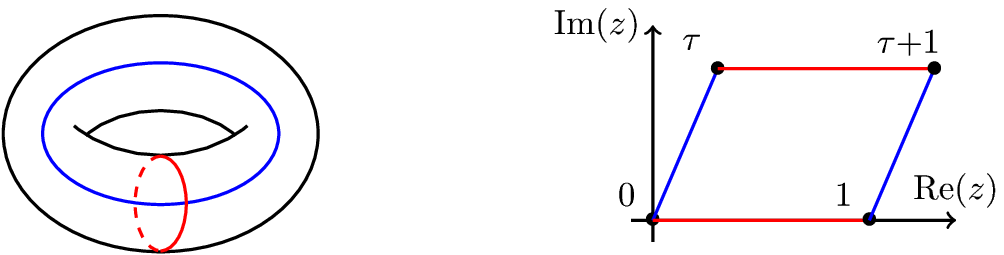}}

The non-zero modes are functionally integrated through
OPE contractions, in particular at genus one we have
\eqn\allOPEs{
\eqalign{
\hat d_\a(z) K(0)     &\rightarrow  D_\a K(0)\, g^{(1)}(z,\tau),\cr
\hat d_\a(z) \hat\Pi^m(0) &\rightarrow  (\ga^m\p\t(0))_\a\, g^{(1)}(z,\tau)\cr
\hat d_\a(z)\t^\b(0)  &\rightarrow  \d^\b_\a\, g^{(1)}(z,\tau)\cr
}\quad\eqalign{
\hat\Pi^m(z) K(0)    &\rightarrow -  k^m K(0)\, g^{(1)}(z,\tau),\cr
\hat d_\a(z)\hat d_\b(0)    &\rightarrow -  \ga^m_{\a\b}\Pi_m(0)\,g^{(1)}(z,\tau),\cr
\hat N^{mn}(z)\l^\a(0) &\rightarrow - \half (\l(0)\ga^{mn})^\a\, g^{(1)}(z,\tau) \ ,\cr
}}
where the worldsheet singularities are captured by \MPS\ ($\p\equiv {\p\over\p z}$)
\eqn\Fzw{
g^{(1)}(z,\tau)
\equiv \p \log \t_1(z,\tau)\,,
}
and the standard odd Jacobi theta function with $q\equiv \exp(2\pi i\tau)$ is given by
\eqn\jactheta{
\theta_{1}(z,\tau) \equiv  2q^{1/8}\sin(\pi z)  \! \prod_{n=1}^{\infty}
(1-q^n)
\bigl(1 - q^ne^{2\pi i z}\bigr)\bigl(1 - q^{n}e^{-2\pi i z}\bigr)\,.
}
In the above OPEs, $K(z)$ represents a generic superfield which depends on
$\t^\a(z)$ but not on any derivative $\p^{n}\t^\a(z)$ for $n{\geq} 1$, and whose
$x$ dependence is entirely contained in the plane-wave factor\foot{To avoid
factors of $i$ in the formulae, we depart from the standard conventions of
real momenta and redefined $ik^m\rightarrow k^m$.} $e^{k\cdot x}$.
The functional integration of the variables $x^m(z,\bar z)$ gives rise to the
so-called Koba--Nielsen factor and will be reviewed in the next subsection.
The above OPEs can be read off from their tree-level counterparts
\refs{\SiegelXJ, \psf} via the substitution ${1\over z}\rightarrow
g^{(1)}(z,\tau)$.

It turns out that the four-point amplitude computed with the pure-spinor
formalism does not involve any OPE contraction; its outcome is based purely on
zero-mode integrations \MPS. Therefore, the $n$-point amplitude admits at most
$n{-}4$ OPE contractions, and it will be explained in the later
sections~\secthree\ and \loopintsec\ that this gives rise to modular forms of weight $n{-}4$
after integrating over the zero modes of $\Pi^m$ in \onepresc.

\newsubsubsec\twotwosec Zero-mode integrations

The non-zero modes of the worldsheet fields are integrated out using OPE
contractions \allOPEs, and a systematic procedure to capture the resulting tensor
structures will be reviewed in section
\SYMsec. Similarly, as in the tree-level correlator, we will assume that the OPE
residues which feature double poles have been absorbed into single-pole residues
using integration-by-parts identities \nptString. The net effect of such
manipulations can be accounted for via multiparticle superfields in the BCJ
gauge \Gauge\ to be reviewed in subsection \multisec.

Once all non-zero modes are integrated out in that manner, the correlator
\onepresc\ will depend only on the zero modes of the worldsheet variables
$[d_\a(z), N^{mn}(z), \Pi^m(z)]$ and $[\l^\a(z),\t^\a(z)]$. In this section we
outline a practical procedure to integrate them using the pure-spinor zero-mode
measures defined in \MPS. Unlike the other worldsheet fields, the zero mode of
$\Pi^m(z)$ is denoted with a different symbol $\ell^m$, called the {\it loop
momentum}. Its integration will be discussed in the context of {\it chiral
splitting}, see subsection~\chiralsplitsec.

As explained in \MPS, in performing the path integral of the prescription
\onepresc\ the role of the picture-changing operators ${\cal Z}$ is to ensure
that all bosonic and fermionic zero modes are absorbed correctly. In doing this
it will be convenient to start integrating out $d_\a$ and $N^{mn}$ while leaving
the zero modes of $\Pi^m$, $\l^\a$ and $\t^\a$ to be dealt with at later stages.
The reason for this is that the integrations over $d_\a$ and $N^{mn}$ can be
performed, under mild assumptions, using group-theory arguments alone. This will
lead to effective rules which will then be used as input on section~\LocalBBsec\
to define local kinematic building blocks based on the multiparticle SYM
superfields of section~\SYMsec.

To see how this comes about, the first thing to note is that in integrating out
the zero modes $d_\a$ and $N^{mn}$ using the pure-spinor measures of \MPS\ one
introduces two pure spinors $\l^\a$ into the rest of the path integral. In
addition, since the picture-changing operators contribute a fixed number of ten
fermionic $d_\a$ zero modes, an additional six zero modes must come from the
$b$-ghost and the external vertices \integratedU. We will consider the
contributions from two classes of terms in the $b$-ghost \bghost, given by
$b^{(2)}\equiv \Pi d^2\d(N)$ and $b^{(4)}\equiv d^4 \d'(N)$.

When the $b$-ghost contributes $b^{(4)} \equiv d^4 \d'(N)$, the external vertices
must provide the remaining two $d_\a$ zero modes. However, there must also be a
$N^{mn}$ zero mode due to the factor of $\d'(N)$ \MPS. Therefore, the
non-vanishing configuration of zero modes from the external vertices must be
proportional to $d_\a d_\b N^{mn}$. Given the expression \integratedU\ for
integrated vertex operators, this contribution is of the schematic form
$U_2U_3U_4 \rightarrow d_\a d_\b N_{mn} W_2^\a W_3^\b F^{mn}_4$, see section
\multisec\ for multiparticle generalizations. Taking into account the
introduction of two pure spinors from the pure-spinor measures, the integration
of the factor $d_\a d_\b N_{mn} W_2^\a W_3^\b F^{mn}_4$ from the external
vertices can be summarized as
\eqn\efrule{
\int d^4 \d'(N)
d_\a d_\b N^{mn}\rightarrow (\l\ga^{[m})_\a(\l\ga^{n]})_\b\,,
}
where the integral sign represents the integration using the
zero-mode measures of \MPS.
Up to an overall coefficient, this is the unique
outcome because there is only one two-form
irreducible representation in the tensor product of two pure spinors and two
Weyl spinors\foot{We acknowledge the use
of the {\tt LiE} program in these decompositions \LiE.}
\eqn\tensorp{
[0,0,0,0,2]\otimes
[0,0,0,0,1]^{\wedge 2} = [0,1,0,0,0] + \cdots.
}
Hence, the net contribution from this sector to the correlator is given by
a unique Lorentz-scalar combination of superfields
\eqn\just{
\int b^{(4)} U_2 U_3 U_4 \rightarrow
(\l\ga_{m}W_2)(\l\ga_{n}W_3)F^{mn}_4 + {\rm cyc}(2,3,4)\,.
}
Similarly, when the $b$-ghost contribution comes from the term $b^{(2)}\equiv
\Pi d^2\d(N)$, the external vertices must provide four
$d_\a$ zero modes, and this time there is no need for an
additional $N^{mn}$. Therefore, $b^{(2)}$ requires the external vertices 
to contribute $d_\a d_\b d_\ga d_\d W_2^\a W_3^\b W_4^\ga W_5^\d$, 
see section \multisec\ for multiparticle generalizations.
One can show that, up to an overall coefficient, the effective rule
for integrating these zero modes is given by
\eqn\efruletwo{
\int \Pi d^2\d(N)
d_\a d_\b d_\ga d_\d \rightarrow
\ell_m (\l\ga^{a})_{[\a}(\l\ga^{b})_\b(\ga^{abm})_{\ga\d]}\,.
}
The argument to see this is similar to \tensorp:
there is a single vector representation
in the tensor product of $\l^2 W^4$,
\eqn\moretensorp{
[0,0,0,0,2]\otimes
[0,0,0,0,1]^{\wedge 4} = [1,0,0,0,0] + \cdots \,.
}
Therefore the
unique Lorentz-invariant overall contribution from this zero-mode
sector can be summarized by the following superfield combination
\eqn\Wfour{
\int b^{(2)} U_2U_3U_4U_5 \rightarrow
\ell_m (\l\ga_a W_2)(\l\ga_b W_3)(W_4 \ga^{abm} W_5) + {\rm perm}(2,3,4,5)\,.
}
In section~\LocalBBsec\ we will see how \Wfour\ motivates the
introduction of tensorial local building blocks that capture
the kinematics of one-loop correlators.

The above rules are readily generalized for additional
instances of zero modes of $\Pi^m$,
\eqnn\efruleloop
\eqnn\efruletwoloop
$$\eqalignno{
\int \Pi^{m_1} \Pi^{m_2} \ldots \Pi^{m_r} d^4 \d'(N)
d_\a d_\b N^{pq} &\rightarrow \ell^{m_1} \ell^{m_2} \ldots \ell^{m_r}
(\l\ga^{[p})_\a(\l\ga^{q]})_\b\,, &\efruleloop\cr
\int \Pi^{m_1}  \ldots \Pi^{m_r}  \Pi d^2\d(N)
d_\a d_\b d_\ga d_\d &\rightarrow \ell^{m_1} \ldots \ell^{m_r} 
\ell_n (\l\ga^{a})_{[\a}(\l\ga^{b})_\b(\ga^{abn})_{\ga\d]}\,.\qquad{}&\efruletwoloop
}$$
The analogues of \just\ and \Wfour\ for the remaining terms of the $b$-ghost \bghost\
besides $b^{(2)}$ and $b^{(4)}$ are currently out of reach. Instead, we will infer their
contributions to one-loop correlators from first principles to be detailed in section
\FinalAssemblysec. Up to integration-by-parts equivalent terms, $b^{(2)}$ and $b^{(4)}$
provide the highest numbers of zero modes of $d_\alpha,N^{mn}$ and therefore
start to contribute at the lowest multiplicities.
Using these zero-mode considerations it follows that the loop integrand for
$n$-point open-string amplitudes \onepresc\ is a {\it polynomial} of degree
$n{-}4$ in the loop momentum $\ell$.

\newsubsubsec\psssec Pure-spinor superspace

The angle brackets $\langle \! \langle\ldots \rangle \! \rangle$ in the
amplitude prescription \onepresc\ represent the complete path integral over all
the worldsheet degrees of freedom.  After integrating the zero modes of $d_\a$,
$N^{mn}$ and all the other variables except for $\l^\a$ and $\t^\a$, these
$\langle \! \langle\ldots \rangle \! \rangle$ are replaced by $\langle \ldots
\rangle $ which represent the remaining functional integration over zero modes
of $\l^\a$ and $\t^\a$. In integrating the variables in this order, the
kinematic factors become expressions in {\it pure-spinor superspace\/} as
defined in \expPSS. Pure-spinor superspace compactly encodes all states in the
supermultiplet, and the components can be extracted using the prescription \psf
\eqn\psfpresc{
\langle (\l \ga^m \t)(\l \ga^n \t)(\l \ga^p \theta) (\t\ga_{mnp} \t)\rangle=2880
}
for integration over zero modes of $\l^\a$ and $\t^\a$. In fact, the amplitudes
exhibit their most convenient form when written in pure-spinor
superspace, i.e.\ {\it without} performing the integration in \psfpresc,
and will be represented as such in this series of papers.

A key feature of the measure \psfpresc\ is its interplay with the BRST operator
\eqn\BRST{
Q \equiv \lambda^\alpha D_\alpha \ , \ \ \ \ \ \ 
D_\alpha \equiv {\partial \over \partial \theta^\alpha}
+ {1\over 2} (\gamma^m \theta)_\alpha {\partial \over \partial x^m} .
}
As pointed out in \psf, the measure \psfpresc\ is only sensitive to the 
cohomology of $Q$: BRST-closed superfields $Q S(x,\theta,\lambda)=0$ are mapped
to gauge invariant and supersymmetric components in $\langle
S(x,\theta,\lambda)\rangle$, whereas BRST-exact terms decouple, i.e.  $\langle E
(x,\theta,\lambda) \rangle =0$ if $E (x,\theta,\lambda) = Q
\Sigma(x,\theta,\lambda)$. This cohomology structure can be exploited to obtain
non-trivial relations among seemingly different superspace expressions including
amplitudes at different loop orders \mafraids.

\newsubsec\chiralsplitsec Chiral splitting of the Koba--Nielsen factor

The zero-mode integrations of the matter variables $x^m(z,\bar z)$ or
equivalently $\Pi^m(z)$ is performed employing the techniques of the {\it
chiral-splitting} formalism of \refs{\verlindes,\DHokerPDL,\xerox}. The idea is
to defer the zero-mode integration for $\Pi^m$ within the path integral in
\onepresc\ to the last step of the amplitude computation\foot{More formally,
chiral splitting is implemented by inserting the integrated delta function
$1=\int d^D \ell \ \delta^D(\ell^m - \oint_{A}dz \, \Pi^m(z))$ into the path
integral.} and to interpret it as a string-theory antecedent of the loop
momentum in Feynman integrals, to be denoted by
\eqn\Pizero{
\ell^m \equiv \oint_{A}dz \, \Pi^m(z)\,.
}
In this setting, the contributions from the plane-wave factors $e^{k\cdot x}$
can be reproduced from the {\it Koba--Nielsen factor}
\eqn\KNKNA{
{\cal I}_n(\ell) \equiv  \exp\Big( \sum^n_{i<j} s_{ij} \log \theta_1(z_{ij},\tau)
+  \sum_{j=1}^n  z_j(\ell\cdot k_j) + {\tau \over 4\pi i} \ell^2 \Big)\,,
}
and our notation ${\cal I}_n(\ell)$ for the Koba--Nielsen factor
omits its dependence on the variables $z_j,k_j,\tau$.
Chiral splitting can be easily undone:

In a closed-string context, the loop integration comprised by the path integral
$\langle\!\langle\ldots\rangle\!\rangle_{\rm closed} $ over left- and
right movers reproduces the more conventional and modular invariant form of the
Koba--Nielsen factor,
\eqnn\KNKNB
$$\eqalignno{
\hat {\cal I}_n &= \Big \langle \! \! \Big \langle
\prod_{j=1}^n e^{k_j \cdot x(z_j,\bar z_j)}
\Big \rangle \! \! \Big \rangle_{\rm closed}   =
\int d^{D} \ell \ \big|{\cal I}_n(\ell)  \big|^2 &\KNKNB \cr
&= {(2\pi i)^D\over (2 \Im \tau)^{{D\over 2}}}
\exp\Big(\sum^n_{i<j}s_{ij} \Big[\log\big| \theta_1(z_{ij},\tau)  \big|^2
- {2\pi \over \Im \tau} (\Im z_{ij})^2 \Big]\Big)\,.
}$$
On the one hand, the combination $\log\big| \theta_1(z,\tau) \big|^2 - {2\pi
\over \Im \tau} (\Im z)^2$ in the exponent exhibits double periodicity under
translations $z\rightarrow z+1$ and $z\rightarrow z+\tau$ around the homology
cycles of the Riemann surface. On the other hand, its second term $\sim {(\Im
z_{ij})^2 \over \Im \tau} $ obstructs holomorphic factorization of the
moduli-space integrand for closed-string amplitudes.

In an open-string context, the path integral $\langle \! \langle \ldots \rangle \! \rangle$ 
in \onepresc\ only comprises half the non-zero modes of $x^m$ as compared to its
closed-string counterpart $\langle \! \langle \ldots \rangle \! \rangle_{\rm closed}$
in \KNKNB. Accordingly, the plane-wave correlator of the open string yields half
of the Koba--Nielsen exponent,
\eqnn\KNKNBop
$$\eqalignno{
\hat {\cal I}^{\rm open}_n &=
 \Big \langle \! \! \Big \langle   \prod_{j=1}^n e^{k_j \cdot x(z_j,\bar z_j)}
 \Big \rangle \! \! \Big \rangle  =
 \int d^{D} \ell \ \big|{\cal I}_n(\ell)  \big| &\KNKNBop
  \cr
&= {(2\pi i)^D \over (\Im \tau)^{{D\over 2}}} \exp \Big( \sum^n_{i<j}s_{ij} \Big[ \log \big|  \theta_1(z_{ij},\tau) \big|
 - { \pi (\Im z_{ij})^2 \over \Im \tau}  \Big]\Big)  \ ,
}$$
while the loop integration is the same as in \KNKNB\ since the loop momentum is
a joint zero mode of left- and right movers in $\langle \! \langle \ldots
\rangle \! \rangle_{\rm closed}$. The purpose of writing the last line of
\KNKNBop\ in terms of $\Im z_j$ and $\Im \tau$ is to have a universal expression
for all the topologies of open-string amplitudes in \onepresc: While the
punctures and modulus of a planar cylinder diagram are accounted for by purely
imaginary choices of $z_j,\tau$, the non-planar cylinder and the M\"obius strip
also introduce real parts for some of $z_j$, $\tau$, see e.g.\
\refs{\AntoniadisVW, \PolchinskiRQ}. Still, one has to keep in mind that there
is no distinction between holomorphic and antiholomorphic variables in an
open-string setup when taking total derivatives of the Koba--Nielsen factor
\KNKNBop. Accordingly, open and closed strings give rise to the same equivalence
classes of correlators with respect to total-derivative relations as discussed
in section \totalderivsec.

Zero-mode integrations at multiplicities higher than four require generalizations
of \KNKNB\ and \KNKNBop\ and will be discussed in sections~\secthree\ and \loopintsec.

\newsubsubsec\corrdefsec Definition of open-string correlators

The main challenge to be addressed in this work is to determine the dependence
of the open-string amplitude \onepresc\ on the polarizations and momenta. The
universal Koba--Nielsen factor \KNKNA\ due to plane waves will be stripped off
from
\eqn\theampA{
\langle \! \langle (\mu, b) \, {\cal Z}\, V_1(z_1)\prod_{j=2}^n   U_j(z_j)\rangle\!\rangle
=  \int d^{D} \ell \ |{\cal I}_n(\ell)| \, \langle {\cal K}_n(\ell) \rangle\,.
}
The residual task is to identify kinematic factors ${\cal K}_n(\ell)$ in
pure-spinor superspace that depend on the loop momentum as well as the zero
modes of $\lambda^\alpha, \theta^\alpha$ and capture the superfield kinematics
arising from the path integral. Given their origin from integrating out all the
non-zero modes as well as the zero modes of $d_\alpha$ and $N^{mn}$, we will
henceforth refer to these kinematic factors $ {\cal K}_n(\ell) $ as {\it
correlators}. They carry the key information on the amplitudes
\eqn\theamp{
{\cal A}_n  =
\sum_{\rm top} C_{\rm top} \int_{D_{\rm top}}\!\!\!\!
d\tau \, dz_2 \, d z_3 \, \ldots \, d z_{n} \, \int d^{D} \ell \ |{\cal I}_n(\ell)| \,
\langle {\cal K}_n(\ell)  \rangle \, ,
}
and the computational methods and organizing principles for
correlators to be developed in this work are tailored to reveal
hidden double-copy structures.

\subsubsec{Closed-string correlators and amplitudes}

By virtue of chiral splitting, left- and right-moving worldsheet
degrees of freedom completely decouple at the level of the loop
integrand, and closed-string correlators are obtained from
holomorphic squares of their open-string instances.  More precisely,
the $n$-point closed-string amplitude reads
\eqn\theclosedamp{
{\cal M}_n  =
\int_{{\cal F}}
d^2\tau \, d^2z_2 \, d^2 z_3 \, \ldots \, d^2 z_{n} \,
\int d^{D} \ell \ |{\cal I}_n(\ell)|^2 \,
\langle {\cal K}_n(\ell)\rangle \, \langle\tilde{\cal K}_n(-\ell)\rangle\,,
}
where ${\cal F}$ denotes the fundamental domain of the modular group
$SL_2(\Bbb Z)$ and the punctures $z_j$ are integrated over a torus
of modular parameter $\tau$. The reflection $\ell \rightarrow -\ell$
in the right-moving correlator is due to our normalization
conventions for external momenta. Finally, the tilde in $\tilde
{\cal K}_n$ instructs to replace the super-Yang--Mills superfields
$[A^i_\a,A^i_m,W_i^\a,F_i^{mn}]$ by another copy, where the Weyl
spinors have the same (opposite) chirality for type-IIB (type-IIA)
superstrings.

In situations where both ${\cal K}_n(\ell)$ and $\tilde {\cal K}_n(-\ell)$
depend on $\ell$, we will see in section~\loopintsec\ that quadratic and higher
terms in the loop momentum introduces vector contractions between left- and
right-moving superfields proportional to $\pi/(\Im\tau)$, see e.g.\
\refs{\Richards, \oneloopMichael} and \refs{\GregoriHI, \BergWUX} in cases of
maximal and reduced supersymmetry, respectively. This exemplifies how the
double-copy structure of the closed-string integrand in \theclosedamp\
disappears after performing the loop integration \OchirovXBA: While $ \langle
{\cal K}_n(\ell) \rangle \, \langle \tilde {\cal K}_n(-\ell) \rangle$ is
evidently a holomorphic square of open-string correlators, its loop integral
over $\int d^{D} \ell \, |{\cal I}_n(\ell)|^2$ no longer factorizes. That is why
chiral splitting is a convenient framework to study the double-copy properties
of gravity amplitudes from a string-theory perspective.

By the appearance of open-string correlators ${\cal K}_n$ in closed-string
amplitudes \theclosedamp, they need to be well-defined functions on the torus, at
least after integration over $\ell$.  In particular -- after stripping off a
global factor of $(\Im \tau)^{-5}$ -- the loop integral over $|{\cal
I}_n|^2{\cal K}_n$ and $|{\cal I}_n|^2{\cal K}_n\tilde {\cal K}_n$ must have
modular weight $(n{-}4,0)$ and $(n{-}4,n{-}4)$, respectively.

\newnewsec\SYMsec Multiparticle SYM superfields

After introducing a convenient notation we review the recursive construction of
multiparticle super-Yang--Mills superfields of \EOMbbs. A special emphasis will
be given to their {\it local} representatives, as they will play an essential
role in the construction of one-loop correlators in later sections.

\newsubsec\wordssec Combinatorics on words

Let us first introduce a notation based on words and review a few associated
results that will be used in the rest of this work. Good introductions to the
combinatorics on words and related subjects can be found in
\refs{\lothaire,\reutenauer}.

In dealing with objects that contain multiple particle labels, one is faced with
many permutations and associated operations acting on the labels of the
participating particles. A convenient framework to handle such things is to use
the notion of words and linear maps acting on them. As such,
permutations\foot{Words with repeated letters do not appear in the context of
scattering amplitudes.} of particle labels referring to the external legs are
encoded in {\it words\/} composed from {\it letters\/} in the {\it alphabet\/}
of natural numbers; $\{1,2,3, \ldots\}$.

Words will be written in upper-case (e.g. $P=134256$) and its letters in
lower-case (e.g. $i=3$). The {\it length\/} of the word $P=p_1 p_2\ldots p_n$ is
the number $n$ of its letters and is denoted by $|P|$. The reversal of the
word $P=p_1p_2 \ldots p_n$ is the word $\tilde P=p_n \ldots p_2p_1$.

The {\it concatenation\/} product of the words $P=p_1\ldots p_n$ and $Q=q_1
\ldots q_m$ is the word $PQ=p_1 \ldots p_nq_1 \ldots q_m$. The {\it empty
word\/} is denoted by $\emptyset$ and it is the unit with respect to the
concatenation, i.e. $P\emptyset = \emptyset P= P$. Unless otherwise noted,
labeled objects are defined to be zero when their label is the empty word (such
as the momentum $k^m_\emptyset\equiv0$).

The {\it deconcatenation\/} of a word $P$ into two words is denoted $P=XY$ and
is given by all pairs of words $X,Y$ such that $P=XY$ under concatenation (with
obvious generalization for $P=XYZ$ etc). For example, the deconcatenation of
$P=XY$ when $P=312$ is given by the the words $(X,Y)$: $(\emptyset, 312),
(3,12), (31,2)$ and $(312,\emptyset)$. The deconcatenation map often occurs as a
summation condition, e.g.
\eqn\decsum{
T_P = \sum_{P=XY}F_X F_Y\quad \Longrightarrow\quad T_{312} = F_\emptyset F_{312}
+ F_3 F_{12} + F_{31}F_2 + F_{312}F_\emptyset
}
for arbitrary labeled objects $T$ and $F$.
The {\it shuffle product} of words of length $n$ and $m$ is
defined recursively by
\eqn\Shrecurs{
\emptyset\shuffle A = A\shuffle\emptyset = A,\qquad
A\shuffle B \equiv a_1(a_2 \ldots a_{n} \shuffle B) + b_1(b_2 \ldots b_{m}
\shuffle A)\,,
}
and it generates all ${(n{+}m)!\over n!m!}$ possible ways to interleave the
letters of $A$ and $B$ without changing their orderings within $A$ and $B$.
For example,
\eqnn\shex
$$\displaylines{
1\shuffle 2 = 12 + 21\,,\quad
12\shuffle 3 = 123 + 132 + 312\,, \hfil\shex\hfilneg\cr
12\shuffle 34 = 1234 + 1324 + 1342 + 3142 + 3124 + 3412\,.
}$$
A word $P$ is said to be {\it a shuffle of\/} $X$ and $Y$
if it appears in their shuffle product, i.e. if $P\in X \shuffle Y$.
From the examples \shex\ it follows that
$3142$ is a shuffle of $12$ and $34$.

The {\it deshuffle\/} of $P$ is denoted $P= X\shuffle Y$ and is the sum of all
pairs of words $X,Y$ such that $P$ is a shuffle of $X$ and $Y$. An efficient
algorithm that generates $X,Y$ in the deshuffle of $P$ follows from the linear
map $\d(P)=X\otimes Y$ defined by
\eqn\deltamap{
\d(a_1a_2 \ldots a_n)\equiv \d(a_1)\d(a_2) \ldots\d(a_n),\quad
\d(a_i) \equiv \emptyset \otimes a_i + a_i\otimes\emptyset\,,\quad
\d(\emptyset) \equiv \emptyset\otimes\emptyset\,.
}
For example,
\eqnn\excupD
$$\eqalignno{
\d(1) &= \emptyset\otimes 1 + 1\otimes\emptyset\,, &\excupD\cr
\d(12) &= \d(1)\d(2) = (\emptyset\otimes 1 +
1\otimes\emptyset)(\emptyset\otimes 2 + 2\otimes\emptyset) =
\emptyset\otimes 12  + 1\otimes 2 + 2\otimes 1 + 12\otimes\emptyset\,,\cr
\d(123) &= \delta(12)\delta(3) =
(\emptyset\otimes 12  + 1\otimes 2 + 2\otimes 1 + 12\otimes\emptyset)
(\emptyset\otimes 3 + 3\otimes\emptyset)\cr
&= \emptyset\otimes123
+ 1\otimes23
+ 2\otimes13
+ 12\otimes3
+ 3\otimes12
+ 13\otimes2
+ 23\otimes1
+ 123\otimes\emptyset\,.
}$$
An alternative characterization is
$\d(P) = \sum_{X,Y}\langle P, X\shuffle Y\rangle\, X\otimes Y$
where $\langle \cdot,\cdot\rangle$ denotes
the scalar product on words defined by
\eqn\AdotB{
\langle A, B\rangle \equiv \d_{A,B},\qquad
\d_{A,B}= \cases{$1$, & if $A=B$\cr
		 $0$, & otherwise}\,.
}
We will see in section~\BRSTvarsec\ that the deshuffle coproduct describes the
BRST variation of local multiparticle superfields just like the deconcatenation
describes the BRST variation of their non-local counterparts.

As words are restricted to be permutations of
the letters $\{1,2,3, \ldots\}$, an explicit sum
over permutations is often represented by a sum over words, e.g.
\eqn\wordsum{
\sum_{P}T_P \equiv \sum_{|P|=1}^{\infty}\sum_{\a\in\{p_1, \ldots, p_{\len{P}}\}}T_\a\,.
}
Furthermore, two common operations on words
are given by the left-to-right bracketing map $\ell(A)$ and the rho map $\rho(A)$.
They are defined recursively as \reutenauer
\eqnn\ellmap
\eqnn\rhomap
$$\eqalignno{
\ell(123 \ldots n)&\equiv \ell(123 \ldots n{-}1)n - n\ell(123 \ldots n{-}1)\,,\quad
\ell(i)\equiv i\,,\quad\ell(\emptyset)\equiv0\,,&\ellmap\cr
\rho(123 \ldots n) &\equiv 1\rho(23 \ldots n) - n\rho(123 \ldots n{-}1),
\hskip27pt\rho(i)\equiv i\,,\quad\rho(\emptyset)\equiv0\,, &\rhomap
}$$
for example
$\ell(123) = 123 - 213 - 312 + 321$
and
$\rho(123)= 123 - 132 - 312 + 321$.
In sections~\multisec\ and \BGmapsec\ these maps will be used, among other
applications, in the discussion of superfields in the BCJ gauge and as a
practical prescription to convert non-local Berends--Giele currents into their
local counterparts. There is a vast literature dealing with these and similar
maps in the context of free Lie algebras, see for instance \reutenauer.

In addition, unless otherwise noted every labeled object considered in this series of
papers is linear on words, e.g. $T_{A+B} \equiv T_A + T_B$. This linearity will
be frequently exploited to avoid unnecessary summation symbols, for instance
\eqn\shortAB{
T_{A\shuffle B}\equiv\sum_{\sigma\in A\shuffle B} T_\sigma\,,\qquad
T_{\ell(A)}\equiv\sum_{\sigma\in \ell(A)} T_\sigma\,.
}
To further illustrate the above points, the Kleiss--Kuijf amplitude relations
\KK\ among Yang--Mills tree amplitudes become $A_{P1Qn} = (-1)^{|P|}A_{1(\tilde
P\shuffle Q)n}$, while the symmetry \refs{\BGSym,\Gauge} obeyed by the Berends--Giele currents
\BGpaper\ is written as $\cK_{A\shuffle B} = 0$.

In this work we use the convention that
whenever words of external-state labels in a subscript are separated through a
comma (rather than a vertical bar), the parental object is understood to by
symmetric under exchange of these words. For example,
\eqn\symABCdef{
T_{A,B,C}=T_{A,C,B}=T_{B,A,C}\,.
}
In addition to denoting a sum over permutations with standard
notations such as
\eqn\standsum{
T_A T_{B,C,D} + (A\leftrightarrow B,C,D)\equiv
T_A T_{B,C,D}+T_B T_{A,C,D}+T_C T_{A,B,D}+T_D T_{A,B,C}\,,
}
more general permutations will be handled with the notation
$+(A,B|A,B,C,D)$; it instructs to sum over all
ordered combinations of the words $A,B$ taken from the set $\{A,B,C,D\}$, for
example
\eqnn\sumbin
$$\eqalignno{
T_A T_B T_{C,D} +(A,B|A,B,C,D)&\equiv T_A T_B T_{C,D}
+ T_A T_C T_{B,D}
+ T_A T_D T_{B,C} &\sumbin\cr
&\quad{}+ T_B T_C T_{A,D}
+ T_B T_D T_{A,C}
+ T_C T_D T_{A,B}\,.
}$$
Generalizations of the form $+(A_1, \ldots A_n|A_1,
\ldots A_{n+m})$ for a total number ${n+m\choose n}$ of terms
follow similarly.

\newsubsec\singlesec Single-particle

A ten-dimensional covariant description of the SYM equations of motion makes use
of four types of superfields seen in the vertex operators \Vvertex\ and
\integratedU: the gluino and gluon potentials $A_\a(x,\t)$, $A^m(x,\t)$ and
their field-strengths $W^\a(x,\t)$, and $F^{mn}(x,\t)$. They satisfy the
following linearized equations of motion \refs{\wittentwistor,\SiegelYI}
\eqn\SYMEOM{
\eqalign{
D_{\a} A_{\b} + D_\b A_\a & = \ga^m_{\a\b} A_m\cr
D_\a F_{mn} & = \p_{[m} (\ga_{n]} W)_\a
}\qquad\eqalign{
D_\a A_m &= (\ga_m W)_\a + \p_m A_\a  \cr
D_\a W^{\b} &= {1\over 4}(\ga^{mn})^{\phantom{m}\b}_\a F_{mn}\ ,
}}
see \BRST\ for the supersymmetric derivative $D_\a$ in $D=10$ superspace\foot{We will
freely swap
$k_m\leftrightarrow \partial_m$ without warning due to our
convention $ik_m\rightarrow k_m$.}.

We will use the collective notation
\eqn\Ksingle{
K_i\in \{A^i_\a(x,\t),A_i^m(x,\t),W_i^\a(x,\t),F_i^{mn}(x,\t)\}\
}
for the four types of superfields describing the $i^{\rm th}$ external leg
in an open-string amplitude \theamp. The superfields $K_i$ will be referred
to as {\it single-particle} superfields.

\newsubsec\twopartsec Two-particle

The vertex operators \Vvertex\ and \integratedU\ for massless states in the
pure-spinor formalism are expanded in terms of single-particle superfields. The
computation of OPEs among the above vertex operators as required by the CFT
amplitude prescription in the pure-spinor formalism leads to a natural
definition of {\it multiparticle superfields} \EOMbbs. In contrast to the
standard description of \SYMEOM, these superfields encompass more than a single
particle label. For example, after absorbing the double poles into total
derivatives, the single pole in the OPE $U_1(z_1)U_2(z_2)$ can be written as \fivetree
\eqn\ranktwoU{
U_{12}\equiv \p\t^\a A^{12}_\a + \Pi^m A^{12}_m + d_\a W^\a_{12} + \half
N^{mn}F^{12}_{mn} \, ,
}
where the two-particle superfields are given by
\eqnn\Atwo
$$\eqalignno{
A^{12}_\a &= - \half\bigl[ A^1_\a (k^1\cdot A^2) + A^1_m (\ga^m W^2)_\a
- (1\leftrightarrow 2)\bigr]\,,&\Atwo \cr
A^{12}_m &=  \half\Bigl[ A^1_p F^2_{pm} - A^1_m(k^1\cdot A^2) + (W^1\ga_m W^2)
- (1\leftrightarrow 2)\Bigr]\,, \cr
W_{12}^\a &= {1\over 4}(\ga^{mn}W^2)^\a F^1_{mn} + W_2^\a (k^2\cdot A^1)
- (1\leftrightarrow 2)\,,\cr
F^{12}_{mn}
& = k^{12}_m A^{12}_n - k^{12}_n A^{12}_m - (k_1\cdot k_2)(A^1_m A^2_n -A^1_n A^2_m)\,.
}$$
The last line involves the two-particle momentum $k^m_{12}=k_1^m {+}k_2^m$, see
\multmom\ for the definition of multiparticle momenta. By virtue of \SYMEOM, one
can check that the two-particle superfields \Atwo\ satisfy the following
equations of motion:
\eqnn\EOMAtwo
$$\eqalignno{
D_{(\a} A^{12}_{\b)} &= \ga^m_{\a\b}A^{12}_m + (k_1\cdot k_2)(A^1_\a A^2_\b + A^1_\b A^2_\a)\,, &\EOMAtwo\cr
D_\a A^{12}_m &= (\ga_m W^{12})_\a + k^{12}_m A^{12}_\a + (k_1\cdot k_2)(A^1_\a A^2_m - A^2_\a A^1_m)\,, \cr
D_\a W^\b_{12} &= {1\over 4}(\ga^{mn})_\a{}^\b F^{12}_{mn} + (k_1\cdot k_2)(A^1_\a W_2^\b - A^2_\a W^\b_1)\,, \cr
D_\a F^{12}_{mn} &= k^{12}_m (\ga_n W^{12})_\a - k^{12}_n (\ga_m W^{12})_\a
+ (k_1\cdot k_2)(A^1_\a F^2_{mn} - A^2_\a F^1_{mn})  \cr
& \ \  + (k_1\cdot k_2)( A^{1}_{n} (\ga_{m} W^2)_\a - A^{2}_{n} (\ga_{m} W^1)_\a
- A^{1}_{m} (\ga_{n} W^2)_\a + A^{2}_{m} (\ga_{n} W^1)_\a)\,,\cr
}$$
which augment the linearized equations of motion in \SYMEOM\ by contact terms
$\sim k_1 \cdot k_2$. Up to BRST exact terms \EOMbbs, the two-particle version
\eqn\Vonetwo{
V_{12} \equiv \l^\a A^{12}_\a
}
of the unintegrated vertex operator also appears
in the OPE $V_1(z_1)U_2(z_2)$. Written in terms of the BRST charge $Q=\l^\a D_\a$, the
equations of motion \EOMAtwo\ become
\eqnn\QTwo
$$\eqalignno{
QV_{12} &= (k_1\cdot k_2)V_1 V_2\,,&\QTwo\cr
QA_{12}^m &= (\l\ga_m W_{12}) + k_{12}^m V_{12} + (k_1\cdot k_2)(V_1
A_2^m - V_2 A_1^m)\,, \cr
Q W^\b_{12} &= {1\over 4}(\l\ga_{mn})^\b F_{12}^{mn} + (k_1\cdot
k_2)(V_1 W_2^\b - V_2 W^\b_1)\,, \cr
Q F_{12}^{mn} &= k_{12}^m (\l\ga^n W_ {12}) - k_{12}^n (\l\ga^m W_{12}) 
+ (k_1\cdot k_2)(V_1 F_2^{mn} - V_2 F_1^{mn})  \cr
& \ \  + (k_1\cdot k_2)( A_{1}^{n} (\l\ga^{m} W_2) - A_{2}^{n}
(\l\ga^{m} W_1)
- A_{1}^{m} (\l\ga^{n} W_2) + A_{2}^{m} (\l\ga^{n} W_1))\ ,
}$$
where the term $(\l\ga^m\l)A^{12}_m$ is absent by the pure-spinor constraint \puresp.

\newsubsec\multisec Multiparticle

Higher-point amplitudes can be elegantly described by {\it multiparticle
superfields\/} that contain information on multiple particles at once. These
superfields not only played a fundamental role in the derivation of the
$n$-point disk amplitude in \nptString\ but will also simplify the description
of one-loop correlators.

\subsubsec Multiparticle vertex operators

\noindent As shown in \EOMbbs, there is a multiparticle generalization of the
the above superfields,
\eqn\Kmulti{
K_P\in \{A_\a^P, A_m^P, W^\a_P, F^{mn}_P\}\,,
}
that is suggested by iterated OPE calculations among vertex
operators. Up to total derivatives and BRST-exact terms, the
results of iterated OPEs boil down to the generalization
\eqn\rankhigherU{
V_P \equiv \l^\a A_\a^P \ , \ \ \ \ \ \ 
U_{P}\equiv \p\t^\a A^{P}_\a + \Pi^m A^{P}_m + d_\a W^\a_{P} + \half
N^{mn}F^{P}_{mn}\,,
}
of \Vonetwo\ and \ranktwoU. The appearance of $d_\a$ and $N_{mn}$
in \rankhigherU\ immediately addresses the generalization of the
zero-mode integration in section \twotwosec\ to higher multiplicity, where
\just\ and \Wfour\ become
\eqnn\justMULT
\eqnn\WfourMULT
$$\eqalignno{
\int b^{(4)} U_A U_B U_C &\rightarrow
(\l\ga_{m}W_A)(\l\ga_{n}W_B)F^{mn}_C + {\rm cyc}(A,B,C)\,, &\justMULT
\cr
\int b^{(2)} U_AU_BU_CU_D &\rightarrow
\ell_m (\l\ga_a W_A)(\l\ga_b W_B)(W_C \ga^{abm} W_D) + {\rm perm}(A,B,C,D)\,,\qquad{}
&\WfourMULT
}$$
see section~\LocalBBsec\ for the systematic construction of tensorial
superfield building blocks.

\subsubsec Lie symmetries of multiparticle superfields

The construction of multiparticle superfields \Kmulti\ is detailed in section 3
of \EOMbbs: Recursive equations following the structure of \Atwo\ are augmented
by certain algorithmic redefinitions, which conspire to total derivatives or
BRST-exact terms and were later identified as standard non-linear gauge
transformations in \Gauge. More importantly, the symmetries resulting from these
redefinitions are characterized by the {\it generalized Jacobi identities\/} or
{\it Lie symmetries\/} (see section 8.6.7 of \reutenauer),
\eqn\genjac{
K_{A\ell(B)C}+K_{B\ell(A)C} = 0 \, ,\quad A,B\neq\emptyset \, ,\quad\forall\,C\,,
}
where $\ell(A)$ is the left-to-right bracketing \ellmap.
These are the same symmetries obtained by the following string of structure
constants,
\eqn\coljacobi{
K_{1234 \ldots p} \leftrightarrow\; f^{12 a_2} \,
f^{a_2 3 a_3} \, f^{a_3 4 a_4} \ldots f^{a_{p-1} p a_p} \, ,
}
and their simplest examples read
\eqnn\liex
$$\eqalignno{
K_{12C} + K_{21C} &= 0 \, ,\quad \forall \, C \cr
K_{123C} + K_{231C} + K_{312C} &= 0 \, ,\quad \forall \, C &\liex\cr
K_{1234C} + K_{2143C} + K_{3412C} + K_{4321C} &= 0 \, ,\quad \forall  \,C\,.
}$$
By the correspondence \coljacobi\ with contracted structure constants, the first
two lines of \liex\ are the kinematic counterparts of the antisymmetry
$f^{12a}=-f^{21a}$ and the Jacobi identities $f^{12a} f^{a3b} + {\rm
cyc}(1,2,3)=0$. More generally, the correspondence \coljacobi\ between the
symmetries of color and kinematic factors lines up with the BCJ duality between
color and kinematics \BernQJ: Multiparticle superfields $K_P$ implement the BCJ
duality in the tree-level subdiagram of \figbcj\ \EOMbbs. Accordingly,
superfields that satisfy the symmetries \genjac\ are said to be in the {\it BCJ
gauge\/} \Gauge.

\ifig\figbcj{The correspondence between local multiparticle
superfields $K_{123\ldots p}=K_P$ and tree-level subdiagrams.}
{\epsfxsize=0.70\hsize\epsfbox{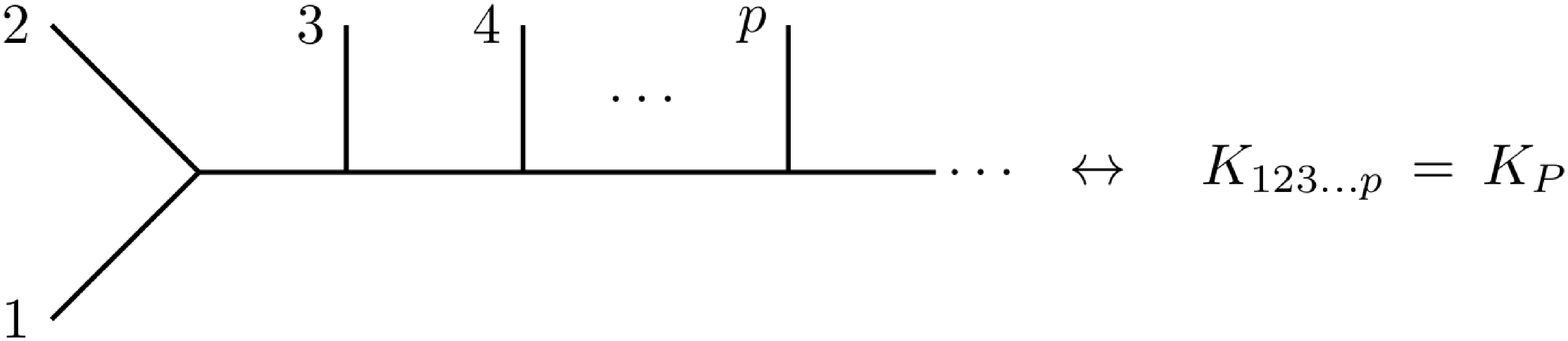}}

Since the symmetries \genjac\ are unchanged for any suffix word $C$,
multiparticle superfields $K_P$ preserve the symmetries of their
lower-multiplicity counterparts. For instance,
the symmetry $K_{12} + K_{21} = 0$ (when $C=\emptyset$)
carries over to $K_{123} + K_{213} = 0$ (when $C=3$) and
so forth for arbitrary $C$.

\subsubsec Nested bracket notation for superfields in BCJ gauge

Since the superfields $K_P$ in the BCJ gauge satisfy the same generalized Jacobi
symmetries as nested brackets $\ell(P)\equiv[[ \ldots[[p_1,p_2],p_3],
\ldots],p_n]$, it is convenient to use a notation where this is manifest. To
this effect, a word $P$ is understood as having a nested bracket structure
$P\to\ell(P)$ and we define\foot{Note, however, that in the definition
$K_P\equiv K_{\ell(P)}$ one must not expand the Dynkin bracket as it would imply
that $K_{\ell(P)} = |P|K_P$ since $K_P$ satisfies the generalized Jacobi
identities. So it is important to stress that \leftbranch\ is a notational
device.}
\eqn\leftbranch{
K_P\equiv K_{\ell(P)}\,,
}
for instance, $K_{12}=K_{[1,2]}$ and $K_{123}=K_{[[1,2],3]}$.
The Jacobi symmetry allows the definition of
local superfields with a even more general bracketing structure such as
$K_{[A,[B,C]]}$.
It then follows from Baker's identity \reutenauer,
\eqn\BakerId{
[\ell(A),\ell(B)] = \ell(A\ell(B))\,,
}
that it is always possible to flatten brackets within
local superfields,
\eqn\Kbrac{
K_{[A,B]} \equiv K_{[\ell(A),\ell(B)]} =  K_{\ell(A\ell(B))} \equiv
K_{A\ell(B)}\,.
}
For example,
\eqnn\Vex
$$\eqalignno{
K_{[1,2]} &= K_{1\ell(2)} = K_{12}, &\Vex\cr
K_{[12,3]} &= K_{12\ell(3))} = K_{123},\cr
K_{[12,34]} &= K_{12\ell(34)} = K_{1234}-K_{1243}\,,\cr
K_{[1,[[2,3],4]]} &= K_{1\ell(234)}=
K_{1234} - K_{1324} - K_{1423} + K_{1432}\,.
}$$
Of course, one can check that the right-hand side of the last identity can also
be written as $-K_{2341}$. The above relations are equivalent to the Jacobi
identities used in the context of kinematic numerators subject to the BCJ
duality.  They can be visualized as flattening out of the planar binary tree
associated with the two branches $A$ and $B$ (see \figbcjtwo). In the context of
the pure-spinor superstring, the identities \Vex\ have been firstly derived in
\refs{\nptFT,\nptString} for the special case $K_P=V_P$ as a consequence of the
BRST algebra obeyed by $V_P$ to be reviewed below.

\ifig\figbcjtwo{The planar binary tree associated with the
multiparticle superfield $K_{[A,B]}$.}
{\epsfxsize=0.70\hsize\epsfbox{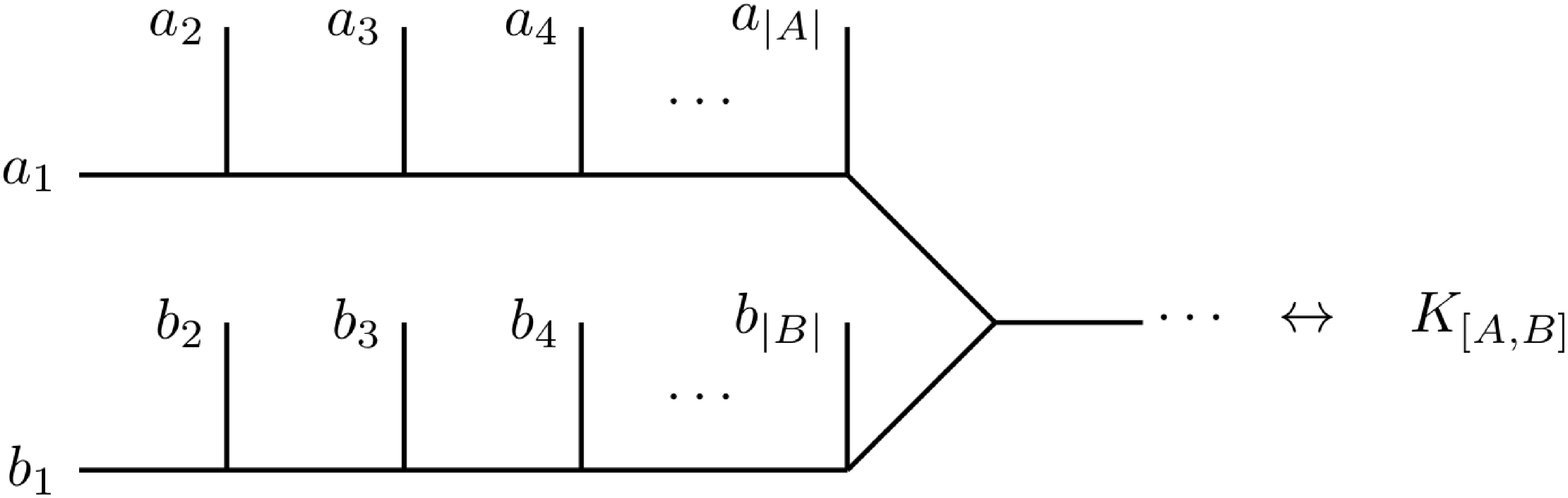}}

\newsubsubsec\BRSTvarsec BRST variation of BCJ-gauge superfields

In evaluating the BRST variations of multiparticle superfields
one is faced with an interesting pattern. Explicit
calculations using the equations of motion of the single-particle
superfields in the generalization of the definitions \Atwo\
to a multiparticle setup reveals the following behavior, for example \EOMbbs
\eqnn\exampOne
$$\eqalignno{
QV_1 &= 0\,, &\exampOne\cr
QV_{12} &= (k_1\cdot k_2)V_1V_2 \, ,\cr
QV_{123} & = (k_1\cdot k_2)\big[V_1 V_{23} + V_{13}V_2\big]
+ (k_{12}\cdot k_3) V_{12} V_3\,, \cr
Q V_{1234} &=(k_1\cdot k_2)\bigl[V_1V_{234}
+ V_{13} V_{24} + V_{14} V_{23} +  V_{134} V_2 \bigr] \cr
&\quad{} + (k_{12}\cdot k_3)\bigl[V_{12} V_{34}
+  V_{124} V_3\bigr]
+ (k_{123}\cdot k_4) V_{123} V_4\,,\cr
Q V_{12345} &=(k^1\cdot k^2)\bigl[
  V_{1} V_{2345}
+ V_{13} V_{245}
+ V_{134} V_{25}
+ V_{1345} V_{2} \cr
&\qquad{}+ V_{135} V_{24}
+ V_{14} V_{235}
+ V_{145} V_{23}
+ V_{15} V_{234}
\bigr]\cr
&\quad{} + (k^{12}\cdot k^3)\bigl[
V_{12} V_{345}
+  V_{124} V_{35}
+  V_{1245} V_{3}
+  V_{125} V_{34}
\bigr]\cr
&\quad{} + (k^{123}\cdot k^4)\bigl[
V_{123} V_{45}
+ V_{1235} V_{4}\bigr]\cr
&\quad{} + (k^{1234}\cdot k^5) V_{1234} V_{5}\,.
}$$
It turns out that the deconcatenation and deshuffle maps defined in
section~\basicsec\ can be used to capture not only these identities for $V_P$
but also for the other multiparticle superfields in a precise manner\foot{In
previous papers these BRST variations were formulated using {\it sets} and the
powerset operation. Since sets are by definition unordered, this
characterization was imprecise. This is rectified in \GeneralQ\ by using {\it
words} together with the deshuffle map.}. That is, one can show that
multiparticle superfields $K_P$ in the BCJ gauge satisfy the following BRST
variations ($k_\emptyset\equiv0$) \EOMbbs
\eqnn\GeneralQ
$$\eqalignno{
QV_{P} &= \!\!\sum_{P=XjY\atop Y=R\shuffle S}\!\!(k_{X} \cdot k_j)\,V_{XR}V_{jS}\,, &\GeneralQ\cr
QA^m_{P} &=  (\l\ga^m  W_{P}) + k^m_{P}  V_{P}\,
+\!\!\sum_{P=XjY\atop Y= R\shuffle S}\!\!(k_{X} \cdot k_j)\big[V_{XR}A^m_{jS} - V_{jR}A^m_{XS}\big]\,,\cr
QW^\b_{P} &=  {1\over 4}(\l\ga^{mn})^\b  F^{P}_{mn}\, +\!\!
\sum_{P=XjY\atop Y= R\shuffle S}\!\!(k_{X}\cdot k_j)
\big[V_{XR}W^\b_{jS} - V_{jR}W^\b_{XS}\big]\,,\cr
QF^{mn}_{P} &= 2 k^{[m}_{P} (\l\ga^{n]}  W_{P})\,
+\!\! \sum_{P=XjY\atop Y= R\shuffle S}\!\!(k_{X} \cdot k_j)\big[
V_{XR}F^{mn}_{jS} + A^{[n}_{XR}(\l\ga^{m]}W_{jS})
- (X \leftrightarrow j)
\big]\,,
}$$
where $P=XjY$ denotes the deconcatenation of the word $P$ into the word $X$, a
single letter $j$, and a word $Y$. Moreover, $Y=R\shuffle S$ denotes the
deshuffle of the word $Y$ into the words $R$ and $S$, see section~\basicsec\ for
more details and examples. To illustrate how these definitions are used, the
first line of $QV{1234}$ in \exampOne\ is generated by one of the
deconcatenation terms of $P{=}1234{=}XjY$, namely $X{=}1,j{=}2,Y{=}34$, and
gives rise to four terms $V_{XR}V_{jS}$ according to the deshuffle $\d(34)=
\{(\emptyset,34),(3,4),(4,3),(34,\emptyset)\}$.

Note that when applying the formula \GeneralQ\ to $QV_1$, the deconcatenation in
$1=XjY$ implies that at least two words among $X$, $j$ and $Y$ are empty. By
defining $k_\emptyset\equiv0$ the momentum contraction $(k_X\cdot k_j)$ vanishes
and we get the correct answer.

\newnewsec\LocalBBsec Pure-spinor superspace: local superfields

In this section we motivate and define a family of {\it local\/} kinematic
building blocks for one-loop open-string amplitudes. On the one hand, they will
be demonstrated to capture the contributions to the correlators \theampA\ in
pure-spinor superspace arising from the $b$-ghost sectors described in section
\twotwosec. On the other hand, these building blocks are intertwined by their
BRST variations: This defining property will be referred to as {\it BRST
covariance\/} and shown to be the suitable starting point for local and BRST
invariant correlators.

The non-local counterparts of the subsequent building blocks in the form of
supersymmetric Berends--Giele currents have been considered in \partI, and we
now complete that discussion by explicitly presenting their {\it local\/}
versions. In the appendix~\brstapp\ we display the BRST variations of every
local building block relevant for correlators up to multiplicity eight.
A subset of these local building blocks has been used in the construction of the four,
five and six-point one-loop amplitudes of ten-dimensional SYM in
\towardsoneloop\ and the six-point string amplitudes in \MafraNWR.

\newsubsec\secTAA Scalars

The zero-mode integrations in the one-loop amplitude prescription \onepresc\
select certain superfields from the vertex operators according to their
associated worldsheet variables. For example, the $b$-ghost zero-mode
contribution of the form $b^{(4)}=d^4\d'(N)$ was argued to require the zero
modes $d_\a d_\b N^{mn}$ from the external vertices, see section \twotwosec.
From the expression \rankhigherU\ for the multiparticle integrated vertex
operator, these zero modes are accompanied by superfields $W_A^\a W_B^\b
F_C^{mn}$. Then, by the resulting effective rule \justMULT, one is naturally led
to the definition \partI
\eqn\TABCdef{
T_{A,B,C} \equiv {1\over 3}(\l\ga_m W_A)(\l\ga_n W_B)F^{mn}_C +
\cyc(A,B,C)\,.
}
Using the BRST variation of the multiparticle superfields \GeneralQ,
it follows that the BRST variation of \TABCdef\ is given by ($k_\emptyset\equiv 0$)
\eqn\QTABC{
QT_{A,B,C} =
\!\!\!\sum_{A=XjY\atop Y=R\shuffle S}\!\!
(k_{X} \cdot k_j)\big[
V_{XR}T_{jS,B,C} - V_{jR}T_{XS,B,C}\big] + (A\leftrightarrow B,C)\,,
}
where the notation for the sums is explained below \GeneralQ.
For example, the BRST variations of all $T_{A,B,C}$ up to multiplicity five
are given by
\eqnn\QTs
$$\eqalignno{
QT_{1,2,3}&= 0\,, &\QTs\cr
QT_{12,3,4} &= (k_1\cdot k_2)\big[V_1T_{2,3,4}-V_2T_{1,3,4}\big]\,,\cr
QT_{123,4,5} &= (k_1\cdot k_2)\big[V_1T_{23,4,5}+ V_{13}T_{2,4,5}
 - V_{2}T_{13,4,5} -V_{23}T_{1,4,5}\big]\cr
&\quad{}+ (k_{12}\cdot k_3)\big[V_{12}T_{3,4,5}-V_{3}T_{12,4,5}\big]\,,\cr
QT_{12,34,5} &= (k_1\cdot
k_2)\big[V_1T_{2,34,5}-V_2T_{1,34,5}\big]+(12\leftrightarrow34)\,,
}$$
while the multiplicity-six and -seven BRST variations will be listed in the
appendix~\brstapp. Since the right-hand side of the BRST variation \QTABC\
involves the same class of objects $T_{B,C,D}$ as seen on the left-hand side,
the family of building blocks \TABCdef\ is said to be {\it BRST covariant}. The
appearance of $V_A$ on the right-hand side is inherited from the multiparticle
equations of motion \GeneralQ\ and an integral part of our notion of BRST
covariance.

Note that $T_{A,B,C}$ is symmetric in $A,B,C$ by its definition \TABCdef, in
agreement with the convention \symABCdef\ adopted throughout this work.

\newsubsec\vecsec Vectors

Vectorial building blocks can be defined from the zero-mode
integrations of correlators that contain a single loop momentum
$\ell^m$ (the zero mode of $\Pi^m$). In this case, there are
two different classes of terms in the correlator \onepresc, see section \twotwosec:
\medskip
\item{i)} the $b$-ghost contributes
$b^{(4)}=d^4\d'(N)$ zero modes and the external vertices $\ell^m d^2 N^{mn}$
\item{ii)} the $b$-ghost contributes $\ell^m d^2\d(N)$ via $b^{(2)}$ and the
external vertices $d^4$
\medskip
\noindent According to the zero-mode integrations \just\ and \Wfour,
the superfield expressions for the two cases above are given by
\eqn\twocases{
{\rm i}) \ A^m_A T_{B,C,D}\quad \hbox{ and }\quad  {\rm ii}) \ W^m_{A,B,C,D} \, ,
}
where the effective rule \WfourMULT\ gives rise to
\eqn\Wmdef{
 W^m_{A,B,C,D}
\equiv {1\over 12}(\l\ga_n W_A)(\l\ga_p W_B)(W_C\ga^{mnp} W_D) + (A,B|A,B,C,D)\,.
}
The relative coefficient of these superfields is uniquely fixed as
\eqn\TmABCDdef{
T^m_{A,B,C,D}\equiv \big[A^m_A T_{B,C,D} + (A\leftrightarrow
B,C,D)\big] + W^m_{A,B,C,D}
}
once we impose the covariant BRST transformation
\eqn\QTm{
QT^m_{A,B,C,D} = k_A^m V_A T_{B,C,D}{} +
\!\!\!\sum_{A=XjY\atop Y=R\shuffle S}\!\!\!
(k_{X}\cdot k_j)\big[
V_{XR}T^m_{jS,B,C,D} - V_{jR}T^m_{XS,B,C,D}\big] + (A\leftrightarrow
B,C,D)\,,
}
for example, the BRST variations of all $T^m_{A,B,C,D}$ up to multiplicity six
are given by
\eqnn\QTms
$$\eqalignno{
QT^m_{1,2,3,4} & = k^m_1 V_1 T_{2,3,4} + (1 \leftrightarrow 2,3,4)\,,
&\QTms\cr
QT^m_{12,3,4,5} & =  \big[ k_{12}^m V_{12}T_{3,4,5}
+ (12\leftrightarrow 3,4,5)\big]
+ (k_1\cdot k_2)\big(V_1 T^m_{2,3,4,5} - V_2 T^m_{1,3,4,5}\big)\,,\cr
QT^m_{123,4,5,6} & = \big[ k_{123}^m V_{123}T_{4,5,6} + (123\leftrightarrow 4,5,6)\big] \cr
&\quad{} + (k_1\cdot k_2)\big[ V_1 T^m_{23,4,5,6} + V_{13}T^m_{2,4,5,6} - (1\leftrightarrow 2)\big]\cr
&\quad{} + (k_{12}\cdot k_3)\big[ V_{12}T^m_{3,4,5,6} -
(12\leftrightarrow3)\big]\,,\cr
QT^m_{12,34,5,6} & = \big[ k_{12}^m V_{12}T_{34,5,6} + (12\leftrightarrow 34,5,6)\big] \cr
&\quad{}+ (k_1\cdot k_2)\big[V_1 T^m_{2,34,5,6} - (1\leftrightarrow2)\big] +
(k_3\cdot k_4)\big[V_3 T^m_{12,4,5,6} - (3\leftrightarrow4)\big]\,,
}$$
while the examples at multiplicity seven are listed in the appendix~\brstapp.
In order to track the origin of BRST covariance, we first compute the BRST
variations of the superfields in \twocases\ to obtain
\eqnn\QAmT
\eqnn\WmQ
$$\eqalignno{
QA^m_A T_{B,C,D} &= k_A^m V_A T_{B,C,D} + (\l\ga^m W_A)T_{B,C,D} &\QAmT\cr
\noalign{\vskip2pt}
&\quad{} +
\!\!\sum_{A=XjY\atop Y=R\shuffle S}\!\!
(k_{X} \cdot k_j)\,\big[
V_{XR}A^m_{jS}  - (X\leftrightarrow j)\big] T_{B,C,D}\cr
\noalign{\vskip2pt}
&\quad{} +\!\!\sum_{B=XjY\atop Y=R\shuffle S}\!\!
(k_{X} \cdot k_j)\,\big[
V_{XR}A^m_A T_{jS,C,D} - (X\leftrightarrow j)\big] +
(B\leftrightarrow C,D)\,,\cr
\noalign{\vskip3pt}
QW^m_{A,B,C,D} &= - (\l\ga^m W_A)T_{B,C,D} + (A\leftrightarrow B,C,D) &\WmQ\cr
&\quad{}+
\sumQ{A}{X}{j}{Y}{R}{S}\big[V_{XR}W^m_{jS,B,C,D}-(X\leftrightarrow j)\big]
+ (A\leftrightarrow B,C,D)\,.
}$$
The linear combination in \TmABCDdef\ is tailored to cancel the non-covariant
term $(\l\ga^m W_A)T_{B,C,D}$ in which the vector index is carried by a gamma
matrix and one arrives at \QTm. The remaining terms in \QTm\ are compatible with
the notion of BRST covariance: deshuffle sums of $T^m_{A,B,C,D}$ itself or terms
of the form $k^m_A V_A T_{B,C,D}$, where the vector index is carried by momenta.
As firstly observed in \refs{\oneloopMichael,\threeloop}, BRST covariant vector
building blocks are crucial for BRST invariance of closed-string amplitudes
that contain vector contractions between left- and right-movers.

The non-local counterparts of $T_{A,B,C}$ and $T^m_{A,B,C,D}$ can be
found in section 2.4 of \partI.

\newsubsec\secTCC Tensors

Local building blocks of higher tensor ranks can be defined from the zero-mode
integrations of correlators that contain higher powers of loop momenta. For
instance, with two loop momenta\foot{Unless otherwise noted or when written
inside $[ \ldots]$, the convention for summing over the permutations
$(B_1\leftrightarrow B_2, \ldots,B_p)$ applies only to the line in which the
permutation appears.}
\eqnn\Tmndef
$$\eqalignno{
T^{mn}_{A,B,C,D,E} &\equiv  A_A^{(m} A_B^{n)} T_{C,D,E} + (A,B|A,B,C,D,E) &\Tmndef\cr
&\quad{} + A_A^{(m} W^{n)}_{B,C,D,E} + (A \leftrightarrow B,C,D,E)\cr
&= A^m_A W^n_{B,C,D,E} + A^n_A T^m_{B,C,D,E} + (A \leftrightarrow B,C,D,E)\,,
}$$
or three loop momenta,
\eqnn\Tmnpdef
$$\eqalignno{
T^{mnp}_{A,B,C,D,E,F} &\equiv  A_A^{(m} A_B^n A_C^{p)}
T_{D,E,F} + (A,B,C|A,B,C,D,E,F) &\Tmnpdef\cr
&\quad{}+A_A^{(m} A_B^n W^{p)}_{C,D,E,F} + (A,B|A,B,C,D,E,F)\,,\cr
}$$
and in general,
\eqnn\Ttensor
$$\eqalignno{
T^{m_1\ldots m_r}_{B_1,B_2,\ldots,B_{r+3}} &\equiv
T_{B_1,B_2,B_3} A_{B_4}^{(m_1} A_{B_5}^{m_2} \ldots
A_{B_{r+3}}^{m_r)} + (B_1,B_2,B_3| B_1,B_2,\ldots, B_{r+3})\qquad{}&\Ttensor\cr
&\quad{} +  W^{(m_1}_{B_1,B_2,B_3,B_4}  A_{B_5}^{m_2} \ldots A_{B_{r+3}}^{m_r)}
+ (B_1,\ldots,B_4| B_1,B_2,\ldots, B_{r+3})\,.
}$$
Similarly as before, the terms in the first line originate from the $\Pi^{m_1}
\Pi^{m_2} \ldots \Pi^{m_{r}} d_\alpha d_\beta N_{pq}$ zero-mode coefficient in
the external vertices under the integration rules \efruleloop. The second line
in turn originates from the $b$-ghost sector linear in $\Pi^m$, see
\efruletwoloop.

Straightforward but tedious calculations using the BRST variations \GeneralQ\ of
multiparticle superfields imply the rank-two variation,
\eqnn\BRSTmn
$$\eqalignno{
Q T^{mn}_{A,B,C,D,E} &=\delta^{mn} Y_{A,B,C,D,E}\cr
&\quad{}+ k_{A}^{(m} V_{A} T^{n)}_{B,C,D,E} +(A \leftrightarrow B,C,D,E)  &\BRSTmn \cr
&\quad{}+ \sumQ{A}{X}{j}{Y}{R}{S}
\big[ V_{XR} T^{mn}_{jS,B,C,D,E} - (X\leftrightarrow j)\big]
+(A \leftrightarrow B,C,D,E) \, ,
}$$
where the anomaly building block $Y_{A,B,C,D,E}$ and its generalizations will
be introduced in the next subsection. Similarly, we find the following variation
at rank three:
\eqnn\QTmnp
$$\eqalignno{
QT^{mnp}_{B_1,B_2,\ldots,B_{6}} &= \delta^{(mn}Y^{p)}_{B_1,\ldots,B_{6}} \cr
&\quad{}+ k_{B_1}^{(m} V_{B_1}  T^{np)}_{B_2,\ldots,B_{6}}
+ (B_1\leftrightarrow B_2,\ldots, B_{6})&\QTmnp \cr
&\quad{}+\!\sumQ{B_1}{X}{j}{Y}{R}{S}
\big[V_{XR} T^{mnp}_{jS,B_2,\ldots,B_{6}} - (X\leftrightarrow j)\big]
+ (B_1\leftrightarrow B_2,\ldots, B_{6})\,. \cr
}$$
In general, the BRST variation is given by
\eqnn\QTensor
$$\eqalignno{
QT^{m_1\ldots m_r}_{B_1,B_2,\ldots,B_{r+3}} &=
\delta^{(m_1 m_2} Y^{m_3 \ldots m_r)}_{B_1,B_2,\ldots,B_{r+3}} \cr
&\quad{}+  k_{B_1}^{(m_1}  V_{B_1} T^{m_2\ldots m_r)}_{B_2,\ldots,B_{r+3}}
+ (B_1\leftrightarrow B_2,\ldots, B_{r+3}) &\QTensor \cr
&\quad{}+\!\sumQ{B_1}{X}{j}{Y}{R}{S}
\bigl[ V_{XR} T^{m_1\ldots m_r}_{jS,B_2,\ldots,B_{r+3}}
- (X\leftrightarrow j) \bigr]
+ (B_1\leftrightarrow B_2,\ldots, B_{r+3})\,.
}$$
For example, the BRST variations of the two- and three-tensor $T^{m
\ldots}_{A,B, \ldots}$ up to multiplicity six are given by
\eqnn\QTexs
$$\eqalignno{
QT^{mn}_{1,2,3,4,5} & =  \d^{mn}Y_{1,2,3,4,5}
+ \big[ k^{(m}_{1}V_1 T^{n)}_{2,3,4,5} + (1\leftrightarrow 2,3,4,5)\big]\,,&\QTexs\cr
QT^{mn}_{12,3,4,5,6} & =  \d^{mn}Y_{12,3,4,5,6} +
\big[ k^{(m}_{12}V_{12} T^{n)}_{3,4,5,6} + (12\leftrightarrow 3,4,5,6)\big]\cr
&\quad{}+ (k_1\cdot k_2) \big[ V_1 T^{mn}_{2,3,4,5,6} - (1\leftrightarrow2)\big]\,,\cr
QT^{mnp}_{1,2,3,4,5,6} & = \d^{(mn}Y^{p)}_{1,2,3,4,5,6}
+ \big[ k_1^{(m} V_1 T^{np)}_{2,3,4,5,6} + (1\leftrightarrow2,3,4,5,6)\big]\,,
}$$
while the multiplicity-seven examples will be listed in the appendix~\brstapp.
Here, our notion of BRST covariance is extended as follows: The admissible terms
on the right-hand sides of the variations are either deshuffle sums of the terms
$T^{m_1\ldots m_r}_{B_1,B_2,\ldots,B_{r+3}}$ on left-hand side along with $V_A$, or
they comprise an anomalous superfield $Y^{m_1\ldots m_r}_{B_1,\ldots,B_{r+5}}$
to be introduced next.

The non-local counterparts of $T^{m_1\ldots m_r}_{B_1,\ldots,B_{r+3}}$ can be 
found in sections 3 and 4 of \partI.

\newsubsubsec\anoBBsecI Anomalous building blocks

One-loop amplitudes of the open superstring at $n\geq 6$ points are known to
exhibit a gauge anomaly before combining the worldsheet topologies \AnomalyGreen,
also see \PolchinskiTU. The supersymmetric kinematic factor of the six-point
anomaly derived with the pure-spinor formalism in \anomalypaper\ was given in
terms of the pure-spinor superspace expression $(\l\ga^m W_2)(\l\ga^n
W_3)(\l\ga^p W_4)(W_5\ga_{mnp}W_6)$. By promoting the $W_i$ to multiparticle
superfields, one arrives at its higher-point extension
\eqnn\Wanondef
$$\eqalignno{
Y_{A,B,C,D,E} &\equiv \half (\l\ga^m W_A)(\l\ga^n W_B)(\l\ga^p
W_C)(W_D\ga_{mnp}W_E)&\Wanondef\cr
&=(\l\ga_m W_A)W^m_{B,C,D,E}\,,
}$$
as well as its tensorial generalization
\eqnn\Ytensordef
$$\eqalignno{
Y^{m_1\ldots m_r}_{B_1,B_2,\ldots,B_{r+5}} &\equiv
Y_{B_1,\ldots,B_5} A_{B_6}^{(m_1} A_{B_7}^{m_2} \ldots A_{B_{r+5}}^{m_r)}
+ (B_1,\ldots,B_5| B_1,\ldots, B_{r+5}) \cr
& = A_{B_1}^{m_1} Y^{m_2 \ldots m_r}_{B_2,B_3,\ldots,B_{r+5}}
+ (B_1\leftrightarrow B_2,B_3,\ldots,B_{r+5})\,,&\Ytensordef\cr
}$$
and their symmetry in $B_1,B_2,\ldots,B_{r+5}$ follows from the pure-spinor
constraint and group-theory arguments \partI. These definitions enter the BRST
variations \BRSTmn\ to \QTensor\ of the higher-rank building blocks introduced
above. By the arguments in appendix B.5 of \partI, the bosonic components
$\langle Y^{m_1\ldots m_r}_{B_1,B_2,\ldots,B_{r+5}} \rangle$ are parity odd,
i.e.\ proportional to the ten-dimensional Levi--Civita symbol.

The BRST variations of the anomaly building blocks \Ytensordef\ themselves are
covariant as well: They follow the structure of $Q T_{A,B,C}$, $Q T^m_{A,B,C,D}$
and $QT^{m_1\ldots m_r}_{B_1,\ldots,B_{r+3}}$ in \QTABC, \QTm\ and \QTensor,
respectively\foot{The symmetry of $Y_{A,B,C,D,E}$ in $A,B,\ldots,E$ and the BRST
variation of $Y^m_{A,B,C,D,E,F}$ rely on the group-theory fact that the tensor
$t_{\alpha_1\ldots \alpha_5}\equiv (\l\ga_m)_{\alpha_1}(\l\ga_n
)_{\alpha_2}(\l\ga_p) _{\alpha_3} \ga^{mnp}_{\alpha_4 \alpha_5}$ is totally
antisymmetric in $\alpha_1$ to $\alpha_5$ and that the vector
$t_{[\alpha_1\ldots \alpha_5} (\lambda \gamma^m)_{\alpha_6]}$ vanishes \partI.},
\eqnn\QYanon
$$\openup2\jot\eqalignno{
Q Y_{A,B,C,D,E} &=
\!\!\!\sum_{A=XjY\atop Y=R\shuffle S}\!\!(k_X\cdot k_j) \Bigl[V_{XR} Y_{jS,B,C,D,E}
- (X\leftrightarrow j)\Bigr] + (A \leftrightarrow B,C,D,E)\,,  \cr
Q Y^m_{A,B,C,D,E,F} &= k_A^m V_A Y_{B,C,D,E,F}  + (A \leftrightarrow
B,C,D,E,F) &\QYanon\cr
&\quad{}+\!\!\! \sum_{A=XjY\atop Y=R\shuffle S}\!\!(k_X\cdot k_j)
\Bigl[V_{XR} Y^m_{jS,B,C,D,E,F} - (X\leftrightarrow j)\Bigr] + (A
\leftrightarrow B,C,D,E,F)\,,\cr
Q Y^{m_1 \ldots m_r} _{B_1,B_2, \ldots,B_{r+5}} &=
k_{B_1}^{(m_1} V_{B_1} Y^{m_2\ldots m_r)}_{B_2,\ldots,B_{r+5}}
+ (B_1 \leftrightarrow B_2, \ldots,B_{r+5})\cr
&\quad{}+\!\!\!\! \sum_{B_1=XjY\atop Y=R\shuffle S}\!\!\!(k_X\cdot k_j)
\Bigl[V_{XR} Y^{m_1\ldots m_r}_{jS,B_2, \ldots,B_{r+5}} - (X\leftrightarrow j)\Bigr]
+ (B_1 \leftrightarrow B_2, \ldots,B_{r+5})\,.
}$$
For example, the BRST variations of the above anomaly building blocks up to
multiplicity six are given by
\eqnn\YsuptoSix
$$\eqalignno{
QY_{1,2,3,4,5} & = 0\,,&\YsuptoSix\cr
QY_{12,3,4,5,6} & = (k_1\cdot k_2)\big[V_1Y_{2,3,4,5,6} -V_2Y_{1,3,4,5,6}\big]\,,\cr
QY^m_{1,2,3,4,5,6} & = k_1^m V_1Y_{2,3,4,5,6} + (1\leftrightarrow2,3,4,5,6)\,,
}$$
while the multiplicity-seven and -eight examples are listed in the
appendix~\brstapp.
The non-local counterparts of $Y^{m_1 \ldots m_r} _{B_1, \ldots,B_{r+5}}$ can be
found in sections 3 and 4 of \partI.

\newsubsec\RefBBsec Refined building blocks

In this section, we extend the system of $T^{m_1\ldots m_r}_{B_1,\ldots,B_{r+3}}$
by additional building blocks that preserve the key property of BRST covariance.
This extension is initiated by the observation that the five-point linear combination
$k_1^m V_1 T^m_{2,3,4,5} +\bigl[ V_{12} T_{3,4,5} + (2 \leftrightarrow 3,4,5) \bigr]$
is BRST closed \oneloopMichael. Indeed, one can identify a local BRST generator,
\eqnn\Jex
$$\eqalignno{
J_{1|2,3,4,5} &\equiv A^m_1 T^m_{2,3,4,5} -\half\big[ (A_1\cdot
A_2)T_{3,4,5} + (2\leftrightarrow3,4,5)\bigr] &\Jex\cr
&=\half A_1^m \bigl( T^m_{2,3,4,5} + W^m_{2,3,4,5}\bigr)\,,
}$$
which reproduces the above terms along with an anomaly building block \Wanondef:
\eqn\QJex{
QJ_{1|2,3,4,5} = k_1^m V_1 T^m_{2,3,4,5}
+\bigl[ V_{12} T_{3,4,5} + (2 \leftrightarrow 3,4,5) \bigr] + Y_{1,2,3,4,5}\,.
}
Although the emergence of the expression \Jex\ for $J_{1|2,3,4,5}$ from the amplitude
prescription is unclear, its independent study is motivated by the connection with the
earlier building blocks via BRST covariance.

We emphasize that label $1$ enters \Jex\ on special footing, i.e\ it does not
participate in the symmetrization of the other labels $2,3,4,5$. That is why the
notation for this {\it refined} label $1$ separates it from the rest by a
vertical bar\foot{Note that $J_{A|B,C,D,E}$ can be interpreted as the refinement
of $T^m_{B,C,D,E}$ and should be denoted by $T_{A|B,C,D,E}$ just like the other
refined building blocks discussed below which share the parental notation, see
e.g. \Yrefdef. This inconsistency in the notation is a hysterical artifact.}.
The refined building block $J_{1|2,3,4,5} $ can be generalized to multiparticle
labels. On top of promoting the superfields on the right-hand side of \Jex\ to
their multiparticle versions $A^m_A$ and $T^m_{B,C,D,E}$, BRST covariance
requires additional corrections $\sim H_{[A,B]}$ in
\eqn\Jdef{
J_{A|B,C,D,E} \equiv
A^m_A T^m_{B,C,D,E}
- \big[\big(H_{[A,B]}+\half(A_A\cdot A_B)\big)T_{C,D,E}+
(B\leftrightarrow C,D,E)\big]\,.
}
A partial list of explicit expressions for local $H_{[A,B]}$ can be found in \Gauge.
The appearance of the {\it redefining} superfields $H_{[A,B]}$ in
\Jdef\ is needed in order to write the following BRST variation in
terms of multiparticle superfields $V_A$ in the BCJ gauge:
\eqnn\QJ
$$\eqalignno{
QJ_{A|B,C,D,E} &= k^m_A V_A T^m_{B,C,D,E} + \bigl[
V_{[A,B]}T_{C,D,E} + (B\leftrightarrow C,D,E)\bigr] +
Y_{A,B,C,D,E}\cr
\noalign{\vskip5pt}
&\quad{}+\!\!\! \sum_{A=XjY\atop Y=R\shuffle S}\!\!(k_X\cdot k_j)\bigl[
V_{XR}J_{jS|B,C,D,E} - (X\leftrightarrow j)\bigr]&\QJ\cr
&\quad{}+\!\!\!
\sum_{B=XjY\atop Y=R\shuffle S}\!\!(k_X\cdot k_j)\bigl[
V_{XR}J_{A|jS,C,D,E} -(X\leftrightarrow j)\bigr] + (B\leftrightarrow C,D,E)\,.
}$$
The brackets of the term $V_{[A,B]}$ in the first line can be flattened via \Kbrac.
For example, the BRST variations of $J_{A|B,C,D,E}$ up to multiplicity six are
given by
\eqnn\QJthree
$$\eqalignno{
QJ_{1|2,3,4,5} &= k_1^m V_1 T^m_{2,3,4,5}
+\bigl[ V_{12} T_{3,4,5} + (2 \leftrightarrow 3,4,5) \bigr]
+ Y_{1,2,3,4,5}\,, &\QJthree\cr
QJ_{12|3,4,5,6} &= k_{12}^m V_{12} T^m_{3,4,5,6}
+\big[V_{123} T_{4,5,6} + (3 \leftrightarrow 4,5,6) \big]
+ Y_{12,3,4,5,6}  \cr
&\quad{} + (k_1\cdot k_2)(V_1 J_{2|3,4,5,6}-V_2 J_{1|3,4,5,6})\,,\cr
QJ_{1|23,4,5,6} &= k_1^m V_1 T^m_{23,4,5,6}-V_{231} T_{4,5,6}
+\big[ V_{14} T_{23,5,6} + (4 \leftrightarrow 5,6) \big] \cr
&\quad{} + Y_{1,23,4,5,6}
+ (k_2\cdot k_3)(V_2 J_{1|3,4,5,6}-V_3J_{1|2,4,5,6})\,,
}$$
while the multiplicity-seven examples will be listed in the appendix~\brstapp.
In checking the BRST variations of the above refined building block
it is convenient to note,
\eqn\QDAB{
QD_{A,B} = \hat V_{(A,B)} +\Big[\,\sumQ{A}{X}{j}{Y}{R}{S}
\big[V_{XR}D_{jS,B}-(X\leftrightarrow j)\big] + (A\leftrightarrow
B)\Big]\,,
}
where
\eqn\DABdef{
D_{A,B} \equiv \half (A_A\cdot A_B)\,,\qquad
\hat V_{(A,B)} \equiv \half\big[
 V_{A} (k_A\cdot  A_B)
+ A_A^m (\l\ga_m  W_B) + (A\leftrightarrow B)\bigr]\,.
}

\subsubsec Higher-rank tensors

One can also define higher-rank tensors following the same logic,
\eqn\Jtensordef{
J^{m_1 \ldots m_r}_{A|B_1, \ldots,B_{r+4}} \equiv
A^p_A T^{pm_1 \ldots,m_r}_{B_1, \ldots,B_{r+4}}
- \big[\big(H_{[A,B_1]}+\half(A_A\cdot A_{B_1})\big)T^{m_1
\ldots,m_r}_{B_2, \ldots,B_{r+4}}+ (B_1\leftrightarrow B_2,
\ldots,B_{r+4})\big]\, .
}
In doing so, the word $A$ separated by a vertical bar is said to be {\it refined}.
Straightforward but long and tedious calculations
show that
\eqnn\QJtensor
$$\eqalignno{
QJ^{m_1 \ldots m_r}_{A|B_1, \ldots,B_{r+4}} &=
k^p_A V_A T^{pm_1 \ldots m_r}_{B_1, \ldots,B_{r+4}}
+ \d^{(m_1m_2} Y^{m_3 \ldots m_r)}_{A|B_1, \ldots,B_{r+4}}
+ Y^{m_1 \ldots m_r}_{A,B_1, \ldots,B_{r+4}}
&\QJtensor\cr
\noalign{\vskip5pt}
&\quad{}+ V_{[A,B_1]}T^{m_1 \ldots m_r}_{B_2, \ldots,B_{r+4}}
+ k_{B_1}^{(m_1}V_{B_1}J^{m_2 \ldots m_r)}_{A|B_2,
\ldots,B_{r+4}}
+ (B_1\leftrightarrow B_2, \ldots,B_{r+4})\cr
\noalign{\vskip5pt}
&\quad{}+\!\!\! \sum_{A=XjY\atop Y=R\shuffle S}\!\!(k_X\cdot k_j)\bigl[
V_{XR}J^{m_1 \ldots m_r}_{jS|B_1, \ldots,B_{r+4}}
- (X\leftrightarrow j)\bigr]\cr
&\quad{} +\!\!\!\!\sum_{B_1=XjY\atop Y=R\shuffle S}\!\!\!(k_X\cdot k_j)\bigl[
V_{XR}J_{A|jS,B_2, \ldots,B_{r+4}}
- (X\leftrightarrow j) \bigr] + (B_1\leftrightarrow B_2, \ldots,B_{r+4})\, ,
}$$
where the additional class of anomaly superfields
$Y^{m_3 \ldots m_r}_{A|B_1, \ldots,B_{r+4}}$ in the first line
will be defined below. For example, the BRST variations of the above
superfields up to multiplicity seven are given by
\eqnn\QJmex
$$\eqalignno{
QJ^m_{1|2,3,4,5,6} &= k_1^p T^{pm}_{2,3,4,5,6} + Y^m_{1,2,3,4,5,6}\cr
&\quad{}+ V_{12}T^m_{3,4,5,6} + k_2^m V_2 J_{1|3,4,5,6} +
(2\leftrightarrow3,4,5,6)\,,&\QJmex\cr
QJ^m_{12|3,4,5,6,7} &= k_{12}^p V_{12}T^{pm}_{3,4,5,6,7} +
Y^m_{12,3,4,5,6,7}\cr
&\quad{}+ V_{123}T^m_{4,5,6,7} + k_3^m V_3 J_{12|4,5,6,7} +
(3\leftrightarrow4,5,6,7)\cr
&\quad{}+(k_1\cdot k_2)\bigl[V_1J^m_{2|3,4,5,6,7} -
(1\leftrightarrow2)\bigr]\,,\cr
QJ^m_{1|23,4,5,6,7} &= k_1^p V_1T^{pm}_{23,4,5,6,7} + Y^m_{1,23,4,5,6,7}\cr
&\quad{}- V_{231}T^m_{4,5,6,7} + k_{23}^m V_{23}J_{1|4,5,6,7} \cr
&\quad{}+ V_{14}T^m_{23,5,6,7} + k_{4}^m V_{4}J_{1|23,5,6,7} +
(4\leftrightarrow5,6,7)\cr
&\quad{}+(k_2\cdot k_3) \big( V_{2} J^m_{1|3,4,5,6,7} - V_{3}
J^m_{1|2,4,5,6,7}\big)\,,\cr
QJ^{mn}_{1|2,3,4,5,6,7} &=
  k_1^pV_{1} T^{mnp}_{2,3,4,5,6,7}
 + \d^{mn} Y_{1|2,3,4,5,6,7}
 +  Y^{mn}_{1,2,3,4,5,6,7}\cr
&\quad{}+  V_{12} T^{mn}_{3 , 4 , 5 , 6 , 7}
+ k_2^{(m} V_{2} J^{n)}_{1|3,4,5,6,7} + (2\leftrightarrow3,4,5,6,7)\,.
}$$
The inclusion of $J^{m_1 \ldots m_r}_{A|B_1, \ldots,B_{r+4}} $ and their
generalizations into our system of ghost-number two building blocks is
essential to rule out local cohomology objects: Up to and including
multiplicity eight, they allow to identify a BRST generator for each local
BRST-invariant at ghost number three which is constructed from the alphabet of
building blocks introduced in this section, see the appendix~\LocalMonapp\
for more details.

The non-local counterparts of $J^{m_1 \ldots m_r}_{A|B_1, \ldots,B_{r+4}}$ can be
found in sections 5 and 6 of \partI.

\newsubsubsec\anoBBsecII Refined anomaly building blocks

One can also repeat the analysis above and define the refinement of the
anomaly building blocks. A covariant BRST variation written in terms
of BCJ-gauge superfields fixes their general definition to be
\eqn\Yrefdef{
Y_{A|B_1, \ldots,B_{r+6}}^{m_1 \ldots m_r} \equiv
\half A^p_A Y^{p m_1 \ldots m_r}_{B_1, \ldots,B_{r+6}}
- \bigl[H_{[A,B_1]}Y^{m_1 \ldots m_r}_{B_2, \ldots, B_{r+6}}
+ (B_1 \leftrightarrow B_2, \ldots, B_{r+6})\bigr]\,.
}
One can show that the BRST variation is of the same structure
as $QJ^{m_1 \ldots m_r}_{A|B_1, \ldots,B_{r+4}}$ in \QJtensor,
\eqnn\qanomA
$$\openup2\jot\eqalignno{
Q Y^{m_1\ldots m_r}_{A|B_1,\ldots,B_{r+6}} &=
k_A^p V_A Y^{pm_1\ldots m_r}_{B_1,\ldots ,B_{r+6}} &\qanomA\cr
&\quad{}+ V_{[A,B_1]} Y^{m_1\ldots m_r}_{B_2,\ldots,B_{r+6}}
+ k_{B_1}^{(m_1}  V_{B_1} Y^{m_2\ldots m_r)}_{A|B_2,\ldots,B_{r+6}}
+ (B_1 \leftrightarrow B_2,\ldots ,B_{r+6})  \cr
&\quad{} +\!\!\!\! \sum_{B_1=XjY\atop Y=R\shuffle S}
\!\!\!(k_X\cdot k_j)\bigl[
 V_{XR} Y^{m_1\ldots m_r}_{A|jS,B_2,\ldots,B_{r+6}}
 - (X\leftrightarrow j)
\bigr]
 + (B_1 \leftrightarrow B_2,\ldots ,B_{r+6})  \cr
&\quad{}+\!\!\! \sum_{A=XjY\atop Y=R\shuffle S}
\!\!(k_X\cdot k_j)\bigl[
V_{XR} Y^{m_1\ldots m_r}_{jS|B_1,\ldots,B_{r+6}}
- (X\leftrightarrow j)\bigr]\,.
}$$
For example, the BRST variations of the above superfields up to multiplicity
eight are
\eqnn\simYsev
$$\eqalignno{
QY_{1|2,3,4,5,6,7} &= k_1^p V_1 Y^p_{2,3,4,5,6,7} + \bigl[
V_{12}Y_{3,4,5,6,7} + (2\leftrightarrow3,4,5,6,7)\bigr]\,,&\simYsev\cr
QY_{12|3,4,5,6,7,8} &=
k_{12}^pV_{12} Y^p_{3,4,5,6,7,8}
+ \big[V_{123}Y_{4,5,6,7,8} + (3\leftrightarrow4,5,6,7,8)\big]\cr
&\quad{}+ (k_1\cdot k_2)\big( V_{1} Y_{2|3,4,5,6,7,8} - (1\leftrightarrow2)\big)\,,\cr
QY_{1|23,4,5,6,7,8} &=
k^p_{1} V_{1} Y^p_{23,4,5,6,7,8} - V_{231}Y_{4,5,6,7,8} + \big[
V_{14}Y_{23,5,6,7,8} + (4\leftrightarrow5,6,7,8)\big]\cr
&\quad{}+(k_2\cdot k_3)\big(V_{2} Y_{1|3,4,5,6,7,8} - (2\leftrightarrow3)\big)\,,\cr
QY^m_{1|2,3,4,5,6,7,8} &=
k_1^p V_{1} Y^{pm}_{2,3,4,5,6,7,8}
+ \big[ V_{12} Y^m_{3,4, \ldots,8} + k_2^m V_2 Y_{1|3,4, \ldots,8} +
(2\leftrightarrow3, \ldots,8)\big]\,.
}$$
The non-local counterparts of $Y^{m_1 \ldots m_r}_{A|B_1, \ldots,B_{r+6}}$ can be 
found in section 6 of \partI.

\subsubsec Higher-refinement building blocks

It is possible to generalize the degree of refinement of multiparticle
building blocks in a straightforward manner by contracting refined
superfields with additional instances of $A^m_B$.

\subsubsubsec Jacobi currents

The first non-trivial instance of an additional refined slot can be
obtained by considering the term $A^1_m J_{2|3,4,5,6,7}^m$. The BRST
covariance principle suggests the definition to be
\eqnn\Jdoubly
$$\eqalignno{
J_{1,2|3,4,5,6,7}&\equiv A_m^1 J^m_{2 | 3,4,5,6,7}
- {1\over2} \bigl[
(A_1 \cdot A_3) J_{2|4,5,6,7} + (3 \leftrightarrow 4,5,6,7)
\bigr]  \, ,&\Jdoubly\cr
}$$
since its $Q$-variation can be expressed in terms of simpler building blocks,
\eqnn\QJdoubly
$$\eqalignno{
QJ_{1,2|3,4,5,6,7} &= k_1^m V_1 J^m_{2|3,4,5,6,7}
+ k_2^m V_2 J^m_{1|3,4,5,6,7} + Y_{1|2,3,4,5,6,7}
+ Y_{2|1,3,4,5,6,7} &\QJdoubly\cr
&\quad{}+\bigl[V_{13}J_{2|4,5,6,7} + (3\leftrightarrow4,5,6,7)\bigr]
+\bigl[V_{23}J_{1|4,5,6,7} + (3\leftrightarrow4,5,6,7)\bigr]
}$$
and therefore generalizes \QJex.
The notation $J_{1,2|3,4,5,6,7}=J_{2,1|3,4,5,6,7}$ reflects a symmetry
in the refined slots $1,2$ which is not manifest from the definition \Jdoubly.

To define a general recursion for arbitrary tensor ranks and
arbitrary degree $d$ of refinement, it will be convenient to introduce
\eqnn\refWa
$$\eqalignno{
W^{m_1 \ldots m_{r-1}| m_r}_{A_1,\ldots,A_d|B_1, \ldots,B_{d+r+3}} &\equiv
\half A_{A_1}^p
W^{pm_1 \ldots m_{r-1}| m_r}_{A_2,\ldots,A_d|B_1, \ldots,B_{d+r+3}} &\refWa\cr
&\quad{}- \big[H_{[A_1,B_1]} W^{m_1 \ldots m_{r-1}|m_r}_{A_2, \ldots, A_d|
B_2, \ldots,B_{d+r+3}} + (B_1\leftrightarrow B_2, \ldots,
B_{d+r+3})\big]\,.
}$$
Using this auxiliary superfield
the recursion for refined currents $J^{m_1 \ldots m_r}_{A_1,
\ldots,A_d|B_1, \ldots, B_{d+r+3}}$ of arbitrary refinement becomes
\eqnn\JacobiDefs
$$\eqalignno{
J^{m_1 \ldots m_r}_{B_1, \ldots, B_{r+3}} &\equiv
T^{m_1 \ldots m_r }_{B_1, \ldots,B_{r+3}}\,, &\JacobiDefs\cr
J^{m_1 \ldots m_r}_{A_1, \ldots,A_d|B_1, \ldots, B_{d+r+3}} &\equiv
\half A_{A_1}^p \Bigl[
J^{pm_1 \ldots m_r }_{A_2,\ldots, A_d|B_1, \ldots,B_{d+r+3}} +
W^{m_1 \ldots m_r| p}_{A_2,\ldots, A_d|B_1, \ldots,B_{d+r+3}}\Bigr]\cr
&\quad{}- H_{[A_1,B_1]}J^{m_1 \ldots m_{r}}_{A_2, \ldots, A_d|B_2, \ldots,B_{d+r+3}}
+ (B_1\leftrightarrow B_2, \ldots, B_{d+r+3})\,.
}$$
In general, for scalars of refinement $d=2$, one finds
\eqnn\HOl
$$\eqalignno{
QJ_{A,B | C,D,E,F,G} &= Y_{A|B,C,D,E,F,G}
+ V_A k_A^m J^m_{B | C,D,E,F,G} + (A\leftrightarrow B) &\HOl\cr
&\quad{}+ V_{[A,C]} J_{B|D,E,F,G}
+ V_{[B,C]} J_{A|D,E,F,G} + (C\leftrightarrow D,E,F,G) \cr
\noalign{\vskip5pt}
&\quad{}+\!\!\! \sum_{A=XjY\atop Y=R\shuffle S}\!\!\!
(k_X\cdot k_j) \bigl[V_{XR} J_{jS,B|C,D,E,F,G}
- (X\leftrightarrow j)\bigr] + (A\leftrightarrow B)\cr
&\quad{}+\!\!\! \sum_{C=XjY\atop Y=R\shuffle S}\!\!\!
(k_X\cdot k_j) \bigl[V_{XR}J_{A,B|jS,D,E,F,G}
- (X\leftrightarrow j) \bigr]
+ (C\leftrightarrow D,\ldots,G)\,,  \cr
}$$
and this will be the maximum degree of refinement present in the
eight-point correlator. For completeness, even higher
degrees of refinement and tensor ranks are possible,
\eqnn\HRJCgen
$$\openup\jot\eqalignno{
Q &J^{m_1\ldots m_r}_{A_1,\ldots,A_d | B_1,\ldots,B_{d+r+3}} =
\d^{(m_1 m_2}
Y^{m_3\ldots m_r)}_{A_1,\ldots,A_d|B_1,\ldots,B_{d+ r+3}} &\HRJCgen\cr
&+ k_{B_1}^{(m_1}
V_{B_1}J^{m_2\ldots m_r)}_{A_1,\ldots,A_d|B_2,\ldots,B_{d+r+3}}
+ (B_1 \leftrightarrow B_2,\ldots,B_{d+r+3}) \cr
&+ V_{[A_1,B_1]}J^{m_1\ldots m_r}_{A_2,\ldots, A_d | B_2,\ldots,B_{d+r+3}}
+ { A_1 \leftrightarrow A_2,A_3,\ldots, A_d \choose
B_1 \leftrightarrow B_2,\ldots ,B_{d+r+3}}  \cr
\noalign{\vskip3pt}
&+ Y^{m_1\ldots m_r}_{A_2,\ldots,A_d |A_1,B_1,\ldots, B_{d+r+3}}
+ k_{A_1}^p V_{A_1} J^{pm_1\ldots m_r}_{A_2,\ldots,A_d |B_1,\ldots ,B_{d+r+3}}
+ (A_1 \leftrightarrow A_2,\ldots, A_d)\cr
&+\!\!\!\sum_{A_1=XjY\atop Y=R\shuffle S}\!\!\!(k_X\cdot k_j)
\Bigl[V_{XR} J^{m_1\ldots m_r}_{jS,A_2,\ldots,A_d |B_1,\ldots,B_{d+r+3}}
- (X\leftrightarrow j)\Bigr] + (A_1 \leftrightarrow A_2,\ldots ,A_{d})\cr
&+\!\!\! \sum_{B_1=XjY\atop Y=R\shuffle S}\!\!\! (k_X\cdot k_j)
\Bigl[V_{XR} J^{m_1\ldots m_r}_{A_1,\ldots,A_d|jS,B_2,\ldots,B_{d+r+3}}
- (X\leftrightarrow j)\Bigr]
+ (B_1 \leftrightarrow B_2,\ldots ,B_{d+r+3})\,,
}$$
where the objects $Y^{m_3\ldots m_r}_{A_1,\ldots,A_d|B_1,\ldots,B_{d+ r+3}}$
in the first line will be defined next. The non-local counterparts of
$W^{m_1 \ldots m_{r-1}|m_r}_{A_1,\ldots,A_d | B_1,\ldots,B_{d+r+3}}$ and
$J^{m_1 \ldots m_r}_{A_1,\ldots,A_d | B_1,\ldots,B_{d+r+3}}$ can be found
in section 6 of \partI.

\newsubsubsubsec\anoBBsecIII Anomalous building blocks

The higher-refinement generalization of the local superfields
discussed above can also be applied to the anomalous building blocks.
However, it turns out that already the simplest scalar building block
with a double refinement can only appear starting at nine points,
\eqn\doblyY{
Y_{1,2|3,4,5,6,7,8,9} \equiv\half A^m_1 Y^m_{2|3,4,5,6,7,8,9}\,.
}
More generally, defining anomaly building blocks with
higher degree of refinement by
\eqnn\refWc
$$\eqalignno{
Y^{m_1 \ldots  m_r}_{A_1,\ldots,A_d|B_1, \ldots,B_{d+r+5}} &\equiv
\half A_{A_1}^p Y^{pm_1 \ldots m_r}_{A_2,\ldots,A_d|B_1, \ldots,B_{d+r+5}}
&\refWc\cr
&\quad{}- \big[H_{[A_1,B_1]}Y^{m_1 \ldots m_{r}}_{A_2, \ldots, A_d|B_2, \ldots,B_{d+r+5}}
+ (B_1\leftrightarrow B_2, \ldots, B_{d+r+5})\big]\,,
}$$
their bosonic components are parity odd (cf.\ appendix B.5 of \partI) and their
BRST variations inherit the structure of
$QJ^{m_1\ldots m_r}_{A_1,\ldots,A_d | B_1,\ldots,B_{d+r+3}}$ in \HRJCgen,
\eqnn\qanomB
$$\eqalignno{
Q Y^{m_1\ldots m_r}_{A_1,\ldots,A_d | B_1,\ldots,B_{d+r+5}}
&= V_{[A_1,B_1]}Y^{m_1\ldots m_r}_{A_2,\ldots, A_d | B_2,\ldots,B_{d+r+5}}
+ { A_1 \leftrightarrow A_2,A_3,\ldots, A_d \choose
B_1 \leftrightarrow B_2,\ldots ,B_{d+r+5}}\cr
&\quad{}+\Bigg[\! \sum_{A_1=XjY\atop Y=R\shuffle S}\!\!\!(k_X\cdot k_j)
\bigl(V_{XR} Y^{m_1\ldots m_r}_{jS,A_2,\ldots,A_d  |B_1,\ldots,B_{d+r+5}}
- (X\leftrightarrow j)\bigr)\cr
&\qquad{}+ k_{A_1}^p V_{A_1} Y^{pm_1\ldots m_r}_{A_2,\ldots,A_d |B_1,\ldots ,B_{d+r+5}}
+ (A_1 \leftrightarrow A_2,\ldots, A_d)\Biggr] &\qanomB\cr
&\quad{}+ \Biggl[\!\sum_{B_1=XjY\atop Y=R\shuffle S}\!\!\!(k_X\cdot k_j)
\bigl(V_{XR} Y^{m_1\ldots m_r}_{A_1,\ldots,A_d|jS,B_2,\ldots,B_{d+r+5}}
- (X\leftrightarrow j)\bigr)\cr
&\quad{}+ k_{B_1}^{(m_1}V_{B_1} Y^{m_2\ldots m_r)}_{A_1,\ldots,A_d|B_2,\ldots,B_{d+r+5}}
+ (B_1 \leftrightarrow B_2,\ldots ,B_{d+r+5})\Biggr] \, .\cr
}$$
For example,
\eqnn\QdoublyY
$$\eqalignno{
QY_{1,2|3,4,5,6,7,8,9} &= k_1^p V_1 Y^p_{2|3,4,5,6,7,8,9}
+ k_2^p V_2 Y^p_{1|3,4,5,6,7,8,9}&\QdoublyY\cr
&\quad{}+V_{13}Y_{2|4,5,6,7,8,9}
+V_{23}Y_{1|4,5,6,7,8,9} +(3\leftrightarrow4,5,6,7,8,9)\,.
}$$
The non-local counterparts of
$Y^{m_1 \ldots m_r}_{A_1,\ldots,A_d | B_1,\ldots,B_{d+r+3}}$ can be
found in section 6 of \partI.

\newsubsubsec\tracesT Trace relations

As an immediate consequence of their definition \Jdef, refined scalar building blocks
are related to traces of unrefined tensors \partI,
\eqn\HOb{
\delta_{mn}T^{mn}_{A,B,C,D,E} =2 \big[ J_{A|B,C,D,E} +(A\leftrightarrow B,C,D,E)\big]\,.
}
By the definitions \Jtensordef\ and \JacobiDefs, this generalizes to higher tensor rank
\eqn\HOd{
\delta_{np}T^{np m_1\ldots m_r}_{B_1,\ldots,B_{r+5}} =
2 \big[ J^{m_1\ldots m_r}_{B_1|B_2,\ldots,B_{r+5}} +(B_1\leftrightarrow
B_2,\ldots,B_{r+5}) \big]\,,
}
and to higher degree of refinement, respectively \partI
\eqn\HOc{
\delta_{np}J^{np m_1\ldots m_r}_{A_1,\ldots,A_d|B_1,\ldots,B_{d+r+5}} =
2 \big[ J^{m_1\ldots m_r}_{A_1,\ldots,A_d,B_1|B_2,\ldots,B_{d+r+5}}
+(B_1{\leftrightarrow} B_2,\ldots,B_{d+r+5})\big]\,.
}
The same structures arise for anomaly building blocks
\eqn\HOe{
\delta_{np}Y^{np m_1\ldots m_r}_{A_1,\ldots,A_d|B_1,\ldots,B_{d+r+7}} =
2 \big[ Y^{m_1\ldots m_r}_{A_1,\ldots,A_d,B_1|B_2,\ldots,B_{d+r+7}}
+(B_1{\leftrightarrow} B_2,\ldots,B_{d+r+7}) \big]\,,
}
and both of \HOc\ and \HOe\ can be straightforwardly iterated to express double
traces such as $\delta_{mn}\delta_{pq}J^{mnpq\ldots}_{A_1,\ldots,A_d|\ldots}$ in
terms of objects with degree of refinement $d{+}2$.

\newnewsec\BRSTsec Pure-spinor superspace: non-local superfields

The goal of this work is to assemble one-loop correlators in pure-spinor
superspace from kinematic building blocks and their associated worldsheet
functions. When comparing these two classes of ingredients, one discovers
surprising parallels in their structures and relations which will be referred to
as a {\it duality between worldsheet functions and kinematics}, see part II. One incarnation
of this duality is based on the BRST pseudo-invariants discussed in \partI\ and
has been pioneered in \MafraIOJ. The purpose of this section is to review the
BRST pseudo-invariants from the perspective of the above local building blocks.
They will be related to their non-local Berends--Giele representations of
\partI\ via the so-called Berends--Giele map.

\newsubsec\BGmapsec The Berends--Giele map

Every local building block discussed in section~\LocalBBsec\ can be mapped to
its non-local counterpart studied in \partI. This mapping is induced by a
relation among the local superfields $K_P$ and their non-local {\it
Berends--Giele superfield\/} $\cK_P$ given\foot{For historic reasons the BG
superfield associated with $V_P$ is denoted $M_P$ rather than $\cV_P$.
Similarly, the BG image of the local building block $T^{m_1 \ldots}_{A,B,
\ldots}$ is denoted by $M^{m_1 \ldots}_{A,B, \ldots}$.} by the Berends--Giele
(BG) map:
\eqn\BGmap{
K_{iA} =  \sum_B S(A|B)_i \cK_{iB}, \qquad
\cK_{iA} =  \sum_B \Phi(A|B)_i K_{iB}\,,
}
where $S(A|B)_i$ is the KLT matrix \oldMomKer\ (also
known as the {\it momentum kernel} \MomKer) and
$\Phi(A|B)_i$ corresponds to its inverse \DPellis,
\eqn\InvSPhi{
\d_{A,B} = \sum_C S(A|C)_i \Phi(C|B)_i\,,
}
where $\d_{A,B}$ is equal to one if $A=B$ and zero otherwise, see \AdotB.
Both matrices $S$ and $\Phi$ are symmetric and subject to the conditions
$\Phi(A|B)_i=S(A|B)_i\equiv 0$
if $A$ is not a permutation of $B$
and they admit
the following recursive forms \refs{\NLSM,\FTlimit}
\eqnn\KLT
$$\displaylines{
S(P,j|Q,j,R)_i = (k_{iQ}\cdot k_j)S(P|Q,R)_i,\quad
S(\emptyset|\emptyset)_i = 1\hfil\KLT\hfilneg\cr
\phi(P|Q) = {1\over s_P}\sum_{XY=P\atop AB=Q}
\big(\phi(X|A)\phi(Y|B) - \phi(Y|A)\phi(X|B)\big), \quad
\phi(i|j) = \d_{ij}\,,
}$$
where $\Phi(A|B)_i \equiv \phi(iA|iB)$.
The first instances are given by,
$$\openup2\jot\displaylines{
S(2|2)_1 = (k_1\cdot k_2),\quad S(23|23)_1 = (k_{12}\cdot
k_3)(k_1\cdot k_2),\quad S(23|32)_1 = (k_1\cdot k_3)(k_1\cdot k_2)\,,\cr
\Phi(2|2)_1 = {1\over s_{12}},\quad
\Phi(23|23)_1 = {1 \over s_{12} s_{123}} + { 1 \over s_{23} s_{123}},
\quad
\Phi(23|32)_1 = - { 1 \over s_{23} s_{123}} \, ,
}$$
where $S(32|32)_1$ and $\Phi(32|32)_1$ follow from $S(23|23)_1$ and $\Phi(23|23)_1$,
respectively, by relabeling $2\leftrightarrow 3$.
In \BGmap\ the notation $\sum_B$ instructs to sum
over all words $B$; the condition that $S(A|B)_i$ and $\Phi(A|B)_i$ are zero if
$B$ is not a permutation of $A$ leading to a finite sum.

The simplest applications of \BGmap\ to $K_P \rightarrow V_P$ are
\eqnn\VToM
$$\displaylines{
M_{12} = \Phi(2|2)_1V_{12} = {V_{12}\over s_{12}},\hfil\VToM\hfilneg\cr
M_{123} = \Phi(23|23)_1 V_{123} + \Phi(23|32)_1 V_{132}
= {V_{123}\over s_{12}s_{123}} + {V_{123}-V_{132} \over s_{23} s_{123}}\,,
}$$
and similarly $V_{12} = S(2|2)_1 M_{12} = s_{12}M_{12}$ as well as
\eqn\MToV{
V_{123} = S(23|23)_1 M_{123} + S(23|32)_1 M_{132}
= (s_{13}+s_{23})s_{12}M_{123} + s_{13}s_{12}M_{132}\,.
}
Consistency of the above relations can be checked by plugging the expressions \VToM\ into
\MToV\ and in general follows from \InvSPhi.

It is interesting to observe that the generalized Jacobi symmetries obeyed by
the local superfields are translated to shuffle symmetries under the BG map.
More explicitly, the BG superfields $\cK_P$ related to $K_P$ by \BGmap\
obey
\refs{\BGSym,\Gauge}
\eqn\shufsym{
\cK_{A\shuffle B} = 0,\quad\forall A,B\neq\emptyset\,.
}
Note that in writing \BGmap\ one needs
to fix the first letter of the word $P$ in both the local $K_P$ and
non-local $\cK_P$ representatives to be the same. This can be done with
\eqn\BGLieBasis{
K_{BiA} = - K_{i\ell(B)A}\,,\qquad
\cK_{BiA} = (-1)^{|B|}\cK_{i({\tilde B}\shuffle A)}\,.
}
The first relation follows from Baker's identity \BakerId\ while
the second was proven in \BGschocker.

In general, applying the BG map to each individual slot in a local building
block gives rise to its non-local Berends--Giele version.
Therefore knowing one representation suffices to obtain the other, for
example
\eqnn\TABCtoMABC
$$\eqalignno{
M_{aA,bB,cC} &\equiv \sum_{A',B',C'}\Phi(A|A')_a\Phi(B|B')_b\Phi(C|C')_c
T_{aA',bB',cC'}\,,&\TABCtoMABC\cr
T_{aA,bB,cC} &\equiv \sum_{A',B',C'}S(A|A')_a S(B|B')_b S(C|C')_c M_{aA',bB',cC'}\,.
}$$
However, it is conceptually simpler (but equivalent to \TABCtoMABC)
to define the BG counterpart of
$T_{A,B,C}$ by directly using non-local multiparticle superfields in \BGmap,
for example \partI
\eqn\MABCdef{
M_{A,B,C} \equiv {1\over 3}(\l\ga_m \cW_A)(\l\ga_n \cW_B)\cF^{mn}_C +
\cyc(A,B,C)\,.
}
The relation between $M_{A,B,C} $ and $T_{A,B,C}$ straightforwardly generalizes
to the tensorial, refined and anomalous kinematic factors in section \LocalBBsec. We
will use parental letters $M^{m_1\ldots m_r}_{B_1,\ldots,B_{r+3}} $,
${\cal J}^{m_1\ldots m_r}_{A_1,\ldots,A_d|B_1,\ldots,B_{d+r+3}} $
and ${\cal Y}^{m_1\ldots m_r}_{A_1,\ldots,A_d|B_1,\ldots,B_{d+r+5}} $ for the Berends--Giele
versions of $T^{m_1\ldots m_r}_{B_1,\ldots,B_{r+3}} $,
$J^{m_1\ldots m_r}_{A_1,\ldots,A_d|B_1,\ldots,B_{d+r+3}}$
and $Y^{m_1\ldots m_r}_{A_1,\ldots,A_d|B_1,\ldots,B_{d+r+5}}$, respectively.
The shuffle symmetry \shufsym\ applies to every slot of a BG current, e.g.\
$M_{R\shuffle S,B,C} = 0$ or $\cY^{mn\ldots}_{A,B,R\shuffle S,D,\ldots} = 0$
for $R,S \neq\emptyset$.

In addition to changing the symmetry properties within each word,
the BG map also modifies the behavior under a BRST variation. The
characteristic terms proportional to the momentum contraction
$(k_X\cdot k_j)$ in the BRST variation of the local superfields become
a simpler deconcatenation sum,
\eqn\QVsQMs{
QV_P = \sumQ{P}{X}{j}{Y}{R}{S}V_{XR}V_{jS}\quad \Longleftrightarrow\quad
QM_P = \sum_{P=XY}M_X M_Y\,,
}
where $M_P$ is related to $V_P$ by \BGmap. In general, one
can show that the BRST variations of $K_P$ in \GeneralQ\ are mapped
to the following variations of their Berends--Giele counterparts (note
$M_P\equiv \cV_P$) \EOMbbs
\eqnn\QBGs
$$\eqalignno{
QM_{12\ldots p} &= \sum^{p-1}_{j=1} M_{12\ldots j}M_{j+1\ldots p}\,,
&\QBGs\cr
Q\cA^m_{12\ldots p} &= (\lambda \gamma^m \cW_{12\ldots p})
+ k_{12\ldots p}^m M_{12\ldots p} + \sum_{j=1}^{p-1}(M_{12\ldots j} \cA^m_{j+1\ldots p}
-  M_{j+1\ldots p} \cA^m_{12\ldots j})\,,\cr
Q\cW^\alpha_{12\ldots p} & ={1\over 4}(\lambda \gamma_{mn})^\alpha \cF^{mn}_{12\ldots p}
+ \sum^{p-1}_{j=1} (M_{12\ldots j}\cW^\alpha_{j+1\ldots p}
- M_{j+1\ldots p}\cW^\alpha_{12\ldots j})\,,\cr
Q\cF^{mn}_{12\ldots p} &= k^{[m}_{12\ldots p} (\lambda \gamma^{n]} \cW_{12\ldots p})
+ \sum^{p-1}_{j=1} (M_{12\ldots j}\cF^{mn}_{j+1\ldots p}-M_{j+1\ldots p}\cF^{mn}_{12\ldots j})\cr
&\quad{} + \sum^{p-1}_{j=1} \big[ \cA^{[n}_{12\ldots j} (\lambda \gamma^{m]} \cW_{j+1\ldots p})
-\cA_{j+1\ldots p}^{[n} (\lambda \gamma^{m]} \cW_{12\ldots j}) \big]\,.
}$$
In the appendix of \EOMbbs\ an alternative relation between the local and the
non-local superfields was given in terms of a diagrammatic map using planar
binary trees. Yet one more relation between these objects will be given below
in terms of the so-called {\it S-map}. In summary, there is a multitude of
perspectives on how these superfields are defined and the relations among
them.

\newsubsubsec\smapsubsec The S-map between local and non-local superfields

In this subsection we will describe the so-called S-map which relates
local and non-local superfield representations. This map originally
appeared in the appendix of \EOMbbs\ as a way to encode the BCJ relations among
tree amplitudes and to rewrite scalar BRST cohomology objects in terms of super
Yang--Mills trees.

After
defining a {\it weighted} concatenation product $\otimes^s$ of Berends--Giele
superfields by
\eqn\wconcdef{
\cK_{Ai}\otimes^s \cK_{jB} \equiv s_{ij}\cK_{AijB}\,,
}
the definition of the S-map can be written as
\eqn\Smap{
\cK_{S[A,B]} \equiv (-1)^{|B|+1} \cK_{\rho(A)}\otimes^s \cK_{\tilde\rho(B)}\,,
}
where $\rho(B)$ is defined in \rhomap\ and $\tilde\rho(B)$ denotes
the reversal of $\rho(B)$.
For example, given $\rho(123)= 123 - 132 - 312 + 321$ and
$\tilde\rho(45) = 54-45$, the
S-map $\cK_{S[123,45]}$ yields
\eqnn\SmapEx
$$\eqalignno{
\cK_{S[123,45]} &= (-1)^3\cK_{(123-132-312+321)} \otimes^s {\cal K}_{(54-45)} =
\cK_{(123-132-312+321)} \otimes^s {\cal K}_{(45-54)} \cr
&=s_{34} \cK_{12345} - s_{35} \cK_{12354} - s_{24}\cK_{13245} + s_{25}\cK_{13254} -
s_{24}\cK_{31245} &\SmapEx\cr
&\quad{}+ s_{25}\cK_{31254} + s_{14}\cK_{32145} - s_{15}\cK_{32154}\,,
}$$
and simpler cases include
\eqn\simpSmap{
{\cal K}_{S[1,2]}=s_{12} {\cal K}_{12} \ , \ \ \ \ 
{\cal K}_{S[12,3]}= s_{23} {\cal K}_{123} - s_{13} {\cal K}_{213} \ .
}
A curious property of the S-map is that its iteration over all
letters in a given word yields a translation between the
Berends--Giele currents and its local counterparts in a way
that preserves the bracketing structure.
More precisely,
\eqnn\iterS
$$\displaylines{
\cK_{S[1,2]} = K_{[1,2]},\quad
\cK_{S[S[1,2],3]} = K_{[[1,2],3]},\quad
\cK_{S[S[S[1,2],3],4]} = K_{[[[1,2],3],4]}\cr
\cK_{S[1,S[S[2,3]],4]} = K_{[1,[[2,3],4]]} \, ,\hfil\iterS\hfilneg
}$$
see \Vex\ for a discussion on the bracketing notation for
local superfields in the BCJ gauge. Note that the S-map plays a key role
in deriving BCJ relations \BernQJ\ of SYM tree amplitudes from the BRST
cohomology \MafraVCA, and that the definition \Smap\ is equivalent\foot{Their
equivalence follows from the identity $\rho(A)=\sum_{XjY=A}(X\shuffle \tilde Y)j
(-1)^{|Y|}$ \michos.} to
\eqn\QEone{
{\cal K}_{S[A,B]} \equiv \sum_{i=1}^{|A|} \sum_{j=1}^{|B|} (-1)^{i-j+|A|-1}
s_{a_i b_j} {\cal K}_{(a_1 a_2\ldots a_{i-1} \shuffle a_{|A|} a_{|A|-1}\ldots a_{i+1})a_ib_j
(b_{j-1}\ldots b_2 b_1 \shuffle b_{j+1} \ldots b_{|B|})}
}
for $A=a_1 a_2\ldots a_{|A|}$ and $B=b_1 b_2\ldots b_{|B|}$.

\newsubsec\BRSTpseudosec BRST pseudo-invariants

Following \partI, let us now consider the non-local versions of the local
building blocks discussed above. As mentioned in the previous section, they are
denoted by $M^{m \ldots}_{A, \ldots}$ or by the calligraphic letter of its local
counterpart. The BRST variations for the simplest cases can be written as
\eqnn\QMs
$$\eqalignno{
Q M_{A,B,C} &=
\sum_{XY=A} \big[ M_{X}  M_{Y,B,C}
- (X\leftrightarrow Y)\big] + (A \leftrightarrow B,C) \,, &\QMs\cr
QM^m_{A,B,C,D} &= k^m_{A} M_A M_{B,C,D}
+\!\!\! \sum_{XY=A}\!\! \big[M_{X} M^m_{Y,B,C,D} - (X\leftrightarrow Y)\bigr]
+(A \leftrightarrow B,C,D) \,,\cr
Q M^{mn}_{A,B,C,D,E} &=\delta^{mn}\cY_{A,B,C,D,E}\cr
&\quad{}+ k_{A}^{(m} M_{A} M^{n)}_{B,C,D,E} +(A \leftrightarrow B,C,D,E)\cr
&\quad{} +\!\!\! \sum_{XY=A}\!\! \big[M_{X} M^{mn}_{Y,B,C,D,E}
- (X\leftrightarrow Y)\bigr]  +(A \leftrightarrow B,C,D,E)\,,
}$$
and mirror the structure of their local counterparts \QTABC, \QTm\ and \BRSTmn, respectively.
For refined building blocks, the appearance of $V_{[A,B]}$ on the right-hand side of $QJ_{A|B,\ldots}$ and its
generalizations translates into $M_{S[A,B]}$ under the Berends--Giele map, for instance
\eqnn\pseudoo
$$\eqalignno{
Q{\cal J}_{A|B,C,D,E} &=  {\cal Y}_{A,B,C,D,E} + k_A^m M_A M^m_{B,C,D,E} &\pseudoo\cr
&\quad{}+ M_{S[A,B]} M_{C,D,E} + (B \leftrightarrow C,D,E) \cr
&\quad{}+\! \sum_{XY=B}\!\! \bigl(M_{X} {\cal J}_{A|Y,C,D,E}  - (X\leftrightarrow Y)\bigr)
+ (B \leftrightarrow C,D,E)\cr
&\quad{}+\! \sum_{XY=A}\!\! \bigl(M_X {\cal J}_{Y|B,C,D,E} - (X\leftrightarrow Y)\bigr)\,.
}$$
See \partI\ for more details and examples.

\newsubsubsec\BRSTinvsec BRST invariants

The Berends--Giele currents were shown in \partI\ to be the natural
building blocks in constructing recursion relations for BRST (pseudo-)invariants.
For instance, it was shown using \QMs\ that the following
definitions are BRST invariant:
\eqnn\scalarCs
$$\eqalignno{
C_{1|2,3,4} &\equiv M_1 M_{2,3,4}\,, &\scalarCs \cr
C_{1|23,4,5} &\equiv
M_1 M_{23,4,5} + M_{12} M_{3,4,5} - M_{13} M_{2,4,5}\,,\cr
C_{1|234,5,6} &\equiv M_1 M_{234,5,6} + M_{12}M_{34,5,6}
+ M_{123}M_{4,5,6} - M_{124}M_{3,5,6}\cr
&\quad{}- M_{14}M_{23,5,6} - M_{142}M_{3,5,6} + M_{143}M_{2,5,6}\,, \cr
C_{1|23,45,6}
&\equiv M_1 M_{23,45,6} + M_{12}M_{45,3,6} - M_{13}M_{45,2,6}
+ M_{14}M_{23,5,6} - M_{15}M_{23,4,6}\cr
&\quad{}-M_{412}M_{3,5,6} + M_{314}M_{2,5,6}
+ M_{215}M_{3,4,6}- M_{315}M_{2,4,6}\,.\cr
}$$
Similarly, their vectorial upgrades are also BRST closed:
\eqnn\vectorCs
$$\eqalignno{
C^m_{1|2,3,4,5} &\equiv
M_1 M^m_{2,3,4,5} + \big[ k_2^m M_{12} M_{3,4,5}
+ (2\leftrightarrow 3,4,5) \big]\,,&\vectorCs\cr
C^m_{1|23,4,5,6}
&\equiv M_1 M^m_{23,4,5,6} + M_{12} M^m_{3,4,5,6} - M_{13} M^m_{2,4,5,6}
+ k^m_3 M_{123}M_{4,5,6} - k^m_2 M_{132}M_{4,5,6} 
\cr
&\quad{}+\big[ k^m_4 M_{14}M_{23,5,6} - k^m_4
M_{214}M_{3,5,6} + k^m_4 M_{314}M_{2,5,6} + (4\leftrightarrow 5,6)\bigr] \, .
}$$
These superfields are in the BRST cohomology and were dubbed {\it BRST
invariants} in \EOMbbs. In general, a recursion relation was written down in
\partI\ for these scalar and vector cohomology elements at arbitrary
multiplicities. An alternative algorithm to generate the above combinations of
Berends--Giele currents (and those of the subsequent tensorial generalizations
$C^{m_1\ldots m_r}_{1|A_1,\ldots,A_{r+3}}$) is described in appendix
\Unifiedclikeapp.

\newsubsubsec\anomaloussec BRST pseudo-invariants

The BRST variations of higher-rank tensors no longer vanish but they are
proportional to superfields with an anomalous factor of ${\cal Y}^{m_1\ldots
m_r}_{A_1,\ldots,A_d| B_1,\ldots,B_{d+r+5}}$ in each term. Superspace expressions
with a purely anomalous $Q$ variation are referred to as {\it BRST pseudo-invariant}. 
Similarly, a general recursion was written down in \partI\
and the first non-trivial instance is given by
\eqnn\tenssixpt
$$\eqalignno{
C^{mn}_{1|2,3,4,5,6} & =
M_1M^{mn}_{2,3,4,5,6} +  \big[ k_2^{m} M_{12}M^n_{3,4,5,6}+k_2^{n}
M_{12}M^m_{3,4,5,6} + (2\leftrightarrow 3,4,5,6) \big] \cr
&\quad{} -  \big[ (k_2^{m} k_3^{n} + k_2^{n}k_3^{m})
M_{213}M_{4,5,6} + (2,3|2,3,4,5,6) \big] &\tenssixpt
}$$
and it satisfies
\eqn\QCmn{
QC^{mn}_{1|2,3,4,5,6} = -\d^{mn}M_1\cY_{2,3,4,5,6}\ .
}
Unrefined pseudo-invariants $C^{m_1\ldots m_r}_{1|A_1,\ldots,A_{r+3}}$ of
arbitrary tensor ranks can be characterized by a leading term $M_1 M^{m_1\ldots
m_r}_{A_1,\ldots,A_{r+3}}+\ldots$ as in \scalarCs\ to \tenssixpt. Then, the
recursions of \partI\ adjoin a tail of completions via $\sim M_{1B}$ with $B\neq
\emptyset$ such that $QC^{m_1\ldots m_r}_{1|A_1,\ldots,A_{r+3}}$ is purely
anomalous. Similarly, one can start from a refined leading term $M_1 {\cal
J}^{m_1\ldots m_r}_{A | B_1,\ldots,B_{r+4}}$, and the recursions of \partI\
generate completions such as
\eqnn\pseudoPsix
$$\eqalignno{
P_{1|2|3,4,5,6} &=  M_1 {\cal J}_{2|3,4,5,6} + M_{12} k_2^m M_{3,4,5,6}^m
+ \bigl[ s_{23} M_{123} M_{4,5,6} + (3 \leftrightarrow 4,5,6)\bigr]\,, &\pseudoPsix
}$$
where refined and unrefined terms are mixed and the BRST variation becomes purely anomalous
\eqnn\pseudoQPsix
$$\eqalignno{
QP_{1|2|3,4,5,6} &= - M_1\cY_{2,3,4,5,6}  \,. &\pseudoQPsix
}$$
The same logic applies to higher degrees $d$ of refinement where
pseudo-invariants $P^{m_1\ldots m_r}_{1|A_1,\ldots,A_d
| B_1,\ldots,B_{d+r+3}}$ are defined from recursively generated completions
of the leading term $M_1 {\cal J}^{m_1\ldots m_r}_{A_1,\ldots,A_d
| B_1,\ldots,B_{d+r+3}}$.

\newsubsubsubsec\symmCPone Symmetries of pseudo-invariants

Following our convention for subscripts with words $A_1,A_2,\ldots$ separated by
commas, the most general pseudo-invariant $P^{m_1\ldots m_r}_{1|A_1,\ldots,A_d
| B_1,\ldots,B_{d+r+3}}$ is separately symmetric in the refined slots $A_i$
and the unrefined slots $B_j$ but not under exchange of $A_i$ with $B_j$. Also,
the shuffle-symmetries \shufsym\ of their leading terms propagate to the
pseudo-invariants, e.g.
\eqn\shufsymCP{
C^{m_1 m_2 \ldots m_r}_{1|A_1,\ldots,R\shuffle S,\ldots,A_{r+3}} = 0,\quad\forall R,S\neq\emptyset\,,
}
and the same is true for both types of slots $A_i$ and $B_j$ of
$P^{m_1\ldots m_r}_{1|A_1,\ldots,A_d | B_1,\ldots,B_{d+r+3}}$.

\newsubsubsec\symmCPtwo $Q$ variations of pseudo-BRST invariants

The $Q$ variations \QCmn\ and \pseudoQPsix\ can be compactly generalized
to higher multiplicity by means of BRST invariant combinations of anomalous
superfields such as
\eqnn\GAMa
$$\eqalignno{
\Gamma_{1|2,3,4,5,6} &\equiv M_1 {\cal Y}_{2,3,4,5,6}\,, &\GAMa \cr
\Gamma_{1|23,4,5,6,7} &\equiv
M_1 {\cal Y}_{23,4,5,6,7} + M_{12} {\cal Y}_{3,4,5,6,7} - M_{13} {\cal Y}_{2,4,5,6,7}
\cr
\Gamma^m_{1|2,3,4,5,6,7} &\equiv
M_1 {\cal Y}^m_{2,3,4,5,6,7} + \big[ k_2^m M_{12} {\cal Y}_{3,4,5,6,7}
+ (2\leftrightarrow 3,4,5,6,7) \big] \cr
\Gamma_{1|2|3,4,5,6,7,8} &\equiv  M_1 {\cal Y}_{2|3,4,5,6,7,8} + M_{12} k_2^m {\cal Y}_{3,4,5,6,7,8}^m
+ \bigl[ s_{23} M_{123} {\cal Y}_{4,5,6,7,8} + (3 \leftrightarrow 4,\ldots,8)\bigr]\,.
}$$
The combinatorics of these expressions is identical to ghost-number three
superfields $C_{1|2,3,4}, C_{1|23,4,5},C^m_{1|2,3,4,5}$ and $P_{1|2|3,4,5,6} $
in \scalarCs, \vectorCs\ and \pseudoPsix. Accordingly, the expansion of refined
and tensorial invariants $\Gamma^{m_1\ldots m_r}_{1|A_1,\ldots,A_d
| B_1,\ldots,B_{d+r+5}}$ in terms of $M_A {\cal Y}^{m_1\ldots}_{\ldots}$ can
be inferred from the analogous $P^{m_1\ldots m_r}_{1|A_1,\ldots,A_d
| B_1,\ldots,B_{d+r+3}}$ after obvious adjustments in the number of unrefined
slots\foot{The mismatches in the numbers of slots, say between $C^m_{1|2,3,4,5}$
and the counterpart $\Gamma^m_{1|2,3,4,5,6,7}$, are accounted for by the more
rigorous definition of $\Gamma_{1|A,\ldots}$ in section~8 of \partI, or in the
appendix~\Unifiedclikeapp.}. These ghost-number four superfields capture the
most general BRST variation
\eqnn\GAMb
$$\eqalignno{
Q C^{m_1\ldots m_r}_{1|A_1,\ldots,A_{r+3}}  &=
-\delta^{(m_1 m_2} \Gamma^{m_3\ldots m_r)}_{1|A_1,\ldots,A_{r+3}}\,, &\GAMb\cr
Q P^{m_1\ldots m_r}_{1|A_1,\ldots,A_d |B_1,\ldots,B_{d+r+3}} &=
-\delta^{(m_1 m_2} \Gamma^{m_3\ldots m_r)}_{1|A_1,\ldots,A_d | B_1,\ldots,B_{d+r+3}}\cr
&\quad{}- \Gamma^{m_1\ldots m_r}_{1|A_2,\ldots,A_d |A_1,B_1,\ldots,B_{d+r+3}}
+ (A_1\leftrightarrow A_2,\ldots, A_d)
}$$
of the above ghost-number three pseudo-invariants.

\newsubsubsec\tracesM Trace relations

The trace relations \HOb\ to \HOc\ of the local building blocks
straightforwardly generalize under the Berends--Giele map. Moreover, the
(pseudo-)invariants inherit the trace relations of their leading term, e.g.
\eqn\pseudoTRA{
\delta_{np} C^{npm_1\ldots m_r}_{1|B_1,B_2,\ldots,B_{r+5}} =
2\big[ P^{m_1\ldots m_r}_{1|B_1|B_2,\ldots,B_{r+5}}
+ (B_1\leftrightarrow B_2,\ldots,B_{r+5})\big]\,,
}
and at generic degree of refinement,
\eqn\pseudoTRB{
\delta_{np} P^{npm_1\ldots m_r}_{1|A_1,\ldots,A_d|B_1,\ldots,B_{d+r+5}} =
2\big[ P^{m_1\ldots m_r}_{1|A_1,\ldots,A_d,B_1|B_2,\ldots,B_{d+r+5}}
+ (B_1{\leftrightarrow} B_2,\ldots,B_{d+r+5})\big]\,.
}
These relations will play a key role for the modular properties of the
correlators after integration over $\ell$.

\newsubsec\Deltasec Anomaly counterparts of BRST invariants

We shall now review an interesting class of anomaly superfields
$\Delta^{\ldots}_{1|\ldots}$ that enter relations between the above
(pseudo-)invariants of different tensor ranks. Similar to the formal operation
$M^{m \ldots}_{B,C,\ldots} \rightarrow \cY^{m \ldots}_{B,C,\ldots}$ that
translates the $C^{\ldots}_{1|\ldots}$ into the ghost-number four objects
$\Gamma^{\ldots}_{1|\ldots}$ in section \symmCPtwo, one can generate anomaly
building blocks of ghost-number three via
\eqn\naivetrans{
M_A M^{m \ldots}_{B,C,\ldots}\rightarrow \cY^{m \ldots}_{A,B,C,\ldots}\,,\qquad
M_A \cJ^{m \ldots}_{B|C,\ldots}\rightarrow \cY^{m \ldots}_{B|A,,C,\ldots}\,,
}
by adjusting the number of slots in the obvious manner (also see the alternative algorithm
in appendix \Unifiedclikeapp). Then, the Berends--Giele expansions of the simplest
instances of the scalars \scalarCs\ and vectors \vectorCs\ translate into,
\eqnn\deltasecA
$$\eqalignno{
\Delta_{1|2,3,4,5} &\equiv {\cal Y}_{1,2,3,4,5}\cr
\Delta_{1|23,4,5,6} &\equiv
\cY_{1,23,4,5,6}
+ \cY_{12,3,4,5,6}
- \cY_{13,2,4,5,6} &\deltasecA\cr
\Delta^m_{1|2,3,4,5,6} &\equiv
{\cal Y}^m_{1,2,3,4,5,6} + \big[ k_2^m {\cal Y}_{12,3,4,5,6}
+ (2\leftrightarrow 3,4,5,6) \big]\,.
}$$
As discussed in \partI, it turns out that all the
unrefined superfields $\Delta_{1|A,\ldots}$ are BRST exact after using
momentum conservation,
\eqn\deltasecB{
k_1+ k_{B_1} + \ldots+ k_{B_{r+4}} = 0 \ \ \ \Rightarrow \ \ \ 
\langle \Delta^{m_1 m_2\ldots m_r}_{1|B_1,B_2,\ldots,B_{r+4}} \rangle = 0\,.
}
However, their refined counterparts are non-zero in the BRST cohomology, i.e.
\eqn\notBexact{
\langle \Delta^{m_1 m_2\ldots m_r}_{1|A_1,\ldots ,A_d|B_1,B_2,\ldots,B_{d+r+4}} \rangle
\neq 0\,, \qquad d \geq1\,.
}
For example, the simplest refined anomaly superfield is given by
\eqn\deltasecC{
\Delta_{1|2|3,4,5,6,7}  = {\cal Y}_{2|1,3,4,5,6,7}
+ k_2^m {\cal Y}_{12,3,4,5,6,7}^m + \bigl[ s_{23} {\cal Y}_{123,4,5,6,7}
+ (3 \leftrightarrow 4,5,6,7)\bigr]\,,
}
and follows the combinatorics of $P_{1|2|3,4,5,6}$ in \pseudoPsix\ according to
the translation \naivetrans. Its BRST variation is easily verified to be
\eqnn\manlocOT
$$\eqalignno{
Q\Delta_{1|2|3,4,5,6,7} &= V_1 k_2^m Y^{m}_{2,3,\ldots,7}
- V_{12} Y_{3,4,\ldots,7} + \big[ V_1 Y_{23,4,5,6,7}
+ (3\leftrightarrow 4,\ldots,7) \big] \cr
&= k_2^m \Gamma^m_{1|2,3,\ldots,7} +  \big[ s_{23}  \Gamma_{1|23,4,5,6,7}
+ (3\leftrightarrow 4,\ldots,7) \big] \, ,
&\manlocOT
}$$
see section \symmCPtwo\ for the ghost-number four invariants $\Gamma_{1|\ldots}^{\ldots}$.

\newsubsubsec\localdel Locality of $\langle\Delta_{1|2|3,4,5,6,7}\rangle$

In contrast to the naive expectation from its slot structure, the component
expansion of \deltasecC\ is a {\it local} expression. The poles in $s_{12}$ and
$s_{12j}$ with $j=3,\ldots,7$ in the Berends--Giele currents of \deltasecC\ do
not propagate to the components for the following reasons:
\medskip
\item{$\bullet$}
The components $\langle {\cal Y}_{123,4,5,6,7} \rangle$ cannot have a pole in
$s_{123}^{-1}$ since this would conflict with the vanishing of $\langle
\Delta_{1|234,5,6,7} \rangle=\langle {\cal Y}_{123,4,5,6,7}+\half {\cal
Y}_{12,34,5,6,7} \rangle+{\rm cyc}(1,2,3,4)$ by \notBexact: The Berends--Giele
currents in the cyclic permutations do not introduce any additional pole in
$s_{123}$ to the right-hand side.

\item{$\bullet$} The trace relations among tensorial $\Delta_{1|\ldots}$ in
section~10 of \partI\ include
\eqn\lemlocal{
\half \langle \Delta^{mm}_{1|2,3,4,5,6,7} \rangle = \langle Y_{1|2,3,4,5,6,7}
+ \big[ \Delta_{1|2|3,4,5,6,7} +(2\leftrightarrow 3,4,5,6,7)\big]\rangle\,.
}
Since the left-hand side vanishes by \notBexact\ and each pole $s_{1j}^{-1}$ can
occur in no term other than $\langle \Delta_{1|j|\ldots} \rangle$ on the
right-hand side, the respective residues must be zero.
\medskip
\noindent
This superspace argument is confirmed by the bosonic components
\eqn\delsevcomp{
\langle\Delta_{1|2|3,4,5,6,7}\rangle =
-\half(e_1\cdot e_2)\epsilon_{10}^{k_3k_4k_5k_6k_7e_3e_4e_5e_6e_7} +
\hbox{ fermions}\,,
}
that have been obtained from an automated calculation using \PSS.
We are using the schoonship convention of writing the contracted vectors
as $\e_{10}^{ \ldots m \ldots}k_m \equiv \e_{10}^{ \ldots k \ldots}$.
The locality of $\langle\Delta_{1|2|3,4,5,6,7}\rangle$ will be important in
section~\sevenwssec\ for asserting that the seven-point correlator comprising
$\Delta_{1|2|3,4,5,6,7}$ (and permutations) is local.

\newsubsubsec\localeigh Eight-point examples

Similarly, the eight-point topologies of $\Delta_{1|A|B,C,D,E,F}$ are obtained
by applying the rules \naivetrans\
to the expressions of the refined superfields $P_{1|A|B,C,D,E}$ at seven points,
see \eightDeltas\ for the expansion of $\Delta_{1|23|4,5,6,7,8} $, $\Delta_{1|2|34,5,6,7,8}$
and $\Delta^m_{1|2|3,4,5,6,7,8}$ in terms of Berends--Giele currents.
It is straightforward to show that their BRST variations are given by \partI
\eqnn\eightDQBRST
$$\eqalignno{
Q \Delta_{1|23|4,5,6,7,8} &= k_{23}^m \Gamma^m_{1|23,4,\ldots,8} + \Gamma_{1|3|2,4,\ldots,8}
-\Gamma_{1|2|3,4,\ldots,8}  \cr
&\quad{}+ \big[  s_{34} \Gamma_{1|234,5,6,7,8} - s_{24}\Gamma_{1|243,5,6,7,8}
+ (4\leftrightarrow 5,6,7,8) \big]\cr
Q \Delta_{1|2|34,5,6,7,8} &= k_{2}^m \Gamma^m_{1|2,34,5,\ldots,8} + s_{23}\Gamma_{1|234,5,\ldots,8} 
- s_{24}\Gamma_{1|243,5,\ldots,8}&\eightDQBRST\cr
&\quad{}+ \big[  s_{25} \Gamma_{1|34,25,6,7,8} + (5\leftrightarrow 6,7,8) \big]\cr
Q \Delta^m_{1|2|3,4,5,6,7,8} &= k_{2}^p \Gamma^{pm}_{1|2,3,4,\ldots,8}
- k_2^m \Gamma_{1|2|3,4,\ldots,8} + \big[  s_{23} \Gamma^m_{1|23,4,\ldots,8}
+ (3\leftrightarrow 4,\ldots,8) \big]
}$$
in the momentum phase space with $k_{12\ldots 8}=0$. Since these superfields
will play a role in the construction of the eight-point correlator later on,
it is important to know about the kinematic poles in their component expansions.
The first hint on non-locality comes from rewriting the BRST variations
above after plugging in the definitions of the superfields $\Gamma$ from \partI.
Unlike the local expression \manlocOT\ at seven points, some kinematic poles survive,
e.g.\ $s_{12}^{-1},s_{13}^{-1}, s_{23}^{-1}$ in $Q \Delta_{1|23|4,5,6,7,8}$,
see \manlocOR\ for the full expressions.

Given the correspondence between BRST variations in pure-spinor superspace and
gauge variations in components discussed in the appendix B of \partI, the above
relations indicate that the gauge variations of the eight-point anomaly
superfields contain $s_{ij}^{-1}$ poles. Therefore, the component expansions of
$\langle\Delta_{1|A|B,C,D,E,F}\rangle$ for the eight-point topologies are not
local, see their explicit expansions in the appendix~\Deltaapp.

\newsubsec\CJACOBIsec BRST cohomology identities

We have seen that the family of anomaly building blocks $\Delta_{1|A,\ldots}$ is
obtained by redistributing the slots of $M_A M_{B,C,\ldots}$ in the
Berends--Giele expansion of (pseudo-)invariants. The same procedure can be
applied to derive BRST generators at ghost-number two whose $Q$-variation gives
rise to relations between (pseudo-)invariants, see section 8 to 10 of \partI.
For instance, translating the combinatorics of $\Delta_{1|2|3,4,5,6,7}$ in
\deltasecC\ into a non-anomalous context (and truncating the number of slots in
the obvious manner) yields
\eqn\deltasecD{
D_{1|2|3,4,5}  = {\cal J}_{2|1,3,4,5} + k_2^m M_{12,3,4,5}^m
+ \bigl[ s_{23} M_{123,4,5} + (3 \leftrightarrow 4,5)\bigr]\,,
}
which generates the kinematic Jacobi identity \towardsoneloop
\eqn\deltasecE{
Q D_{1|2|3,4,5}  = k_2^m C^m_{1|2,3,4,5}
+ \big[ s_{23} C_{1|23,4,5} + (3\leftrightarrow 4,5) \big]+
\Delta_{1|2,3,4,5}\,.
}
In a five-point momentum phase space, the last term $\Delta_{1|2,3,4,5}$ drops
out from the cohomology, cf.~\deltasecB. Of course, one could have obtained
\deltasecD\ directly by applying specializations like $M_A
M^{m_1\ldots}_{B,C,\ldots} \rightarrow M^{m_1\ldots}_{A,B,C,\ldots}$ to each
term in the Berends--Giele expansion \pseudoPsix\ of $P_{1|2|3,4,5,6}$. In a
similar manner, one can infer generalizations of $D_{1|2|3,4,5}$ by
redistributing the slots of appropriate $P^{m_1\ldots}_{1|A_1,\ldots,
A_d|B_1,\ldots}$, see section 8 of \partI\ for details. The resulting Jacobi identities
include the six-point cases
\eqnn\sixptQDs
$$\eqalignno{
Q D_{1|4|23,5,6} &= k^m_4 C^m_{1|23,4,5,6} +s_{24} C_{1|423,5,6} -s_{34}
C_{1|432,5,6}   &\sixptQDs\cr
&\quad{}+s_{45} C_{1|23,45,6} +s_{46}C_{1|23,46,5} + \Delta_{1|23,4,5,6}\,, \cr
Q D_{1|23|4,5,6} &= k_{23}^m C^m_{1|23,4,5,6}  + 
\big[  s_{34} C_{1|234,5,6} - s_{24} C_{1|324,5,6}
+ (4\leftrightarrow 5,6) \big]\cr
&\quad{}+ P_{1|3|2,4,5,6} - P_{1|2|3,4,5,6} + \Delta_{1|23,4,5,6}\,, \cr
Q D^n_{1|2|3,4,5,6} &= k^m_2 C^{mn}_{1|2,3,4,5,6} + \big[
s_{23} C^n_{1|23,4,5,6} + (3\leftrightarrow 4,5,6) 
\big] - k_2^n P_{1|2|3,4,5,6}  + \Delta^n_{1|2,3,4,5,6}\,,
}$$
and closed formulae for arbitrary multiplicity and tensor rank 
can be written down using the S-map \Smap, e.g.
\eqnn\JRal
$$\eqalignno{
Q D^{m_1 m_2 \ldots m_r}_{1|A|B_1,\ldots,B_{r+3}} &=
  k_A^p C^{pm_1\ldots m_r}_{1|A,B_1,\ldots,B_{r+3}}
  +\big[ C^{m_1\ldots m_r}_{1| S[A,B_1],B_2,\ldots,B_{r+3}}
  + (B_1 \leftrightarrow B_2,\ldots,B_{r+3}) \big]  \cr
& \quad{}+ (k_{1AB_1\ldots B_{r+3}}^{(m_1}
-k_A^{(m_1} ) P^{m_2\ldots m_r)}_{1|A|B_1,\ldots,B_{r+3}}
+ \delta^{(m_1 m_2} \Delta^{m_3\ldots m_r)}_{1|A|B_1,\ldots,B_{r+3}} &\JRal\cr
&\quad{}+\Delta^{m_1 m_2 \ldots m_r}_{1 | A,B_1,\ldots,B_{r+3}}
+\!\!\sum_{XY=A}\!\! ( P^{m_1\ldots m_r}_{1|Y| X,B_1,\ldots,B_{r+3}}
-P^{m_1\ldots m_r}_{1| X| Y,B_1,\ldots,B_{r+3}} )\,.
}$$
One can also derive kinematic Jacobi identities for refined (pseudo-)invariants
\eqnn\deltasecF
$$\eqalignno{
Q(\ldots) &= k_3^m P^m_{1|2|3,4,5,6,7} - s_{23} P_{1|23|4,5,6,7}
+ \big[ s_{34} P_{1|2|34,5,6,7} +(4\leftrightarrow 5,6,7) \big]+ \Delta_{1|3|2,4,5,6,7} \cr
Q(\ldots) &= k_4^m P^m_{1|23|4,\ldots,8} + s_{24} P_{1|324|5,6,7,8}
- s_{34} P_{1|234|5,6,7,8} \cr
&\quad{}+ \big[ s_{45} P_{1|23|45,6,7,8} +(5\leftrightarrow 6,7,8) \big]
+ \Delta_{1|4|23,5,6,7,8} \cr
Q(\ldots) &= k_{23}^m P^m_{1|4|23,5,6,7,8}
+s_{34} P_{1|234|5,6,7,8}  - s_{24} P_{1|324|5,6,7,8}
- P_{1|2,4|3,5,6,7,8}
&\deltasecF  \cr
&\quad{}+ P_{1|3,4|2,5,6,7,8}+\big[ s_{35} P_{1|4|235,6,7,8}
- s_{25} P_{1|4|325,6,7,8} +(5\leftrightarrow 6,7,8) \big]+ \Delta_{1|23|4,\ldots,8} \cr
Q(\ldots) &= k_{5}^m P^m_{1|4|23,5,6,7,8} - s_{45} P_{1|45|23,6,7,8}
+ s_{25} P_{1|4|325,6,7,8} - s_{35} P_{1|4|235,6,7,8} \cr
&
\quad{}+\big[  s_{56} P_{1|4|23,56,7,8} +(6\leftrightarrow 7,8) \big]
+ \Delta_{1|5|23,4,6,7,8} \cr
Q(\ldots) &= k_3^m P^{mn}_{1|2|3,\ldots,8} - s_{23} P^n_{1|23|4,\ldots,8}
+ \big[ s_{34} P^n_{1|2|34,5,6,7,8} +(4\leftrightarrow 5,6,7,8) \big]  \cr
&\quad{}- k_3^n P_{1|2,3|4,\ldots,8}+ \Delta^n_{1|3|2,4,\ldots,8}
}$$
by taking suitable combinations of $D^{m_1\ldots}_{1|A_1,\ldots,A_d|B_1,\ldots}$
at various degrees $d$ of refinement as a BRST ancestor on the left-hand side.

Finally, the generalization of the Jacobi identity \JRal\ to (pseudo-)invariants
of arbitrary degree of refinement reads \partI
\eqnn\allJacv
$$\eqalignno{
& QD^{m_1\ldots m_r}_{1|A_1,\ldots,A_d|B_1,\ldots,B_{r+d+2}} =
\big[ \Delta^{m_1\ldots m_r}_{1|A_2,\ldots,A_d|A_1,B_1,\ldots,B_{r+d+2}}
+ (A_1\leftrightarrow A_2,\ldots,A_d) \big]\cr
&+   \delta^{(m_1 m_2}   \Delta^{m_3\ldots m_r)}_{1|A_1,\ldots,A_d|B_1,\ldots,B_{r+d+2}}
- k^{(m_1}_{A_1A_2\ldots A_d} P^{m_2\ldots m_r)}_{1|A_1,\ldots,A_d|B_1,\ldots,B_{r+d+2}}
&\allJacv \cr
&+ \Big( k_{A_1}^p P^{pm_1\ldots m_r}_{1|A_2,\ldots,A_d|A_1,B_1,\ldots,B_{r+d+2}}
+ \big[P^{m_1\ldots m_r}_{1|A_2,\ldots,A_d| S[A_1,B_1] , B_2,\ldots,B_{r+d+2} }
+ (B_1 \leftrightarrow B_2,\ldots,B_{r+d+2}) \big]\cr
&-\!\! \sum_{XY=A_1}\!\! \big( P^{m_1\ldots m_r}_{1|X,A_2,\ldots,A_d | Y,B_1,\ldots,B_{r+d+2}}
- (X\leftrightarrow Y)\big)
+(A_1\leftrightarrow A_2,\ldots,A_d) \Big)\,,
}$$
and the simplest example at degree of refinement $d= 2$ is
\eqnn\HOae
$$\eqalignno{
Q D_{1|2,3|4,5,6,7}  &= \Delta_{1|2|3,4,5,6,7}+ \Delta_{1|3|2,4,5,6,7}
+ k_3^m P^m_{1|2|3,4,5,6,7}+k_2^m P^m_{1|3|2,4,5,6,7}  \cr
&+ \big[ s_{34} P_{1|2|34,5,6,7} + s_{24} P_{1|3|24,5,6,7}
+ (4 \leftrightarrow 5,6,7) \big]\,.
&\HOae
}$$

\newsec Conclusions

In this first part of the series of papers \wipI\ towards the derivation of one-loop
correlators in string theory, several aspects related to the description of the
massless string states via superfield kinematics have been thoroughly discussed.

The whole setup starts with the standard superfields describing
super-Yang--Mills states in ten-dimensions \wittentwistor\ contained in the
massless vertex operators of the pure-spinor superstring. We then used a
combination of OPEs, zero-mode integration rules and covariance under BRST
transformations to derive several compositions of superfields with the correct
properties to describe higher-point amplitudes in the pure-spinor formalism, in
both the anomalous and non-anomalous sectors. The comprehensive description of
the local representatives is both new and relevant to the derivation of local
$n$-point one-loop correlators in part III of this series\foot{Some of
their properties were implicit in previous works (most notably in \partI) while
in \towardsoneloop\ it was realized that the one-loop amplitudes (up to six
points) in field theory could be written using subsets of the definitions in
this work.}. The emphasis on these local superfield building blocks is warranted
because the one-loop correlators are local objects prior to integration by
parts, where only the OPE contractions and zero-mode integrations are performed.

We then reviewed their non-local representatives from \partI\ with special
emphasis on the multitude of relations valid in the cohomology of the pure-spinor
BRST charge. In this first part of the series, these relations represent
cohomological identities among {\it superfields}. In the sequel part II,
it will be shown that these same identities are realized by a set of objects
completely different in nature: {\it functions} on the genus-one worldsheet!
The pursue of this unexpected connection dubbed ``duality between worldsheet functions
and kinematics'' will lead us to a detailed study of
so-called ``generalized elliptic integrands'' (GEIs), which were briefly
introduced in \MafraIOJ.

Finally, in part III the numerous definitions as well as surprising relationships and
identities uncovered in parts I and II will pave the way to the assembly of
one-loop correlators in many different representations.

\bigskip
\noindent{\bf Acknowledgements:}
We are indebted to the IAS Princeton and to Nima Arkani-Hamed for kind
hospitality during an inspiring visit which initiated this project. This
research was supported by the Munich Institute for Astro- and Particle Physics
(MIAPP) of the DFG cluster of excellence ``Origin and Structure of the
Universe'', and we are grateful to the organizers for creating a stimulating
atmosphere. CRM is supported by a University Research Fellowship from the Royal
Society. The research of OS was supported in part by Perimeter Institute for
Theoretical Physics. Research at Perimeter Institute is supported by the
Government of Canada through the Department of Innovation, Science and Economic
Development Canada and by the Province of Ontario through the Ministry of
Research, Innovation and Science.

\appendix{A}{A new algorithm for the combinatorics of shuffle invariants}
\applab\clikeapp

\noindent
In this appendix, a multi-word generalization of the standard map $\rho(A)$ in
\rhomap\ is introduced. It will lead to a systematic unification of the
combinatorics of all $C$-like pseudo-invariants as well as their
canonicalization identities. Among other things, this simplifies the discussion
of the master recursion from section~8 of \partI\ and allows to write down a
closed formula for the scalar functions $\cZ_{A,B,C,D}$ of part II, see
appendix \bootstrapapp.

\subsec Multi-word generalization of the rho map

\subsubsec Scalar multi-word rhomap

Let us define a recursive two-word version of \rhomap\ by
\eqn\twoslot{
\rho(A|iBj)\equiv
(Ai|Bj) - (Aj|iB) + \rho(Ai|Bj) - \rho(Aj|iB)\,,\quad
\rho(A|i)\equiv 0\,.
}
For example,
\eqnn\twoex
$$\eqalignno{
\rho(342|56) &= (3425|6) - (3426|5)&\twoex\cr
\rho(342|567) &=
(3425|67) - (3427|56)
+ \rho(3425|67) - \rho(3427|56) \cr
&=(3425|67) - (3427|56)
+ (34256|7) - (34257|6)
- (34275|6) + (34276|5)\,.
}$$
The asymmetry $\rho(A|B)\neq\rho(B|A)$ motivates the vertical bar notation
(rather than a comma) for separating
the two words, in accordance with the convention for building blocks.
In addition, the definition \twoslot\ will be generalized for an arbitrary number
of words using the following recursion (with $\rho(A|B)\otrh\emptyset\equiv
\rho(A|B)$)
\eqnn\mrhomap
\eqnn\twowordrec
$$\eqalignno{
\rho(A|B,C, \ldots) &\equiv \rho(A|B)\otrh C\otrh \cdots
+ (B\leftrightarrow C,\ldots)\,&\mrhomap\cr
(A|B)\otimes_\rho C\otimes_\rho D\cdots &\equiv
(A|B,C,D, \ldots) + \big[\rho(A|C)\otrh B\otrh D\otrh\cdots + (C\leftrightarrow D,
\ldots)\big]\qquad{}&\twowordrec
}$$
Note that the word $B$ is excluded from the permutations in \twowordrec. It is
also important to notice that the recursion \twowordrec\ eventually stops due
to the condition $\rho(A|i) = 0$. To illustrate this last point, consider
$(234|56)\otrh78 = (234|56,78) + \rho(234|78)\otrh56$ where the second term by
itself requires further usage of \twowordrec:
\eqnn\simplerect
$$\eqalignno{
\rho(234|78)\otrh56 &= (2347|8)\otrh56 - (2348|7)\otrh56 &\simplerect\cr
&=(2347|8,56) + \rho(2347|56)\otrh8
-(2348|7,56) - \rho(2348|56)\otrh7\,.
}$$
Since the words attached to $\otrh$ end up becoming letters,
the recursion eventually stops due to $\rho(A|i)\equiv0$.

\subsubsec Tensorial multi-word rhomap

In order to upgrade the recursions above to a tensorial setting we
modify the end point of the recursive definition \twoslot\ to
\eqn\generalrho{
\rho^m(A|i) \equiv k_i^m (Ai|\emptyset)\,, \ \ \ \ 
\rho^{m_1 m_2 m_3 \ldots}(A|i) \equiv k_i^{(m_1} (Ai|\emptyset)^{m_2 m_3 \ldots)}\,.
}
All other definitions are kept unchanged except for having
extra vector indices; e.g.,
\eqnn\multivector
$$\eqalignno{
\rho^m(A|iBj) &\equiv (Ai|Bj)^m - (Aj|iB)^m +\rho^m(Ai|Bj) -
\rho^m(Aj|iB)\,,&\multivector\cr
\rho^m(A|B,C, \ldots) &\equiv
\rho^m(A|B)\otrh C\otrh \cdots
+ (B\leftrightarrow C,\ldots)\cr
(A|B)^m\otimes_\rho C\otrh D\cdots &\equiv
(A|B,C,D, \ldots)^m + \big[\rho^m(A|C)\otrh B\otrh D\otrh\cdots + (C\leftrightarrow D,
\ldots)\big]\,,
}$$
and the generalization to multiple vector indices is straightforward.

\newsubsubsec\InvariantMap Tensorial word-invariant maps

The multi-word generalization of the rhomap can be used to define
the following {\it word invariants},
\eqnn\wordinvs
$$\eqalignno{
\cI(A|B,C, \ldots) &\equiv (A|B,C, \ldots) + \rho(A|B,C, \ldots)&\wordinvs\cr
\cI^{m_1m_2 \ldots}(A|B,C, \ldots) &\equiv (A|B,C, \ldots)^{m_1m_2 \ldots}
+ \rho^{m_1m_2 \ldots}(A|B,C, \ldots)\,.
}$$
The reason for dubbing them ``word invariants'' will become clear shortly.

For an example application of the scalar three-word-invariant,
let us consider
\eqnn\threeEx
$$\eqalignno{
\cI(2|34,56) &= (2|34,56) + \rho(2|34,56)&\threeEx\cr
&= (2|34,56) + \rho(2|34)\otrh 56 + \rho(2|56)\otrh34\,.
}$$
Applying the recursion \twowordrec\ in the right-hand side gives
\eqnn\interm
$$\eqalignno{
\rho(2|34)\otrh 56 &= (23|4)\otrh 56 - (24|3)\otrh56 &\interm\cr
&= (23|56,4) + \rho(23|56)\otrh4 - (24|56,3) - \rho(24|56)\otrh3\cr
&= (23|56,4) + (235|6,4) - (236|5,4)
- (24|56,3) - (245|6,3) + (246|5,3)\,,\cr
\rho(2|56)\otrh34  &= (25|34,6) + (253|4,6) - (254|3,6)
- (26|34,5) - (263|4,5) + (264|3,5)\,.
}$$
And finally collecting everything from \threeEx\ yields the final result:
\eqnn\exFone
$$\eqalignno{
\cI(2|34,56) &= (2|34,56) + (23|56,4) + (235|6,4) - (236|5,4)
- (24|56,3)\cr
&\quad{}- (245|6,3) + (246|5,3)
 +(25|34,6) + (253|4,6) - (254|3,6)\cr
&\quad{}- (26|34,5) - (263|4,5) + (264|3,5)\,. &\exFone
}$$

\newsubsec\Unifiedclikeapp Unifying all $C$-like building blocks

The word invariants \wordinvs\ can be used to provide an alternative
derivation of the Berends--Giele expansions of tensorial $C$-like
BRST invariants defined in \partI.
More explicitly,
the observation is
\eqn\Clike{
\eqalign{
C_{1|A_1, \ldots,A_3} &= \cI_C(1|A_1, \ldots,A_3)\,, \cr
D_{1|A_1, \ldots,A_2} &= \cI_D(1|A_1, \ldots,A_2)\,, \cr
L_{1|A_1, \ldots,A_4} &= \cI_L(1|A_1, \ldots,A_4)\,,
}\qquad\eqalign{
\Delta_{1|A_1, \ldots,A_4} &= \cI_\Delta(1|A_1, \ldots,A_4)\,,\cr
\Lambda_{1|A_1, \ldots,A_6} &= \cI_\Lambda(1|A_1, \ldots,A_6)\,,\cr
\Gamma_{1|A_1, \ldots,A_5} &= \cI_\Gamma(1|A_1, \ldots,A_5)\,,
}}
where
\eqnn\MasterUs
$$\eqalignno{
\cI_C(A_1| \ldots,A_4)&\equiv \cI(A_1| \ldots,A_4)\hbox{ with }
(P_1| \ldots,P_4)\rightarrow M_{P_1}M_{P_2, P_3,P_4}\,,
&\MasterUs\cr
\cI_D(A_1| \ldots,A_3)&\equiv \cI(A_1| \ldots,A_3)\hbox{ with }
(P_1| \ldots,P_3)\rightarrow M_{P_1,P_2,P_3}\,,\cr
\cI_L(A_1| \ldots,A_5)&\equiv \cI(A_1| \ldots,A_5)\hbox{ with } (P_1|
\ldots,P_5)
\rightarrow \cJ_{P_1|P_2, \ldots,P_5}\,,\cr
\cI_\Delta(A_1| \ldots,A_5)&\equiv \cI(A_1| \ldots,A_5)\hbox{ with } (P_1|
\ldots,P_5)
\rightarrow \cY_{P_1, \ldots,P_5}\,,\cr
\cI_\Lambda(A_1| \ldots,A_7)&\equiv \cI(A_1| \ldots,A_7)\hbox{ with }
(P_1| \ldots,P_7)
\rightarrow \cY_{P_1|P_2, \ldots,P_7}\,,\cr
\cI_\Gamma(A_1| \ldots,A_6)&\equiv \cI(A_1| \ldots,A_6)\hbox{ with }
(P_1| \ldots,P_6)
\rightarrow M_{P_1}\cY_{P_2, \ldots,P_6}\,,\cr
}$$
with obvious tensorial generalizations,
\eqn\BRSTinv{
C^{m}_{1|A,B,C,D} = \cI^{m}_C(1|A,B,C,D)\,, \ \ \ \ 
C^{m_1\ldots m_r}_{1|A_1,\ldots,A_{r+3}} = \cI^{m_1\ldots m_r}_C(1|A_1,\ldots,A_{r+3})\, ,
}
where
$(A|B,C,D,E)^{m}\to M_AM^m_{B,C,D,E}$ and
$(A|B_1,\ldots,B_{r+3})^{m_1\ldots m_r}\to M_AM^{m_1\ldots m_r}_{B_1,\ldots ,B_{r+3}}$.
For the simplest example, consider:
\eqn\Dex{
\cI(1|23,4) = (1|23,4) + \rho(1|23)\otrh4 = (1|23,4) + (12|3,4) - (13|2,4)\,.
}
Hence, we get
$D_{1|23,4} = \cI_D(1|23,4) = M_{1,23,4} + M_{12,3,4} - M_{13,2,4}$ from
the second definition in \MasterUs, in accordance with equation (8.16) from \partI.
Also note that the expansion of unrefined GEIs $E_{1|\ldots}$ in terms of $\cZ$-functions 
(cf.\ part II) can be obtained from
\eqnn\GEIcZ
$$\eqalignno{
E_{1|A_1, \ldots,A_3} &\equiv \cI(1|A_1, \ldots,A_3)\hbox{ with }
(P_1| \ldots,P_4)\rightarrow \cZ_{P_1, \ldots,P_4}  &\GEIcZ \cr
E^{m_1\ldots m_r}_{1|A_1, \ldots,A_{r+3}} &\equiv
\cI^{m_1 \ldots m_r}(1,A_1, \ldots,A_{r+3})\hbox{ with }
(P_1| \ldots,P_{r+4})^{m_1 \ldots m_r}\rightarrow \cZ^{m_1\ldots m_r}_{P_1,
\ldots,P_{r+4}}\,.
}$$

\newsubsec\changebasisapp Change-of-basis identities

The word invariants also give rise to a simple algorithm to obtain various
identities for the change of basis in BRST invariants.

\subsubsec Scalar BRST invariants

Using the shuffle symmetries within the words $A,B$ and $C$ of $C_{j|A,B,C}$,
one can always rewrite an arbitrary $C_{j|PiQ,R,S}$ as a linear combination
of the form $C_{j|iA,B,C}$ (with a given label $i$ in the first position). So, 
without loss of generality, the change of
basis of scalar BRST invariants can be restricted to the case $C_{j|iA,B,C}$.
The change of basis from $j$ to $i$ follows from
\eqn\BRSTcan{
C_{j|iA,B,C}
= C_{i|\cI(Aj|B,C)}\,,
}
which is equivalent to $\cI_C(j|iA,B,C)=\cI_C(i|\cI(Aj|B,C))$.
For example, from $C_{2|1,34,56} = C_{1|\cI(2|34,56)}$ and \exFone\ we get
\eqnn\repr
$$\eqalignno{
C_{2|1,34,56}
&= C_{1|2,34,56} + C_{1|23,56,4} + C_{1|235,6,4} - C_{1|236,5,4}
- C_{1|24,56,3} - C_{1|26,34,5} - C_{1|263,4,5}\cr
&\quad{}- C_{1|245,6,3} + C_{1|246,5,3}
 +C_{1|25,34,6} + C_{1|253,4,6} - C_{1|254,3,6}
+ C_{1|264,3,5}\,,&\repr
}$$
which reproduces the expression (F.1) from \partI.

\subsubsec Tensor BRST invariants

A straightforward generalization of
the algorithm for the change of basis of scalar BRST invariants leads to the
vectorial identity
\eqn\vectorBasis{
C^m_{j|iA,B,C,D} = C_{i|\cI^m(Aj|B,C,D)}\,,
}
where the vectorial word invariant map was defined in \wordinvs. However, the
tensorial identity
requires trace corrections proportional to $\cI_\Delta$ from \MasterUs,
\eqn\tensorBasis{
C^{mn}_{j|iA,B,C,D,E} = C_{i|\rho^{mn}(Aj|B,C,D,E)} + \d^{mn}\cI_\Delta(iAj|B,C,D,E)\,.
}
In summary, the word invariant map can be applied to any number of words (slots) and
it unifies all $C$-like building block recursions from \partI, as well as
their change-of-basis identities.

As a final example of the unifying power of the word invariants, consider
the monodromy derivation of $\cZ_{12,34,56,78}$ to be defined by \protoC. 
We will need to change the GEI $E_{2|1,34,56,78}$ in \protoC\ to a basis of
$E_{1|A,B,C,D}$. Since with the exception of the ``basis'' letters $1$ and $2$
all words are multiparticle, there is no instance of an analogous identity for
$C_{2|A,B,C}$ that can be slot extended. However, it is easy to use the
word invariant map to obtain
\eqnn\snterms
$$\eqalignno{
&\hskip100pt{} E_{2|1,34,56,78} = \cI_E(1|\cI(2|34,56,78))\cr
&=
  E_{1| 23 , 56 , 78 , 4}
 +  E_{1| 235 , 78 , 4 , 6}
 +  E_{1| 2357 , 4 , 6 , 8}
 -  E_{1| 2358 , 4 , 6 , 7}
 -  E_{1| 236 , 78 , 4 , 5} &\snterms\cr
&{} -  E_{1| 2367 , 4 , 5 , 8}
 +  E_{1| 2368 , 4 , 5 , 7}
 +  E_{1| 237 , 56 , 4 , 8}
 +  E_{1| 2375 , 4 , 6 , 8}
 -  E_{1| 2376 , 4 , 5 , 8}
 -  E_{1| 238 , 56 , 4 , 7} \cr
&{} -  E_{1| 2385 , 4 , 6 , 7}
 +  E_{1| 2386 , 4 , 5 , 7} 
 -  E_{1| 24 , 56 , 78 , 3} 
 -  E_{1| 245 , 78 , 3 , 6} 
 -  E_{1| 2457 , 3 , 6 , 8} 
 +  E_{1| 2458 , 3 , 6 , 7} \cr
&{} +  E_{1| 246 , 78 , 3 , 5} 
 +  E_{1| 2467 , 3 , 5 , 8} 
 -  E_{1| 2468 , 3 , 5 , 7} 
 -  E_{1| 247 , 56 , 3 , 8} 
 -  E_{1| 2475 , 3 , 6 , 8} 
 +  E_{1| 2476 , 3 , 5 , 8} \cr
&{} +  E_{1| 248 , 56 , 3 , 7} 
 +  E_{1| 2485 , 3 , 6 , 7} 
 -  E_{1| 2486 , 3 , 5 , 7} 
 +  E_{1| 25 , 34 , 78 , 6} 
 +  E_{1| 253 , 78 , 4 , 6} 
 +  E_{1| 2537 , 4 , 6 , 8} \cr
&{} -  E_{1| 2538 , 4 , 6 , 7} 
 -  E_{1| 254 , 78 , 3 , 6} 
 -  E_{1| 2547 , 3 , 6 , 8} 
 +  E_{1| 2548 , 3 , 6 , 7} 
 +  E_{1| 257 , 34 , 6 , 8} 
 +  E_{1| 2573 , 4 , 6 , 8} \cr
&{} -  E_{1| 2574 , 3 , 6 , 8} 
 -  E_{1| 258 , 34 , 6 , 7} 
 -  E_{1| 2583 , 4 , 6 , 7} 
 +  E_{1| 2584 , 3 , 6 , 7} 
 -  E_{1| 26 , 34 , 78 , 5} 
 -  E_{1| 263 , 78 , 4 , 5} \cr
&{} -  E_{1| 2637 , 4 , 5 , 8} 
 +  E_{1| 2638 , 4 , 5 , 7} 
 +  E_{1| 264 , 78 , 3 , 5} 
 +  E_{1| 2647 , 3 , 5 , 8} 
 -  E_{1| 2648 , 3 , 5 , 7} 
 -  E_{1| 267 , 34 , 5 , 8} \cr
&{} -  E_{1| 2673 , 4 , 5 , 8} 
 +  E_{1| 2674 , 3 , 5 , 8} 
 +  E_{1| 268 , 34 , 5 , 7} 
 +  E_{1| 2683 , 4 , 5 , 7} 
 -  E_{1| 2684 , 3 , 5 , 7} 
 +  E_{1| 27 , 34 , 56 , 8} \cr
&{} +  E_{1| 273 , 56 , 4 , 8} 
 +  E_{1| 2735 , 4 , 6 , 8} 
 -  E_{1| 2736 , 4 , 5 , 8} 
 -  E_{1| 274 , 56 , 3 , 8} 
 -  E_{1| 2745 , 3 , 6 , 8} 
 +  E_{1| 2746 , 3 , 5 , 8} \cr
&{} +  E_{1| 275 , 34 , 6 , 8} 
 +  E_{1| 2753 , 4 , 6 , 8} 
 -  E_{1| 2754 , 3 , 6 , 8} 
 -  E_{1| 276 , 34 , 5 , 8} 
 -  E_{1| 2763 , 4 , 5 , 8} 
 +  E_{1| 2764 , 3 , 5 , 8} \cr
&{} -  E_{1| 28 , 34 , 56 , 7} 
 -  E_{1| 283 , 56 , 4 , 7} 
 -  E_{1| 2835 , 4 , 6 , 7} 
 +  E_{1| 2836 , 4 , 5 , 7} 
 +  E_{1| 284 , 56 , 3 , 7} 
 +  E_{1| 2845 , 3 , 6 , 7} \cr
&{} -  E_{1| 2846 , 3 , 5 , 7} 
 -  E_{1| 285 , 34 , 6 , 7} 
 -  E_{1| 2853 , 4 , 6 , 7} 
 +  E_{1| 2854 , 3 , 6 , 7} 
 +  E_{1| 286 , 34 , 5 , 7} 
 +  E_{1| 2863 , 4 , 5 , 7} \cr
&{} -  E_{1| 2864 , 3 , 5 , 7} 
 +  E_{1| 34 , 56 , 78 , 2}\,,
}$$
where $\cI_E$ is defined in analogy with \MasterUs.
This example demonstrates that the large number of terms in such identities
can be generated by a simple set of combinatorial rules.

\appendix{B}{Empty BRST cohomology for manifestly local expressions}
\applab\localapp

\noindent
Using the alphabet of local building blocks presented in section~\LocalBBsec,
in this appendix we will demonstrate that their most general BRST-closed
linear combinations are also BRST-exact when more than four and up to
eight particles are involved; i.e., the BRST cohomology is empty for manifestly local
expressions of multiplicities five to eight.

At four points, the expression $V_1T_{2,3,4}$ is clearly local, and it is not
BRST-exact in the four-point phase where $s_{ijk}=0$ (its components reproduce
the one-loop four-point amplitude \fourptpaper). Consequently, there is a local
cohomology at four points. In the following we study the higher-point
generalizations and find that the manifestly local BRST cohomology is empty\foot{In absence
of five-point momentum conservation $\langle Y_{1,2,3,4,5} \rangle \sim
\epsilon_{10}(f_1,f_2,\ldots,f_5)$ \anomalypaper, and one could argue that it
constitutes a local element of the BRST cohomology. But this is no longer true
once we invoke $k_{12345}{=}0$ since $Y_{1,2,3,4,5} $ becomes BRST exact in this
case (see section 9.1 of \partI).} for up to eight points among the combinations
of building blocks in section~\LocalBBsec.

\subsec BRST-closed expressions without momentum conservation

In the first series of checks, we treat all the momenta $k_1,k_2,\ldots,k_n$ in an
$n$-point superspace expression as independent, i.e.\ temporarily relax momentum
conservation. An automated brute-force scan of all
possible linear combinations of building blocks from section~\LocalBBsec\ 
using {\tt FORM} \FORM\ led to the following unique BRST-closed
expressions\foot{Note that $QY_{1,2,3,4,5}=0$ is BRST closed by itself but we
add it to ${\rm loc}^{\rm 5pt}_1$ for convenience.}
\eqnn\localClosed
\eqnn\localClosedSix
\eqnn\localClosedSeven
$$\eqalignno{
\loc51 &= k_1^m V_1 T^m_{2,3,4,5}
+ \big[ V_{12} T_{3,4,5} + (2\leftrightarrow 3,4,5) \big] + Y_{1,2,3,4,5}\,,&\localClosed\cr
\loc61 &=
k_1^m Y^m_{1,2,3,4,5,6} +\big[ Y_{12,3,4,5,6} + (2\leftrightarrow 3,4,5,6) \big]
+ k_1^m k_1^n V_1 T^{mn}_{2,3,4,5,6} &\localClosedSix\cr
&+ \big[ (2k_1^m +k_2^m) V_{12} T^m_{3,4,5,6} + s_{12} V_{1} J_{2|3,4,5,6}
+ (2\leftrightarrow 3,4,5,6) \big]\cr
&+ \big[ (V_{123}+V_{132}) T_{4,5,6} + (2,3|2,3,4,5,6) \big]\,,\cr
\loc71 &=
 k_1\cdot k_{234567} Y_{1|2,3,4,5,6,7} + k_1^m k_1^n Y^{mn}_{1,2,3,4,5,6,7}\cr
& + \big[ (2k_1^m + k_2^m) Y^{m}_{12,3,4,5,6,7} 
+  s_{12} Y_{2|1,3,4,5,6,7} +(2\leftrightarrow 3,4,5,6,7)  \big]
&\localClosedSeven\cr
&  + \big[ Y_{123,4,5,6,7}+ Y_{132,4,5,6,7} +(2,3|2, 3,4,5,6,7)  \big] \cr
& +  k_1^m k_1^n k_1^p V_1 T^{mnp}_{2, \ldots,7} + \big[ (3k_1^m k_1^n+ 3 k_1^m k_2^n + k_2^m k_2^n)
V_{12} T^{mn}_{3, \ldots,7} +(2\leftrightarrow 3, \ldots,7) \big]\cr
& + \big[
  \big((3k_1^m +2 k_2^m +k_3^m) V_{123}
+ (3k_1^m +2 k_3^m +k_2^m) V_{132}\big)T^{m}_{4, \ldots,7}
+ (2,3|2, \ldots,7)\big]\cr
& + \big[ (V_{1234} + {\rm perm}(2,3,4)) T_{5,6,7} + (2,3,4|2, \ldots,7)\big]\cr
& +  \big[ s_{12} V_{1} (3k_1^m +k_2^m) J^{m}_{2|3,4,5,6,7} + (2\leftrightarrow 3,4,5,6,7)\big]\cr
& + \big[ (3s_{13}{+}s_{23})V_{12} J_{3|4,5,6,7}+(3s_{12}{+}s_{23})V_{13}
J_{2|4,5,6,7}\cr
&\hskip20pt+(s_{12}{-}s_{13}) V_1 J_{23|4,5,6,7}+ (2,3|2,\ldots,7) \big]\,,
}$$
as well as a much bigger expression at eight points,
\eqnn\manlocK
$$\eqalignno{ 
&\hskip5pt\loc81 = Y^m_{1|2,3,\ldots,8} \Big( 3k_1^m (k_1\cdot k_{23\ldots 8}) + \big[ s_{12} k_2^m + (2\leftrightarrow 3,4,\ldots,8) \big] \Big) &\manlocK \cr
&+ \big[ (s_{12}-s_{13}) Y_{1|23,4,\ldots,8} + (2,3|2,3,\ldots,8) \big] \cr
& + \big[  Y_{12|3,4,\ldots,8}   (3(k_1\cdot k_{23\ldots 8}) + (k_2\cdot k_{34\ldots 8})) + (2\leftrightarrow 3,4,\ldots,8) \big] \cr
&+\big[ s_{12} (3k_1^m + k_2^m) Y^m_{2|1,3,4,\ldots,8}  + (2\leftrightarrow 3,4,\ldots,8) \big] \cr
&+ \big[ (3 s_{12}+s_{23}) Y_{2|13,4,\ldots,8} + (3 s_{13}+s_{23}) Y_{3|12,4,\ldots,8}\cr
& \ \ \ \ +  (s_{12}-s_{13} ) Y_{23|1,4,5,\ldots,8}+ (2,3|2,3,4,\ldots,8) \big] \cr
&+ k_1^m k_1^n k_1^p Y^{mnp}_{1,2,\ldots,8}+ \big[ Y^{mn}_{12,3,\ldots,8}(3 k_1^m k_1^n + 3 k_1^m k_2^n +k_2^m k_2^n) + (2\leftrightarrow 3,4,\ldots, 8)\big] \cr
&+ \big[ Y^m_{123,4,\ldots,8}(3k_1^m + 2 k_2^m + k_3^m)+Y^m_{132,4,\ldots,8}(3k_1^m + 2 k_3^m + k_2^m)+ (2,3|2,3,\ldots,8) \big] \cr
&+ \big[ (Y_{1234,5,6,7,8} + {\rm perm}(2,3,4))
+(2,3,4|2,3,4,\ldots,8) \big] \cr
&+ k_1^m k_1^n k_1^p k_1^q V_1 T^{mnpq}_{2,3,\ldots,8} + \big[  s_{12} V_1 J^{mn}_{2|3,4,\ldots,8} (6k_1^m k_1^n+4k_1^m k_2^n+k_2^m k_2^n)+ (2\leftrightarrow 3,4,\ldots,8) \big] \cr
&+ \big[ (6s_{12}s_{13}+s_{12}s_{23}+s_{13}s_{23}) V_1 J_{2,3|4,5,\ldots,8}+ (2,3|2,3,4,\ldots,8) \big] \cr
&+ \big[  V_1 J^m_{23|4,5,\ldots,8}( 4k_1^m (s_{12}{-}s_{13}) + k_2^m(2 s_{12}{-}s_{13}) + k_3^m(s_{12}{-}2s_{13})) + (2,3|2,3,4,\ldots,8) \big] \cr
&+ \big[ (s_{12}{-}2s_{13}{+}s_{14}) V_1 J_{234|5,6,7,8} + (s_{12}{+}s_{13}{-}2s_{14}) V_1 J_{243|5,6,7,8} 
+ (2,3,4|2,3,4,\ldots,8) \big] \cr
&+ \big[ V_{12} T^{mnp}_{3,4,\ldots,8}(4k_1^mk_1^n k_1^p + 6 k_1^mk_1^nk_2^p +4 k_1^m k_2^n k_2^p + k_2^m k_2^n k_2^p) +(2\leftrightarrow 3,4,\ldots,8) \big] \cr
&+ \big[ V_{12} J^m_{3|4,5,6,7,8} (k_3^m s_{23}{+}4 k_3^m s_{13} {+} 3 k_2^m s_{23} {+} 6 k_2^m s_{13} {+} 4k_1^m s_{23} {+} 12 k_1^m s_{13}) \cr
& \ \ \ \ + V_{13} J^m_{2|4,5,6,7,8} (k_2^m s_{23}{+}4 k_2^m s_{12} {+} 3 k_3^m s_{23} {+} 6 k_3^m s_{12} {+} 4k_1^m s_{23} {+} 12 k_1^m s_{12}) + (2,3|2,3,4,\ldots,8) \big] \cr
&+ \big[   V_{12} J_{34|5,6,7,8}( 4 s_{13}-4s_{14}+ s_{23}-s_{24}) +  V_{13} J_{24|5,6,7,8}( 4 s_{12}-4s_{14}+ s_{23}-s_{34}) \cr
& \ \ \ \  +  V_{14} J_{23|5,6,7,8}( 4 s_{12}-4s_{13}+ s_{24}-s_{34})  + (2,3,4|2,3,4,5,6,7,8) \big] \cr
&+ \big[ V_{123} ( 6 k_1^m k_1^n + 8 k_1^mk_2^n+4 k_1^m k_3^n + 3 k_2^m k_2^n+ 3 k_2^m k_3^n + k_3^m k_3^n)T^{mn}_{4,5,6,7,8} \cr
& \ \ \ \ + V_{132} ( 6 k_1^m k_1^n + 8 k_1^mk_3^n+4 k_1^m k_2^n + 3 k_3^m k_3^n+ 3 k_2^m k_3^n + k_2^m k_2^n)T^{mn}_{4,5,6,7,8} + (2,3|2,3,\ldots,8) \big] \cr
&+ \big[ (V_{123} (s_{34}+3 s_{24}+6 s_{14})J_{4|5,6,7,8} + {\rm perm}(2,3,4)) + (2,3,4|2,3,4,5,6,7,8) \big] \cr
&+ \big[ (V_{1234} (k_4^m + 2 k_3^m + 3 k_2^m+ 4 k_1^m)T^m_{5,6,7,8} + {\rm perm}(2,3,4)) + (2,3,4|2,3,4,5,6,7,8) \big] \cr
&+ \big[ (V_{12345} T_{6,7,8} + {\rm perm}(2,3,4,5)) + (2,3,4,5|2,3,4,5,6,7,8) \big] \ ,
}$$
and these are the only BRST-closed expressions with manifestly local building
blocks up to eight points. But,
as alluded to in the beginning of this appendix,
the above BRST-closed expressions are also BRST-exact, and the local BRST
cohomology is trivial in absence of momentum conservation. In fact,
\eqnn\localExactFive
\eqnn\localExactSix
\eqnn\localExactSeven
\eqnn\localExactEight
$$\eqalignno{
\loc51 &= QL_{1|2,3,4,5}\,,&\localExactFive\cr
\loc61 &= Q\bigl(k_1^m L^m_{1|2,3,4,5,6}\bigr)\,,&\localExactSix\cr
\loc71 &= Q\bigl( k_1^m k_1^n L^{mn}_{1|2,3,4,5,6,7}
+\big[ s_{12} L_{1|2|3,4,5,6,7} + (2\leftrightarrow3,4,5,6,7)\big]\bigr)\,,&\localExactSeven\cr
\loc81 &\equiv Q\bigl(
k_1^m k_1^n k_1^p L^{mnp}_{1|2,3,\ldots,8}
+ \big[ s_{12} (3k_1^m + k_2^m) L^m_{1|2|3,4,\ldots,8}
+ (2\leftrightarrow 3,\ldots,8)\big]&\localExactEight\cr
& \hskip10pt + \big[ (s_{12}-s_{13}) s_{23} L_{1|23|4,\ldots,8} +
(2,3|2,3,4,\ldots,8) \big]\bigr)\,,
}$$
where the building blocks $L$ were defined in \partI, e.g.\ $L_{1|2,3,4,5}= {\cal J}_{1|2,3,4,5}$
as well as $L^m_{1|2,\ldots,6}= {\cal J}^m_{1|2,\ldots,6}
+\big[ k_2^m {\cal J}_{12|3,\ldots,6}+(2{\leftrightarrow} 3,\ldots,6)\big]$.
Their unrefined instances $L^{m_1\ldots m_r}_{1|A_1,\ldots, A_{r+4}}$ can also be explicitly
obtained using the alternative algorithm in \Clike.

Although straightforward to check, the relations
\localExactFive\ to \localExactEight\ contain a hidden
systematics noteworthy of uncovering.
To this effect, we rewrite the
BRST-exactness solution of $\loc51$ in terms of BRST invariants (see equation (10.1) of \partI):
\eqn\manlocD{
\loc51 = Q L_{1|2,3,4,5}  = \Delta_{1|2,3,4,5} + k_1^m C^m_{1|2,3,4,5}\,,
}
which manifests BRST invariance rather than locality. We can interpret \manlocD\
as the statement that the non-localities in the Berends--Giele expansion of $k_1^m
C^m_{1|2,3,4,5}$ in \vectorCs\ are in fact spurious (which is easy to verify).
Since the combinatorics of the Berends--Giele expansion of
the building blocks $L$ and $C$ is the same (see \Clike), the conclusion is
that the ghost-number-two expression $k^1_m L^m_{1|2,3,4,5,6}$ is also
a {\it local} expression. Even though it was not guaranteed to be the case,
computing its BRST variation leads to the manifestly local expression $\loc61$.

Similarly, the BRST-closed expression $\loc61$ can be rewritten in terms of
non-local BRST pseudo-invariants
\eqn\manlocE{
Q\bigl( k_1^mL^m_{1|2,3,4,5,6}\bigr)
= k_1^m  \Delta^m_{1|2,3,4,5,6} + k_1^m k_1^n C^{mn}_{1|2,3,4,5,6}
+\big[ s_{12} P_{1|2|3,4,5,6} + (2\leftrightarrow 3,4,5,6) \big]\,,
}
where again the locality of the left-hand side (i.e., $\loc61$) is obscured by the representation
with BRST (pseudo-)invariants on the right-hand side.

However, the same logic can be applied again:
when promoting the BRST descendant on the right-hand side of
\manlocE\ to a BRST generator via $C^{mn\ldots}_{1|A,B,C,D,E}\rightarrow L^{mn\ldots}_{1|A,B,C,D,E}$
and $\Delta^{mn\ldots}_{1|A,B,C,D,E}\rightarrow 0$, locality follows from the equivalence of the respective
expansions in terms of Berends--Giele currents. We therefore obtain
the ghost-number-two terms $ k_1^m k_1^n L^{mn}_{1|2,3,4,5,6,7}
+\big[ s_{12} L_{1|2|3,4,5,6,7} + (2\leftrightarrow 3,4,5,6,7)\big]$ that are
guaranteed to generate a local ghost-number-three expression upon BRST variation.
The fact that it exactly reproduces the unique expression for $\loc71$ obtained by
a brute-force search demonstrates that the manifestly local BRST cohomology is 
empty at seven points (using the set of building blocks from section~\LocalBBsec).

Similarly, we use the promotion $C\rightarrow L$ and $\Delta\rightarrow 0$ in
the BRST variation of the seven-point identity (see \partI\ or \Clike\ for the anomaly
superfields $\Lambda_{1|\ldots}$)
\eqnn\manlocG
$$\eqalignno{
Q\big( k_1^m &k_1^n L^{mn}_{1|2,3,4,5,6,7}
+\big[ s_{12} L_{1|2|3,4,5,6,7} + (2\leftrightarrow 3,4,5,6,7) \big]  \big) =&\manlocG\cr
&+ k_1\cdot k_{1234567} \Lambda_{1|2,3,4,5,6,7}+ k_1^m k_1^n
\Delta^{mn}_{1|2,3,4,5,6,7}\cr
& +\big[ s_{12} \Delta_{1|2|3,4,5,6,7} + (2\leftrightarrow 3,4,5,6,7) \big]  \cr
& + k_1^m k_1^n k_1^p C^{mnp}_{1|2, \ldots,7}
+ \big[ s_{12} (3k_1^m + k_2^m) P^m_{1|2|3, \ldots,7} + (2\leftrightarrow 3, \ldots,7)  \big] \cr
& + \big[ (s_{12}-s_{13}) s_{23} P_{1|23|4,5,6,7} + (2,3|2,3,4,5,6,7) \big]  \cr
}$$
to obtain a local eight-point expression. Surprisingly, as already stated in
\localExactEight, the outcome of its BRST
variation exactly matches the expression \manlocK\ and demonstrates that
the manifestly local BRST cohomology is empty at eight points.

For completeness, the local expression \manlocK\ for $\loc81$ takes the following
form in terms of (pseudo-)invariants 
\eqnn\manlocI
$$\eqalignno{ 
&\quad{}\loc81 = \Lambda^m_{1|2,3,\ldots,8} \big( 3k_1^m (k_1\cdot k_{12\ldots 8})
+ \big[ k_2^m s_{12} + (2\leftrightarrow 3,4,\ldots,8) \big] \Big)  &\manlocI
\cr
&\ \ \ \ + \big[ (s_{12}-s_{13}) s_{23} \Lambda_{1|23,4,\ldots,8} + (2,3|2,3,4,\ldots,8) \big] 
\cr
&\ \ \ \ +k_1^m k_1^n k_1^p \Delta^{mnp}_{1|2,3,\ldots,8}
+ \big[ s_{12} (3k_1^m + k_2^m) \Delta^m_{1|2|3,4,\ldots,8}
+ (2\leftrightarrow 3,4,\ldots,8)  \big] \cr
& \ \ \ \ + \big[ (s_{12}-s_{13}) s_{23} \Delta_{1|23|4,\ldots,8}
+ (2,3|2,3,4,\ldots,8) \big] \cr
& \ \ \ \ + k_1^m k_1^n k_1^p k_1^q C^{mnpq}_{1|2,3,\ldots,8 }
+ \big[ s_{12} ( 6 k_1^m k_1^n + 4 k_1^m k_2^n + k_2^m k_2^n )
P^{mn}_{1|2|3,4,\ldots,8} + (2\leftrightarrow 3,4,\ldots,8)\big] \cr
& \ \ \ \ +\big[ (6 s_{12}s_{13}+s_{12}s_{23}+s_{13}s_{23} ) P_{1|2,3|4,\ldots,8}
+ (2,3|2,3,4,\ldots,8)\big]   \cr
& \ \ \ \ +\big[ s_{23} ( 4 (s_{12}{-}s_{13}) k_1^m + (2s_{12}{-}s_{13}) k_2^m 
+ (s_{12}{-}2s_{13}) k_3^m) P^m_{1|23|4,5,6,7,8} 
+ (2,3|2,3,\ldots,8)
\big] \cr
& \ \ \ \ + \big[  
(s_{12}-2s_{13}+s_{14})s_{23}s_{34} P_{1|234|5,6,7,8}
+(s_{12}-2s_{14}+s_{13})s_{24}s_{34} P_{1|243|5,6,7,8} \cr
& \ \ \ \ \ \ \ \ \ \
+(s_{13}-2s_{12}+s_{14})s_{23}s_{24} P_{1|324|5,6,7,8}
+(2,3,4|2,3,4,5,6,7,8) \big]\,,
}$$
and it can be used to obtain the (tentatively unique) BRST-closed
manifestly local combination at nine points.

\subsec BRST-closed expressions using momentum conservation

We shall now repeat the above analysis in presence of momentum conservation 
and count the number of manifestly local BRST invariants in an $n$-particle phase space.
At five points, $k_{12345}=0$ gives rise to four independent local BRST invariants obtained from
permutations of
\eqnn\manlocL
$$\eqalignno{
QD_{1|2|3,4,5} &=\Delta_{1|2,3,4,5} + k_2^m C_{1|2,3,4,5}^m
+ \big[s_{23} C_{1|23,4,5}+(3\leftrightarrow 4,5) \big]  &\manlocL \cr
&= Y_{1,2,3,4,5} + k_2^m V_1 T^m_{2,3,4,5} - V_{12} T_{3,4,5}
+ \big[ V_1 T_{23,4,5}+(3\leftrightarrow 4,5) \big]
}$$
in $2\leftrightarrow 3,4,5$. The earlier solution \localExactFive\ then follows
from a sum over the $2\leftrightarrow 3,4,5$ permutations of \manlocL\ via
momentum conservation. 

The non-obvious locality of $ k_2^m C_{1|2,3,4,5}^m + \big[s_{23}
C_{1|23,4,5}+(3\leftrightarrow 4,5) \big] $ is not altered when promoting
$C_{1|\ldots} \rightarrow L_{1|\ldots}$, and we can identify a six-point BRST
generator from the first line of \manlocL,
\eqnn\manlocM
$$\eqalignno{ 
& Q( k_2^m L^m_{1|2,3,4,5,6} + \big[s_{23} L_{1|23,4,5,6} +(3\leftrightarrow 4,5,6) \big] ) \cr
&=k_2^m \Delta^m_{1|2,3,4,5,6} + \big[s_{23} \Delta_{1|23,4,5,6} +(3\leftrightarrow 4,5,6) \big] \cr
& \ \ \ \ + k_1^m k_2^n C^{mn}_{1|2,3,4,5,6} - s_{12} P_{1|2|3,4,5,6} +  \big[s_{23} k_1^m C^m_{1|23,4,5,6}  +(3\leftrightarrow 4,5,6) \big]
 \cr
&= k_2^m Y^m_{1,2,\ldots,6} - Y_{12,3,4,5,6} + \big[ Y_{23,1,4,5,6} + (3\leftrightarrow 4,5,6) \big] &\manlocM \cr
&+ V_1 k_1^m k_2^n T^{mn}_{2,3,4,5,6} - s_{12} V_1 J_{2|3,4,5,6} - V_{12} k_1^m T^m_{3,4,5,6}
 \cr
&+ \big[ V_1 k_1^m T^m_{23,4,5,6} + V_{13} k_2^m T^m_{2,4,5,6} +V_{312} T_{4,5,6}+ (3\leftrightarrow 4,5,6) \big] \cr
&+ \big[ V_{13}T_{24,5,6} + V_{14} T_{23,5,6} +(3,4|3,4,5,6) \big] \ .
}$$
The locality of the right-hand side is clear from the locality of the BRST
generator. Note that the first line $k_2^m Y^m_{1,2,\ldots,6} - Y_{12,3,4,5,6} +
\big[ Y_{23,1,4,5,6} + (3\leftrightarrow 4,5,6) \big]$ is separately BRST closed
under $k_2 \cdot k_{13456}=0$. They are expressible in terms of the BRST exact
$\Delta_{1|\ldots}$ and can be viewed as the anomaly analogue of the
second line of \manlocL. The five permutations of the anomalous and the
non-anomalous terms on the right-hand side of \manlocM\ under $2\leftrightarrow
3,4,5,6$ exhaust the $5+5$ manifestly local BRST invariants at six points as
obtained via a brute-force search with {\tt FORM}.

At seven points, the BRST descendant in the third line of \manlocM\ can be promoted to a BRST generator
whose locality is guaranteed by the last three lines of \manlocM,
\eqnn\manlocN
$$\eqalignno{
&Q(k_1^m k_2^n L^{mn}_{1|2,3,4,5,6,7} - s_{12} L_{1|2|3,4,5,6,7} +  \big[s_{23} k_1^m L^m_{1|23,4,5,6,7}  +(3\leftrightarrow 4,5,6,7) \big] )
\cr
&= k_1^m k_2^n \Delta^{mn}_{1|2,3,4,5,6,7} - s_{12} \Delta_{1|2|3,4,5,6,7} +  \big[s_{23} k_1^m \Delta^m_{1|23,4,5,6,7}  +(3\leftrightarrow 4,5,6,7) \big]  \cr
&\ \ \ \ + k_1^m k_1^n k_2^p C^{mnp}_{1|2,3,\ldots,7}+ \big[ s_{23}k_1^m k_1^n C^{mn}_{1|23,4,5,6,7}+ (3\leftrightarrow 4,5,6,7) \big] - 2 s_{12} k_1^m P^m_{1|2|3,4,5,6,7} 
 \cr
& \ \ \ \ + \big[ s_{13} k_2^m P^m_{1|3|2,4,5,6,7}  +s_{13}s_{23} P_{1|23|4,5,6,7}+ (3\leftrightarrow 4,5,6,7) \big] 
 &\manlocN\cr
 & \ \ \ \ + \big[ s_{14}s_{23} P_{1|4|23,5,6,7} +
 s_{13}s_{24} P_{1|3|24,5,6,7} + (3,4|3,4,5,6,7) \big] \ .
}$$
Hence, the right-hand side has to be local as well, as can be verified by explicit computation,
\eqnn\manlocO
$$\eqalignno{
&Q(k_1^m k_2^n L^{mn}_{1|2,3,4,5,6,7} - s_{12} L_{1|2|3,4,5,6,7} +  \big[s_{23} k_1^m L^m_{1|23,4,5,6,7}  +(3\leftrightarrow 4,5,6,7) \big] )
\cr
&= k_1^m k_2^n Y^{mn}_{1,2,3,4,5,6,7} - s_{12} Y_{2|1,3,4,5,6,7} -  k_1^m Y^m_{12,3,4,5,6,7}
 \cr
&+ \big[ k_1^m Y^m_{1,23,4,5,6,7} +  k_2^m Y^m_{13,2,4,5,6,7} +Y_{312,4,5,6,7}+ (3\leftrightarrow 4,5,6,7) \big] \cr
&+ \big[ Y_{13,24,5,6,7} + Y_{14,23,5,6,7} +(3,4|3,4,5,6,7) \big]
\cr
&+  k_1^m k_1^n k_2^p V_1 T^{mnp}_{2,3,4,5,6,7} - 2 s_{12}V_1 k_1^m J^m_{2|3,4,5,6,7} + V_1 k_2^m \big[ s_{13} J^m_{3|2,4,5,6,7} + (3\leftrightarrow 4,5,6,7) \big] \cr
&+  \big[ V_{13} k_2^m (2k_1^n + k_3^n) T^{mn}_{2,4,5,6,7} 
+ k_1^m k_1^n V_1 T^{mn}_{23,4,5,6,7} 
- 2 V_{13} s_{12} J_{2|4,5,6,7} + (3\leftrightarrow 4,5,6,7) \big] \cr
& -  V_{12} k_1^m k_1^n T^{mn}_{3,4,5,6,7}   + \big[ s_{13} (V_1 J_{23|4,5,6,7}- V_{12} J_{3|4,5,6,7})-s_{23} V_{13} J_{2|4,5,6,7}
+ (3\leftrightarrow 4,5,6,7) \big] \cr
&+ \big[ V_1(s_{13} J_{3|24,5,6,7}{+}(3{\leftrightarrow} 4))   + (3,4|3,4,5,6,7) \big]
- \big[ V_{132} (2k_1^m {+} k_3^m) T^m_{4,5,6,7} +  (3 {\leftrightarrow} 4,5,6,7) \big] \cr
&+ \big[ V_{14}(2k_1^m + k_4^m) T^m_{23,5,6,7}
+V_{13}(2k_1^m + k_3^m) T^m_{24,5,6,7} + (3,4|3,4,5,6,7) \big]
&\manlocO  \cr
&+ \big[  (V_{134}+ V_{143})k_2^m T^m_{2,5,6,7}   - (V_{1342}+V_{1432}) T_{5,6,7}   + (3,4|3,4,5,6,7) \big]  \cr
&+ \big[ (V_{145}+V_{154}) T_{23,6,7} + (V_{135}+V_{153}) T_{24,6,7} + (V_{134}+V_{143}) T_{25,6,7} + (3,4,5|3,4,5,6,7) \big] \ .
}$$
In contrast to the six-point analogue \manlocM, the anomalous terms
\eqnn\manlocOA
$$\eqalignno{
Y^{(7)}_{1|2|3,4,5,6,7} &\equiv k_1^m k_2^n Y^{mn}_{1,2,3,4,5,6,7} - s_{12} Y_{2|1,3,4,5,6,7} -  k_1^m Y^m_{12,3,4,5,6,7}
 \cr
&+ \big[ k_1^m Y^m_{1,23,4,5,6,7} +  k_2^m Y^m_{13,2,4,5,6,7} +Y_{312,4,5,6,7}+ (3\leftrightarrow 4,5,6,7) \big] \cr
&+ \big[ Y_{13,24,5,6,7} + Y_{14,23,5,6,7} +(3,4|3,4,5,6,7) \big] &\manlocOA
}$$
in the first three lines on the right-hand side of \manlocO\ are not BRST invariant by themselves.
Instead, we have
\eqnn\manlocOB
$$\eqalignno{
QY^{(7)}_{1|2|3,4,5,6,7} &= - s_{12} \Big( V_1 k_2^m Y^m_{2,3,4,5,6,7} - V_{12} Y_{3,4,5,6,7}
+ \big[ V_1 Y_{23,4,5,6,7} + (3\leftrightarrow4,5,6,7) \big] \Big)
\cr
&= - s_{12}Q \Delta_{1|2|3,4,5,6,7} \ ,
 &\manlocOB
}$$
consistent with \manlocN\ and the fact that $Q \Delta^{mn}_{1|2,3,\ldots ,7 } =Q \Delta^{m}_{1|23,\ldots ,7 } =0$.
Accordingly, $\langle Y^{(7)}_{1|2|3,4,5,6,7} \rangle = 
-s_{12} \langle \Delta_{1|2|3,4,5,6,7} \rangle$ in the cohomology. The
six permutations of \manlocO\ under $2\leftrightarrow 3,\ldots,7$ are the 
only manifestly local BRST invariants at seven points which can be built 
from our alphabet of building blocks. By the availability of the BRST generator
on the left-hand side of \manlocO, all of them are excluded from the cohomology.

At eight points, a brute-force search with {\tt FORM} identified seven BRST 
invariant linear combinations of manifestly local building blocks, namely 
the $2\leftrightarrow 3,\ldots,8$ permutations of
\eqnn\manlocOC
$$\eqalignno{
&Q \Big(  k_1^m k_1^n k_2^p L^{mnp}_{1|2,3,\ldots,8}+ \big[ s_{23}k_1^m k_1^n L^{mn}_{1|23,4,\ldots,8}+ (3\leftrightarrow 4,5,6,7,8) \big] - 2 s_{12} k_1^m L^m_{1|2|3,4,\ldots,8} 
 \cr
& \ \ \ \ + \big[ s_{13} k_2^m L^m_{1|3|2,4,\ldots,8}  +s_{13}s_{23} L_{1|23|4,\ldots,8}+ (3\leftrightarrow 4,\ldots,8) \big] 
 &\manlocOC \cr
 & \ \ \ \ + \big[ s_{14}s_{23} L_{1|4|23,5,6,7,8} +
 s_{13}s_{24} L_{1|3|24,5,6,7,8} + (3,4|3,4,5,6,7,8) \big]  \Big) \cr
 &= - s_{12}  ( k_2^m \Lambda^m_{1|2,3,\ldots,8} +  \big[ s_{23}  \Lambda_{1|23,4,\ldots,8}  + (3\leftrightarrow 4,\ldots,8) \big] )  \cr
 & \ \ \ \ + k_1^m k_1^n k_2^p \Delta^{mnp}_{1|2,3,\ldots,8}+ \big[ s_{23}k_1^m k_1^n \Delta^{mn}_{1|23,4,\ldots,8}+ (3\leftrightarrow 4,5,6,7,8) \big] - 2 s_{12} k_1^m \Delta^m_{1|2|3,4,\ldots,8} 
 \cr
& \ \ \ \ + \big[ s_{13} k_2^m \Delta^m_{1|3|2,4,\ldots,8}  +s_{13}s_{23} \Delta_{1|23|4,\ldots,8}+ (3\leftrightarrow 4,\ldots,8) \big] 
\cr
 & \ \ \ \ + \big[ s_{14}s_{23} \Delta_{1|4|23,5,6,7,8} +
 s_{13}s_{24} \Delta_{1|3|24,5,6,7,8} + (3,4|3,4,5,6,7,8) \big] \cr
 & \ \ \ \ + k_1^m k_1^n k_1^p k_2^q C^{mnpq}_{1|2,3,\ldots,8} + \big[ s_{23} k_1^m k_1^n k_1^p C^{mnp}_{1|23,4,\ldots,8} + (3\leftrightarrow 4,\ldots,8) \big] \cr
 & \ \ \ \ - 3 s_{12} k_1^m k_1^n P^{mn}_{1|2|3,4,\ldots,8} + \big[ s_{13} k_2^m (3k_1^n + k_3^n) P^{mn}_{1|3|2,4,\ldots,8} + (3\leftrightarrow 4,\ldots,8) \big] \cr
 & \ \ \ \ + \big[ s_{13}s_{23} (3k_1^m + k_3^m)P^m_{1|23|4,5,\ldots,8} - s_{13}(3 s_{12}+s_{23}) P_{1|2,3|4,5,6,7,8}+(3\leftrightarrow 4,\ldots,8) \big] \cr
 & \ \ \ \ + \big[ (s_{13}{-}s_{14}) (s_{34} k_2^m P^m_{1|34|2,5,6,7,8} {+} s_{23} s_{34} P_{1|234|5,6,7,8}
  {-} s_{24} s_{34} P_{1|243|5,6,7,8}) {+} (3,4|3,4,\ldots,8) \big] \cr
 & \ \ \ \ + \big[ s_{14} s_{23} (3k_1^m {+} k_4^m) P^m_{1|4|23,5,6,7,8}
 + s_{13} s_{24} (3k_1^m {+} k_3^m) P^m_{1|3|24,5,6,7,8} + (3,4|3,4,\ldots,8) \big] \cr
 &\ \ \ \ + \big[ s_{23} s_{45} (s_{14} -s_{15}) P_{1|45|23,6,7,8}
 +s_{24} s_{35} (s_{13} -s_{15}) P_{1|35|24,6,7,8} \cr
 & \ \ \ \ \ \ \ \ 
 +s_{25} s_{34} (s_{13} -s_{14}) P_{1|34|25,6,7,8}
 + (3,4,5|3,4,5,6,7,8) \big] \ .
}$$
The BRST generator inherits its locality from the seven-point analogue \manlocN.
Again, the anomalous terms on the right-hand side of \manlocOC\ are not by themselves
BRST invariant. Hence, by the BRST generator on the left-hand side of \manlocOC, none of the
seven local BRST invariants at eight points can belong to the cohomology. As long as we do not
consider superspace combinations beyond the building blocks of section~\LocalBBsec, the 
manifestly local BRST cohomology is therefore demonstrated to be empty at five to eight points.

\appendix{C}{Eight-point anomalous building blocks}
\applab\Deltaapp

\noindent
This appendix is dedicated to the refined anomalous building
blocks at eight points that enter the discussion of the eight-point
correlator via \deltaJ\ and do not vanish in the BRST cohomology,
\eqnn\eightDeltas
$$\eqalignno{
\Delta_{1|23|4,5,6,7,8} &=
\cY_{23|1,4,5,6,7,8}
+ \cY_{3|12|4,5,6,7,8}
- \cY_{2|13,4,5,6,7,8} &\eightDeltas\cr
&\quad{}+\Big( \cY_{123,4,5,6,7,8}^m k_3^m +
\big[s_{34}\cY_{1234,5,6,7,8}+(4\leftrightarrow5,6,7,8)\big] -
(2\leftrightarrow3)\Big)\cr
\Delta_{1|2|34,5,6,7,8} &=
\cY_{2|1,34,5,6,7,8}
+ \cY_{2|13,4,5,6,7,8}
- \cY_{2|14,3,5,6,7,8}\cr
&\quad{}+k_2^m \cY^m_{12,34,5,6,7,8} + \big[s_{25}\cY_{125,34,6,7,8}+
(5\leftrightarrow6,7,8)\big]\cr
&\quad{}+\Big(k_2^m \cY_{124,3,5,6,7,8}^m  +
s_{24}(\cY_{1234,5,6,7,8}+\cY_{1324,5,6,7,8})\cr
&\qquad\quad{}+
\big[s_{25}\cY_{4125,3,6,7,8}+(5\leftrightarrow6,7,8)\big] -
(3\leftrightarrow4)\Big)\cr
\Delta^m_{1|2|3,4,5,6,7,8}&=
\cY^m_{2|1,3,4,5,6,7,8}
+ k_2^p \cY^{pm}_{12,3,4,5,6,7,8}\cr
&\quad{}
+ \Big[ s_{23}\cY^m_{123,4,5,6,7,8} + k_3^m \big(\cY_{2|13,4,5,6,7,8}
- k_2^p \cY^p_{213,4,5,6,7,8} \big)\cr
&\qquad{} - \big[k_4^m s_{23}\cY_{4123,5,6,7,8} + (4\leftrightarrow5,6,7,8)\big]
+ (3\leftrightarrow 4,5,6,7,8) \Big]\,.
}$$
Their BRST variations \eightDQBRST\ can be rewritten such as to expose the 
kinematic poles 
\eqnn\manlocOR
$$\eqalignno{
Q \Delta_{1|23|4,5,6,7,8} &={ k_{23}^m V_1 Y^m_{23,4,\ldots,8}
+ V_{231} Y_{4,5,6,7,8} + \big[ V_1 Y_{234,5,6,7,8}
+(4\leftrightarrow 5,6,7,8) \big]
\over s_{23}} &\manlocOR \cr
&\quad{}+  { k_3^m V_{12} Y^m_{3,4,\ldots,8} - V_{123} Y_{4,5,6,7,8} + \big[
V_{12} Y_{34,5,6,7,8}+(4\leftrightarrow 5,6,7,8) \big]\over s_{12}} \cr
&\quad{} - {k_2^m V_{13} Y^m_{2,4,\ldots,8} - V_{132} Y_{4,5,6,7,8} + \big[
V_{13} Y_{24,5,6,7,8} +(4\leftrightarrow 5,6,7,8) \big] \over s_{13}} \cr
&\quad{} - V_1 Y_{2|3,4,\ldots,8} + V_1 Y_{3|2,4,\ldots,8}\cr
Q \Delta_{1|2|34,5,6,7,8} &= {
k_2^m V_{1} Y^m_{2,34,5,\ldots,8} - V_{12} Y_{34,5,6,7,8}- V_1 Y_{342,5,6,7,8} + \big[
V_{1} Y_{34,25,6,7,8} +(5\leftrightarrow 6,7,8) \big] \over s_{34}}\cr
&\quad{}+{
k_2^m V_{13} Y^m_{2,4,\ldots,8} - V_{132} Y_{4,5,6,7,8} + \big[
V_{13} Y_{24,5,6,7,8} +(4\leftrightarrow 5,6,7,8) \big] \over s_{13}} \cr
&\quad{}-{
k_2^m V_{14} Y^m_{2,3,\ldots,8} - V_{142} Y_{3,5,6,7,8}
+ \big[ V_{14} Y_{23,5,6,7,8} +(3\leftrightarrow 5,6,7,8) \big] \over s_{14}}\cr
Q \Delta^m_{1|2|3,4,5,6,7,8} &= V_1 k_2^p Y^{mp}_{2,3,\ldots,8}
- V_{12} Y^m_{3,4,\ldots,8}
- k_2^m V_1 Y_{2|3,4,\ldots,8}+ \big[ V_1 Y^m_{23,4,\ldots,8}
+ (3\leftrightarrow 4,\ldots,8) \big]\cr
&\quad{}+\Big[ k_3^m {
k_2^p V_{13} Y^p_{2,4,\ldots,8} - V_{132} Y_{4,5,6,7,8} + \big[
V_{13} Y_{24,5,6,7,8}
+(4\leftrightarrow 5,6,7,8) \big]
\over s_{13}}\cr
&\hskip50pt\qquad{}+ (3\leftrightarrow 4,\ldots,8)\Big]
}$$
which should therefore be present in the components of the $\Delta_{1|\ldots}$ superfields
themselves.

Indeed, we shall now write down the bosonic components of the eight-point
topology $\Delta_{1|2|34,5,6,7,8}$ in terms of Berends--Giele currents in the
BCJ gauge, see \Gauge\ for more details.
\eqnn\longDelt
$$\eqalignno{
\langle \Delta_{1|2|34,5,6,7,8} \rangle& =
 +  {1 \over 2}\, (\cA_{12}^{k^{2}} \cF_{34}^{mn} + \cA_{34}^{k^{2}}\cF_{12}^{mn}) \e^{k^{5}k^{6}k^{7}k^{8}e^5e^6e^7e^8mn}
 +  {1 \over 2}\, \cA_{34}^{e^2} \e^{k^{1}k^{5}k^{6}k^{7}k^{8}e^1e^5e^6e^7e^8}\cr
& - {1 \over 4}\,\big[
   (\cF_{12}^{mn} \cF_{34}^{pq}(k^2 \cdot e^5)
 + \cF_{125}^{mn} \cF_{34}^{pq} s_{25}) \e^{k^{6}k^{7}k^{8}e^6e^7e^8mnpq}
+ (5\leftrightarrow6,7,8)\big]\cr
&  +  {1 \over 2}\,\big[(\cF_{214}^{mn}  (k^2 \cdot e^5) + \cF_{4125}^{mn} s_{25}) \e^{k^{3}k^{6}k^{7}k^{8}e^3e^6e^7e^8mn}
  + (5\leftrightarrow6,7,8)\big] - (3\leftrightarrow4)\cr
&  +  {1 \over 2}\,\big[
     (\cF_{214}^{mn} (k^2 \cdot e^3)+ \cF_{1324}^{mn} s_{24} +  \cF_{1234}^{mn} s_{24}) \e^{k^{5}k^{6}k^{7}k^{8}e^5e^6e^7e^8mn}
  - (3\leftrightarrow4)\big]\cr
&  +  {1 \over 2}\,\big[
       (\cA_{13}^{e^2} -  2\cA_{213}^{k^{2}}) \e^{k^{4}k^{5}k^{6}k^{7}k^{8}e^4e^5e^6e^7e^8}
 - (3\leftrightarrow4)\big]\cr
& +  {1 \over 12}\big[{1\over s_{13}}\big(
     ((k^1 \cdot e^2) -  (k^3 \cdot e^2)) (e^1 \cdot e^3)
  +  ((k^2 \cdot e^3) -  (k^1 \cdot e^3)) (e^1 \cdot e^2)\cr
&\hskip20pt  + ((k^3 \cdot e^1) -  (k^2 \cdot e^1)) (e^2 \cdot e^3)\big)
  \e^{k^{4}k^{5}k^{6}k^{7}k^{8}e^4e^5e^6e^7e^8}
- (3\leftrightarrow4)\big]\cr
&  +  {1 \over 12 s_{34}}\Big[\big[
    (k^3 \cdot e^2)(e^3 \cdot e^4)
  + ((k^2 \cdot e^4) -  (k^3 \cdot e^4)) (e^2 \cdot e^3)
  \big]
 \e^{k^{1}k^{5}k^{6}k^{7}k^{8}e^1e^5e^6e^7e^8}\cr
&\hskip40pt - (3\leftrightarrow4)\Big]\cr
&  +  {1 \over 4}\,\big[ \cF_{13}^{mn} (e^2 \cdot e^5) \e^{k^{4}k^{6}k^{7}k^{8}e^4e^6e^7e^8mn}
  + (5\leftrightarrow6,7,8)\big] - (3\leftrightarrow4)\cr
&  +  {1 \over 4}\,\big[ \cF_{34}^{mn}(e^2 \cdot e^5) \e^{k^{1}k^{6}k^{7}k^{8}e^1e^6e^7e^8mn}
+ (5\leftrightarrow6,7,8)\big]\cr
&  +  {1 \over 4}\, \big[\cF_{13}^{mn} (e^2 \cdot e^4) \e^{k^{5}k^{6}k^{7}k^{8}e^5e^6e^7e^8mn}
  - (3\leftrightarrow4)\big]\cr
&  +  {1 \over 4}\, \cF_{34}^{mn} (e^1 \cdot e^2)
\e^{k^{5}k^{6}k^{7}k^{8}e^5e^6e^7e^8mn} \, \Big|_{\theta=0} + {\rm fermions}\,.
&\longDelt
}$$
Given that $\theta$ and fermionic wavefunctions are suppressed on the right-hand side,
the superfields ${\cal A}^m_P$ and ${\cal F}^{mn}_P$ reduce to the bosonic Berends--Giele 
currents in the BCJ gauge \Gauge.
The other topologies in \eightDeltas\ have similar expansions but were omitted and can be found
as computer-readable files attached to the arXiv submission.
Although it is not manifest in the form presented above, all four-channel $s_{ijkl}$ poles turn out to
be absent\foot{This can be seen by inserting ${\cal F}^{mn}_{1234}=k_{1234}^m {\cal A}^n_{1234}
-{\cal A}^m_{123}{\cal A}^n_{4}-{\cal A}^m_{12}{\cal A}^n_{34}-{\cal A}^m_{1}{\cal A}^n_{234}-(m\leftrightarrow n)$ 
into $ \cF_{1234}^{mn}  \e^{k^{5}k^{6}k^{7}k^{8}e^5e^6e^7e^8mn}$ and noting the consequence
$k_{1234}^m {\cal A}^n_{1234}\e^{k^{5}k^{6}k^{7}k^{8}e^5e^6e^7e^8mn}=0$ of momentum conservation.} 
in every eight-point topology of $\Delta$. The three-channel
poles $s_{ijk}$ are however present (despite being absent in the BRST variation).

\appendix{D}{The BRST variations of local building blocks}
\applab\brstapp

\noindent In this appendix we list the BRST variations of every local building
block appearing in the one-loop correlators up to eight points that have not
appeared as examples in the preceding sections. The list below can be generated
from the formulas given in the main text but are presented here for convenience.
Together with the examples from the main text, these equations allow one to
verify all claims related to BRST variations in this series of papers.

\subsec Scalar $T_{A,B,C}$

Similarly, the BRST variations given in \QTs\ are appended by
\eqnn\scT
$$\eqalignno{
QT_{1234,5,6} &= (k_1\cdot k_2)\big[
V_1T_{234,5,6}
+ V_{13}T_{24,5,6}
+ V_{134}T_{2,5,6}
+ V_{14}T_{23,5,6} - (1\leftrightarrow2)\big]&\scT\cr
&\quad{}  + (k_{12}\cdot k_3)\big[
V_{12}T_{34,5,6}
+ V_{124}T_{3,5,6}
- (12\leftrightarrow3)\big]\cr
&\quad{} + (k_{123}\cdot k_4)\big[
V_{123}T_{4,5,6} - (123\leftrightarrow4)\big]\,,\cr
QT_{123,45,6} &= (k_1\cdot k_2)\big[V_1T_{23,45,6}+ V_{13}T_{2,45,6}
-V_{23}T_{1,45,6} - V_{2}T_{13,45,6}\big]\cr
&\quad{} + (k_{12}\cdot k_3)\big[V_{12}T_{3,45,6}-V_{3}T_{12,45,6}\big]\cr
&\quad{} + (k_4\cdot k_5)\big[V_4T_{123,5,6}-V_5T_{123,4,6}\big]\,,\cr
QT_{12,34,56} &= (k_1\cdot k_2)\big[V_1T_{2,34,56} - V_{2}T_{1,34,56}\big] +
(12\leftrightarrow 34,56)\,,\cr
QT_{12345,6,7} &=
(k_1\cdot k_2)\big[V_{1} T_{2345 , 6 , 7}
 +  V_{13} T_{245 , 6 , 7} 
 +  V_{134} T_{25 , 6 , 7} 
 +  V_{1345} T_{2 , 6 , 7}\cr
&\qquad{} +  V_{135} T_{24 , 6 , 7}
 +  V_{14} T_{235 , 6 , 7} 
 +  V_{145} T_{23 , 6 , 7} 
 +  V_{15} T_{234 , 6 , 7}
- (1\leftrightarrow2)\big]\cr
&\quad{}  + (k_{12}\cdot k_3)   \big[
  V_{12} T_{345 , 6 , 7} 
 +  V_{124} T_{35 , 6 , 7} 
 +  V_{1245} T_{3 , 6 , 7} 
 +  V_{125} T_{34 , 6 , 7} 
- (12\leftrightarrow3)\big]\cr
&\quad{}  +  (k_{123}\cdot k_4)   \big[
  V_{123} T_{45 , 6 , 7}
 +  V_{1235} T_{4 , 6 , 7} 
- (123\leftrightarrow4)\big]\cr
&\quad{}  +  (k_{1234}\cdot k_5)   \big[
  V_{1234} T_{5 , 6 , 7}
- (1234\leftrightarrow5)\big]\,,\cr
QT_{1234,56,7} &=
 (k_1\cdot k_2)   \big[
  V_{1} T_{234 , 56 , 7} 
 +  V_{13} T_{24 , 56 , 7} 
 +  V_{134} T_{56 , 2 , 7} 
 +  V_{14} T_{23 , 56 , 7} 
- (1\leftrightarrow2)
\big] \cr
&\quad{}  +  (k_{12}\cdot k_3)   \big[ 
   V_{12} T_{34 , 56 , 7} 
 +  V_{124} T_{56 , 3 , 7} 
	- (12\leftrightarrow3)
\big] \cr
&\quad{}  +  (k_{123}\cdot k_4)   \big[
   V_{123} T_{56 , 4 , 7} 
 -  (123\leftrightarrow4)
\big]\cr
&\quad{}  +  (k_5\cdot k_6)   \big[
   V_{5} T_{1234 , 6 , 7}
- (5\leftrightarrow6)
\big]\cr
QT_{123,456,7} &=
  (k_{1}\cdot k_{2})   \big[ 
  V_{1} T_{456 , 23 , 7} 
 +  V_{13} T_{456 , 2 , 7} 
- (1\leftrightarrow2)
\big]\cr
&\quad{}  +  (k_{12}\cdot k_{3})   \big[ 
  V_{12} T_{456 , 3 , 7} 
- (12\leftrightarrow3)
\big]\cr
&\quad{}  +  (k_{4}\cdot k_{5})   \big[ 
  V_{4} T_{123 , 56 , 7} 
 +  V_{46} T_{123 , 5 , 7} 
- (4\leftrightarrow5)
\big]\cr
&\quad{}  +  (k_{45}\cdot k_{6})   \big[ 
  V_{45} T_{123 , 6 , 7} 
 - (45\leftrightarrow6)
\big]\cr
QT_{123,45,67} &=
  (k_{1}\cdot k_{2})   \big[ 
   V_{1} T_{23 , 45 , 67} 
 +  V_{13} T_{45 , 67 , 2} 
- (1\leftrightarrow2)
\big] \cr
&\quad{}  +  (k_{12}\cdot k_{3})   \big[
   V_{12} T_{45 , 67 , 3} 
- (12\leftrightarrow3)
\big]\cr
&\quad{}  +  (k_{4}\cdot k_{5})   \big[ 
   V_{4} T_{123 , 67 , 5} 
- (4\leftrightarrow5)
\big]\cr
&\quad{}  +  (k_{6}\cdot k_{7})   \big[ 
   V_{6} T_{123 , 45 , 7} 
- (6\leftrightarrow7)
\big]\, .
}$$

\subsec Tensorial $T^{m \ldots}_{A,B,C, \ldots}$

\def\kk#1#2{(k_{#1}\cdot k_{#2})}
\noindent
The vectorial BRST variations in \QTms\ generalize to
\eqnn\vecTs
$$\eqalignno{
QT^m_{1234,5,6,7} &= 
\big[k^m_{1234}V_{1234}T_{5,6,7} + (1234\leftrightarrow5,6,7)\big] &\vecTs\cr
&\quad{} + (k_1\cdot k_2)\big[V_1T^m_{234,5,6,7} + V_{13}T^m_{24,5,6,7} +
V_{14}T^m_{23,5,6,7} + V_{134}T^m_{2,5,6,7} -(1\leftrightarrow2)\big]\cr
&\quad{} + (k_{12}\cdot k_3)\big[V_{12}T^m_{34,5,6,7} + V_{124}T^m_{3,5,6,7} -
(12\leftrightarrow3)\big]\cr
&\quad{} + \kk{123}{4}\big[V_{123}T^m_{4,5,6,7} - (123\leftrightarrow4)\big]
\cr
QT^m_{123,45,6,7} &=
\big[k^m_{123}V_{123}T_{45,6,7} + (123\leftrightarrow45,6,7)\big]\cr
&\quad{} + (k_1\cdot k_2)\big[V_1T^m_{23,45,6,7} + V_{13}T^m_{2,45,6,7} -(1\leftrightarrow2)\big]\cr
&\quad{} + (k_{12}\cdot k_3)\big[V_{12}T^m_{3,45,6,7} - (12\leftrightarrow3)\big]\cr
&\quad{} + \kk{4}{5}\big[V_{4}T^m_{123,5,6,7} - (4\leftrightarrow5)\big]
\cr
QT^m_{12,34,56,7} &=
\big[k^m_{12}V_{12}T_{34,56,7} + (12\leftrightarrow34,56,7)\big]\cr
&\quad{} + \big[\kk12(V_1T^m_{2,34,56,7}-(1\leftrightarrow2)) +
(12\leftrightarrow34,56)\big]\, ,
}$$
while the higher-multiplicity analogues of the tensorial
BRST variations \QTexs\ read
\eqnn\highTs
$$\eqalignno{
QT^{mn}_{123,4,5,6,7} &=
\d^{mn}Y_{123,4,5,6,7} +
\big[k^{(m}_{123}V_{123}T^{n)}_{4,5,6,7} +
(123\leftrightarrow4,5,6,7)\big]&\highTs\cr
&\quad{}  + \kk12\big[V_1T^{mn}_{23,4,5,6,7} +  V_{13}T^{mn}_{2,4,5,6,7} -
(1\leftrightarrow2)\big]\cr
&\quad{}  +\kk{12}{3}\big[V_{12}T^{mn}_{3,4,5,6,7} -
(12\leftrightarrow3)\big]\cr
QT^{mn}_{12,34,5,6,7} &= 
\d^{mn}Y_{12,34,5,6,7} +
\big[k^{(m}_{12}V_{12}T^{n)}_{34,5,6,7} + (12\leftrightarrow34,5,6,7)\big]\cr
&\quad{}  + \kk12\big[V_1T^{mn}_{2,34,5,6,7} - (1\leftrightarrow2)\big]\cr
&\quad{}  + \kk34\big[V_3T^{mn}_{12,4,5,6,7} - (3\leftrightarrow4)\big]\cr
QT^{mnp}_{12,3,4,5,6,7} &= \d^{(mn}Y^{p)}_{12,3,4,5,6,7} +
\big[k_{12}^{(m}V_{12}T^{np)}_{3,4,5,6,7} + (12\leftrightarrow34,5,6,7)\big]\cr
&\quad{}  + \kk12\big[V_1T^{mnp}_{2,3,4,5,6,7} - (1\leftrightarrow2)\big]\cr
QT^{mnpq}_{1,2,3,4,5,6,7} &= \d^{(mn}Y^{pq)}_{1,2,3,4,5,6,7} +
\big[k_1^{(m}V_1T^{npq)}_{2,3,4,5,6,7} + (1\leftrightarrow2,3,4,5,6,7)\big] \, .
}$$

\subsec Refined $J^{m \ldots}_{A|B,C, \ldots}$

The BRST variations of the refined building blocks $J_{A|B,C,D,E}$
that appear in the eight-point correlator read,
\eqnn\QJseight
$$\eqalignno{
QJ_{123|4,5,6,7} &=  k_{123}^m V_{123} T^m_{4,5,6,7}
+\bigr[ V_{[123,4]} T_{5,6,7} + (4 \leftrightarrow 5,6,7) \bigl]
+ Y_{123,4,5,6,7} &\QJseight\cr
 & \quad{} + (k^1\cdot k^2) \bigl[V_1 J_{23|4,5,6,7}+ V_{13} J_{2|4,5,6,7}
 - (1\leftrightarrow 2)\bigr] \cr
 & \quad{} + (k^{12}\cdot k^3) \bigl[V_{12} J_{3|4,5,6,7}
 - (12\leftrightarrow 3)\bigr]\cr
QJ_{12|34,5,6,7} &= k_{12}^m V_{12} T^m_{34,5,6,7}+ \big[V_{[12,34]}T_{5,6,7} + (34\leftrightarrow 5,6,7)\big] + Y_{12,34,5,6,7}  \cr
&\quad{} + (k^1\cdot k^2)\big[ V_1 J_{2|34,5,6,7}- (1\leftrightarrow 2)\big]
+ (k^3\cdot k^4)\big[ V_3 J_{12|4,5,6,7}- (3\leftrightarrow 4)\big] \cr
QJ_{1|23,45,6,7} &=  k_{1}^m V_{1} T^m_{23,45,6,7} + \big[ V_{[1,23]} T_{45,6,7} + (23\leftrightarrow 45,6,7)\big]  + Y_{1,23,45,6,7} \cr
&\quad{} + (k^2\cdot k^3)\big[V_2 J_{1|3,45,6,7}- (2\leftrightarrow 3)\big] + (k^4\cdot k^5)\big[ V_4 J_{1|23,5,6,7}-
(4\leftrightarrow 5)\big]  \cr
QJ_{1|234,5,6,7} &= k_1^m V_1 T^m_{234,5,6,7} + \big[V_{[1,234]} T_{5,6,7} + (234 \leftrightarrow 5,6,7) \big] + Y_{1,234,5,6,7}\cr
& \quad{} + (k^2\cdot k^3)\big[V_2 J_{1|34,5,6,7} + V_{24} J_{1|3,5,6,7} - (2\leftrightarrow 3)\big] \cr
& \quad{} + (k^{23}\cdot k^4) \big[ V_{23} J_{1|4,5,6,7}- (23\leftrightarrow 4)\big] \ ,
}$$
see \QJthree\ for examples at lower multiplicity.
The BRST variation of every tensorial $J^{m \ldots}$ relevant to one-loop correlators up to
eight points has been spelled out in \QJmex.

\subsec Anomaly building blocks $Y^{m \ldots}_{A,B,C, \ldots}$

The BRST variations \YsuptoSix\ of anomaly building blocks generalize
as follows to multiplicity seven and eight,
\eqnn\anonYs
$$\eqalignno{
QY_{123,4,5,6,7} &=
\kk{1}{2}\big[V_1Y_{23,4,5,6,7} + V_{13}Y_{2,4,5,6,7} -
(1\leftrightarrow2)\big]&\anonYs\cr
&\quad{} + \kk{12}{3}\big[V_{12}Y_{3,4,5,6,7} - (12\leftrightarrow3)\big]\cr
QY_{12,34,5,6,7} &= \kk12\big[V_1Y_{2,34,5,6,7}-(1\leftrightarrow2)\big] +
(12\leftrightarrow34)\cr
QY_{1234,5,6,7,8} &=
\kk12\big[V_{1}Y_{234,5,6,7,8}  {+} V_{13}Y_{24,5,6,7,8} {+} V_{14}Y_{23,5,6,7,8} {+}
V_{134}Y_{2,5,6,7,8} - (1{\leftrightarrow}2)\big]\cr
&\quad{} +\kk{12}{3}\big[V_{12}Y_{34,5,6,7,8} + V_{124}Y_{3,5,6,7,8} -
(12\leftrightarrow3)\big]\cr
&\quad{} +\kk{123}{4}\big[V_{123}Y_{4,5,6,7,8} - (123\leftrightarrow4)\big]\cr
QY_{123,45,6,7,8} &=
\kk12\big[V_1Y_{23,45,6,7,8} + V_{13}Y_{2,45,6,7,8} -(1\leftrightarrow2)\big]\cr
&\quad{} +\kk{12}{3}\big[V_{12}Y_{3,45,6,7,8} - (12\leftrightarrow3)\big]\cr
&\quad{} +\kk45\big[V_4Y_{123,5,6,7,8} - (4\leftrightarrow5)\big]\cr
QY_{12,34,56,7,8} &=
\kk12\big[V_1Y_{2,34,56,7,8} -(1\leftrightarrow2)\big] + (12\leftrightarrow34,56)\cr
QY^m_{12,3,4,5,6,7} &= \big[k_{12}^m V_{12}Y_{3,4,5,6,7} +
(12\leftrightarrow3,4,5,6,7)\big]\cr
&\quad{} + \kk12\big[V_1Y^m_{2,3,4,5,6,7}-(1\leftrightarrow2)\big]\cr
QY^m_{123,4,5,6,7,8} &= \big[k^m_{123}V_{123}Y_{4,5,6,7,8} +
(123\leftrightarrow4,5,6,7,8)\big]\cr
&\quad{} + \kk12\big[V_{1}Y^m_{23,4,5,6,7,8} +
V_{13}Y^m_{2,4,5,6,7,8}-(1\leftrightarrow2)\big]\cr
&\quad{} + \kk{12}{3}\big[V_{12}Y^m_{3,4,5,6,7,8}-(12\leftrightarrow3)\big]\cr
QY^m_{12,34,5,6,7,8} &= \big[k^m_{12}V_{12}Y_{34,5,6,7,8} +
(12\leftrightarrow34,5,6,7,8)\big]\cr
&\quad{} + \kk12\big[V_1Y^m_{2,34,5,6,7,8} - (1\leftrightarrow2)\big]
+ \kk34\big[V_3Y^m_{12,4,5,6,7,8} - (3\leftrightarrow4)\big] \cr
QY^{mn}_{1,2,3,4,5,6,7} &= k_1^{(m} V_1Y^{n)}_{2,3,4,5,6,7} +
(1\leftrightarrow2,3,4,5,6,7)\cr
QY^{mn}_{12,3,4,5,6,7,8} &= \big[k_{12}^{(m} V_{12}Y^{n)}_{3,4,5,6,7,8} +
(12{\leftrightarrow}3,4,5,6,7,8)\big] + \kk12\big[V_1Y^{mn}_{2,3,4,5,6,7,8} -
(1{\leftrightarrow}2)\big]\cr
QY^{mnp}_{1,2,3,4,5,6,7,8} &= k_1^{(m}V_1Y^{np)}_{2,3,4,5,6,7,8} +
(1\leftrightarrow2,3,4,5,6,7,8) \, .
}$$
The BRST variations of refined anomaly building blocks relevant to $(n\leq 8)$-point
correlators are spelled out in \simYsev.

\listrefs

\bye

%% file: harvmacMv2.tex


\input amssym.tex 

\def\unredoffs{}
\tolerance=1000\hfuzz=2pt
\catcode`\@=11 
\ifx\hyperdef\UNd@FiNeD\def\hyperdef#1#2#3#4{#4}\def\hyperref#1#2#3#4{#4}\def\href#1#2{#2}\fi
\magnification=1200\unredoffs\baselineskip=16pt plus 2pt minus 1pt
\def\Date#1{\vfill\leftline{#1}\tenpoint\supereject%
\footline={\hss\tenrm\hyperdef\hypernoname{page}\folio\folio\hss}}%

{\count255=\time\divide\count255 by 60 \xdef\hourmin{\number\count255}
 \multiply\count255 by-60\advance\count255 by\time
 \xdef\hourmin{\hourmin:\ifnum\count255<10 0\fi\the\count255}
}
\def\date{\number\day.\number\month.\number\year\ at \hourmin}


\def\nolabels{\def\wrlabeL##1{}\def\eqlabeL##1{}\def\reflabeL##1{}}
\def\writelabels{\def\wrlabeL##1{\leavevmode\vadjust{\rlap{\smash%
{\line{{\escapechar=` \hfill\rlap{\sevenrm\hskip.03in\string##1}}}}}}}%
\def\eqlabeL##1{{\escapechar-1\rlap{\sevenrm\hskip.05in\string##1}}}%
\def\reflabeL##1{\noexpand\llap{\noexpand\sevenrm\string\string\string##1}}}
\nolabels

\global\newcount\secno \global\secno=0
\global\newcount\meqno \global\meqno=1
\def\s@csym{}

\def\newsec#1\par{\global\advance\secno by1%
{\toks0{#1}\message{(\the\secno. \the\toks0)}}%
\global\subsecno=0\eqnres@t\let\s@csym\secsym\xdef\secn@m{\the\secno}\noindent
{\bf\hyperdef\hypernoname{section}{\the\secno}{\the\secno.} #1}%
\writetoca{{\string\hyperref{}{section}{\the\secno}{\bf \the\secno\quad}} {\bf #1}}\par%
\nobreak\medskip\nobreak\noindent\ignorespaces}
\def\eqnres@t{\xdef\secsym{\the\secno.}\global\meqno=1\bigbreak\bigskip}
\def\sequentialequations{\def\eqnres@t{\bigbreak}}\xdef\secsym{}

\global\newcount\subsecno \global\subsecno=0
\def\subsec#1\par{\global\advance\subsecno by1%
{\toks0{#1}\message{(\s@csym\the\subsecno. \the\toks0)}}%
\global\subsubsecno=0%
\ifnum\lastpenalty>9000\else\bigbreak\fi
\noindent{\it\hyperdef\hypernoname{subsection}{\secn@m.\the\subsecno}%
{\secn@m.\the\subsecno.} #1}\writetoca{\string\hskip1.45cm
{\string\hyperref{}{subsection}{\secn@m.\the\subsecno}{\secn@m.\the\subsecno.}}
{#1}}\par\nobreak\medskip\nobreak\noindent\ignorespaces}

\global\newcount\subsubsecno \global\subsubsecno=0
\def\subsubsec#1\par{\global\advance\subsubsecno by1%
{\toks0{#1}\message{(\secn@m.\the\subsecno.\the\subsubsecno. \the\toks0)}}%
\global\subsubsubsecno=0%
\ifnum\lastpenalty>9000\else\bigbreak\fi
\noindent{\it\hyperdef\hypernoname{subsubsection}{\secn@m.\the\subsecno\the\subsubsecno}%
{\secn@m.\the\subsecno.\the\subsubsecno.} #1}
\par\nobreak\medskip\nobreak\noindent\ignorespaces}

\global\newcount\subsubsubsecno \global\subsubsubsecno=0
\def\subsubsubsec#1\par{\global\advance\subsubsubsecno by1%
{\toks0{#1}\message{(\secn@m.\the\subsecno.\the\subsubsecno.\the\subsubsubsecno \the\toks0)}}%
\ifnum\lastpenalty>9000\else\bigbreak\fi
\noindent{\it\hyperdef\hypernoname{subsubsection}{\secn@m.\the\subsecno\the\subsubsecno\the\subsubsubsecno}%
{\secn@m.\the\subsecno.\the\subsubsecno.\the\subsubsubsecno.} #1}%
\par\nobreak\medskip\nobreak\noindent\ignorespaces}


\def\newnewsec#1#2\par{\global\advance\secno by1%
{\toks0{#2}\message{(\secn@m. \the\toks0)}}%
\global\subsecno=0\eqnres@t\let\s@csym\secsym\xdef\secn@m{\the\secno}\noindent
\ifnum\lastpenalty>9000\else\bigbreak\fi
\noindent{\bf\hyperdef\hypernoname{section}{\secn@m}{\secn@m.} #2}%
\writetoca{{\string\hyperref{}{section}{\the\secno}{\bf \the\secno\quad}} {\bf #2}}
\DefWarn#1%
\xdef#1{\noexpand\hyperref{}{section}{\the\secno}%
{\the\secno}}\writedef{#1\leftbracket#1}\wrlabeL{#1=#1}%
\par\nobreak\medskip\nobreak\noindent\ignorespaces}

\def\newsubsec#1#2\par{\global\advance\subsecno by1%
{\toks0{#2}\message{(\secn@m.\the\subsecno. \the\toks0)}}%
\global\subsubsecno=0%
\ifnum\lastpenalty>9000\else\bigbreak\fi
\noindent{\it\hyperdef\hypernoname{subsection}{\secn@m.\the\subsecno}%
{\secn@m.\the\subsecno.} #2}
\DefWarn#1%
\xdef#1{\noexpand\hyperref{}{subsection}{\secn@m.\the\subsecno}%
{\secn@m.\the\subsecno}}\writedef{#1\leftbracket#1}\wrlabeL{#1=#1}%
\writetoca{\string\hskip1.45cm
{\string\hyperref{}{subsection}{\secn@m.\the\subsecno}{\secn@m.\the\subsecno.}}
{#2}}%
\par\nobreak\medskip\nobreak\noindent\ignorespaces}

\def\newsubsubsec#1#2\par{\global\advance\subsubsecno by1%
{\toks0{#2}\message{(\secn@m.\the\subsecno.\the\subsubsecno. \the\toks0)}}%
\global\subsubsubsecno=0%
\ifnum\lastpenalty>9000\else\bigbreak\fi
\noindent{\it\hyperdef\hypernoname{subsubsection}{\secn@m.\the\subsecno\the\subsubsecno}%
{\secn@m.\the\subsecno.\the\subsubsecno.} #2}
\DefWarn#1%
\xdef#1{\noexpand\hyperref{}{subsubsection}{\secn@m.\the\subsecno.\the\subsubsecno}%
{\secn@m.\the\subsecno.\the\subsubsecno}}\writedef{#1\leftbracket#1}\wrlabeL{#1=#1}%
\par\nobreak\medskip\nobreak\noindent\ignorespaces}

\def\newsubsubsubsec#1#2\par{\global\advance\subsubsubsecno by1%
{\toks0{#2}\message{(\secn@m.\the\subsecno.\the\subsubsecno.\the\subsubsubsecno \the\toks0)}}%
\ifnum\lastpenalty>9000\else\bigbreak\fi
\noindent{\it\hyperdef\hypernoname{subsubsection}{\secn@m.\the\subsecno\the\subsubsecno\the\subsubsubsecno}%
{\secn@m.\the\subsecno.\the\subsubsecno.\the\subsubsubsecno.} #2}
\DefWarn#1%
\xdef#1{\noexpand\hyperref{}{subsubsubsection}{\secn@m.\the\subsecno.\the\subsubsecno.\the\subsubsubsecno}%
{\secn@m.\the\subsecno.\the\subsubsecno.\the\subsubsubsecno}}\writedef{#1\leftbracket#1}\wrlabeL{#1=#1}%
\par\nobreak\medskip\nobreak\noindent\ignorespaces}

\def\appendix#1#2{\global\meqno=1\global\subsecno=0\global\subsubsecno=0\xdef\secsym{\hbox{#1.}}%
\bigbreak\bigskip\noindent{\bf Appendix \hyperdef\hypernoname{appendix}{#1}%
{#1.} #2}{\toks0{(#1. #2)}\message{\the\toks0}}%
\xdef\s@csym{#1.}\xdef\secn@m{#1}%
\writetoca{{\string\hyperref{}{appendix}{#1}{\bf {#1}\quad}} {\bf #2}}%
\par\nobreak\medskip\nobreak}

%
\def\checkm@de#1#2{\ifmmode{\def\f@rst##1{##1}\hyperdef\hypernoname{equation}%
{#1}{#2}}\else\hyperref{}{equation}{#1}{#2}\fi}
\def\eqnn#1{\DefWarn#1\xdef #1{(\noexpand\relax\noexpand\checkm@de%
{\s@csym\the\meqno}{\secsym\the\meqno})}%
\wrlabeL#1\writedef{#1\leftbracket#1}\global\advance\meqno by1}
\def\f@rst#1{\c@t#1a\em@ark}\def\c@t#1#2\em@ark{#1}
\def\eqna#1{\DefWarn#1\wrlabeL{#1$\{\}$}%
\xdef #1##1{(\noexpand\relax\noexpand\checkm@de%
{\s@csym\the\meqno\noexpand\f@rst{##1}1}{\hbox{$\secsym\the\meqno##1$}})}
\writedef{#1\numbersign1\leftbracket#1{\numbersign1}}\global\advance\meqno by1}
\def\eqn#1#2{\DefWarn#1%
\xdef #1{(\noexpand\hyperref{}{equation}{\s@csym\the\meqno}%
{\secsym\the\meqno})}$$#2\eqno(\hyperdef\hypernoname{equation}%
{\s@csym\the\meqno}{\secsym\the\meqno})\eqlabeL#1$$%
\writedef{#1\leftbracket#1}\global\advance\meqno by1}
\def\xeqn{\expandafter\xe@n}\def\xe@n(#1){#1}
\def\xeqna#1{\expandafter\xe@n#1}
\def\eqns#1{(\e@ns #1{\hbox{}})}
\def\e@ns#1{\ifx\UNd@FiNeD#1\message{eqnlabel \string#1 is undefined.}%
\xdef#1{(?.?)}\fi{\let\hyperref=\relax\xdef\next{#1}}%
\ifx\next\em@rk\def\next{}\else%
\ifx\next#1\xeqn#1\else\def\n@xt{#1}\ifx\n@xt\next#1\else\xeqna#1\fi
\fi\let\next=\e@ns\fi\next}
\def\DefWarn#1{}
%
\newskip\footskip\footskip14pt plus 1pt minus 1pt 
\def\footnotefont{\ninepoint}\def\f@t#1{\footnotefont #1\@foot}
\def\f@@t{\baselineskip\footskip\bgroup\footnotefont\aftergroup\@foot\let\next}
\setbox\strutbox=\hbox{\vrule height9.5pt depth4.5pt width0pt}
\global\newcount\ftno \global\ftno=0
\def\foot{\global\advance\ftno by1\def\foot@rg{\hyperref{}{footnote}%
{\the\ftno}{\the\ftno}\xdef\foot@rg{\noexpand\hyperdef\noexpand\hypernoname%
{footnote}{\the\ftno}{\the\ftno}}}\footnote{$^{\foot@rg}$}}
%
%
%
\global\newcount\refno \global\refno=1
\newwrite\rfile
\def\ref{[\hyperref{}{reference}{\the\refno}{\the\refno}]\nref}
\def\nref#1{\DefWarn#1%
\xdef#1{[\noexpand\hyperref{}{reference}{\the\refno}{\the\refno}]}%
\writedef{#1\leftbracket#1}%
\ifnum\refno=1\immediate\openout\rfile=\jobname.refs\fi
\chardef\wfile=\rfile\immediate\write\rfile{\noexpand\item{[\noexpand\hyperdef%
\noexpand\hypernoname{reference}{\the\refno}{\the\refno}]\ }%
\reflabeL{#1\hskip.31in}\pctsign}\global\advance\refno by1\findarg}
\def\findarg#1#{\begingroup\obeylines\newlinechar=`\^^M\pass@rg}
{\obeylines\gdef\pass@rg#1{\writ@line\relax #1^^M\hbox{}^^M}%
\gdef\writ@line#1^^M{\expandafter\toks0\expandafter{\striprel@x #1}%
\edef\next{\the\toks0}\ifx\next\em@rk\let\next=\endgroup\else\ifx\next\empty%
\else\immediate\write\wfile{\the\toks0}\fi\let\next=\writ@line\fi\next\relax}}
\def\striprel@x#1{} \def\em@rk{\hbox{}}
\def\lref{\begingroup\obeylines\lr@f}
\def\lr@f#1#2{\DefWarn#1\gdef#1{\let#1=\UNd@FiNeD\ref#1{#2}}\endgroup\unskip}
\def\semi{;\hfil\break}
\def\addref#1{\immediate\write\rfile{\noexpand\item{}#1}} 
\def\listrefs{\vfill\supereject\immediate\closeout\rfile\writestoppt
\baselineskip=\footskip\centerline{{\bf References}}\bigskip{\parindent=20pt%
\frenchspacing\escapechar=` \input \jobname.refs\vfill\eject}\nonfrenchspacing}
\def\startrefs#1{\immediate\openout\rfile=\jobname.refs\refno=#1}
\def\xref{\expandafter\xr@f}\def\xr@f[#1]{#1}
\def\refs#1{\count255=1[\r@fs #1{\hbox{}}]}
\def\r@fs#1{\ifx\UNd@FiNeD#1\message{reflabel \string#1 is undefined.}%
\nref#1{need to supply reference \string#1.}\fi%
\vphantom{\hphantom{#1}}{\let\hyperref=\relax\xdef\next{#1}}%
\ifx\next\em@rk\def\next{}%
\else\ifx\next#1\ifodd\count255\relax\xref#1\count255=0\fi%
\else#1\count255=1\fi\let\next=\r@fs\fi\next}
%

%
\newwrite\ffile\global\newcount\figno \global\figno=1
\def\fig{fig.~\hyperref{}{figure}{\the\figno}{\the\figno}\nfig}
\def\nfig#1{\DefWarn#1%
\xdef#1{fig.~\noexpand\hyperref{}{figure}{\the\figno}{\the\figno}}%
\writedef{#1\leftbracket fig.\noexpand~\xfig#1}%
\ifnum\figno=1\immediate\openout\ffile=\jobname.figs\fi\chardef\wfile=\ffile%
{\let\hyperref=\relax
\immediate\write\ffile{\noexpand\medskip\noexpand\item{Fig.\ %
\noexpand\hyperdef\noexpand\hypernoname{figure}{\the\figno}{\the\figno}. }
\reflabeL{#1\hskip.55in}\pctsign}}\global\advance\figno by1\findarg}
\def\xfig{\expandafter\xf@g}\def\xf@g fig.\penalty\@M\ {}
\def\figs#1{figs.~\f@gs #1{\hbox{}}}
\def\f@gs#1{{\let\hyperref=\relax\xdef\next{#1}}\ifx\next\em@rk\def\next{}\else
\ifx\next#1\xfig #1\else#1\fi\let\next=\f@gs\fi\next}
%
\def\figin{\epsfcheck\figin}\def\figins{\epsfcheck\figins}
\def\epsfcheck{\ifx\epsfbox\UnDeFiNeD
\message{(NO epsf.tex, FIGURES WILL BE IGNORED)}
\gdef\figin##1{\vskip2in}\gdef\figins##1{\hskip.5in}
\else\message{(FIGURES WILL BE INCLUDED)}%
\gdef\figin##1{##1}\gdef\figins##1{##1}\fi}
\def\figinsert{\goodbreak\topinsert}
\def\ifig#1#2#3{\DefWarn#1\xdef#1{fig.~\the\figno}
\writedef{#1\leftbracket fig.\noexpand~\the\figno}%
\figinsert\figin{\centerline{#3}}
\smallskip
\leftskip=0pt \rightskip=0pt
\baselineskip12pt\noindent
{{\bf Fig.~\the\figno}\ \ninepoint #2}
\medskip
\global\advance\figno by1\par\endinsert}
\newwrite\lfile
{\escapechar-1\xdef\pctsign{\string\%}\xdef\leftbracket{\string\{}
\xdef\rightbracket{\string\}}\xdef\numbersign{\string\#}}
\def\writedefs{\immediate\openout\lfile=label.defs \def\writedef##1{%
{\let\hyperref=\relax\let\hyperdef=\relax\let\hypernoname=\relax
 \immediate\write\lfile{\string\checkdef\string##1\rightbracket}}}}%
\def\writestop{\def\writestoppt{\immediate\write\lfile{\string\pageno
 \the\pageno\string\startrefs\leftbracket\the\refno\rightbracket
 \string\def\string\secsym\leftbracket\secsym\rightbracket
 \string\secno\the\secno\string\meqno\the\meqno}\immediate\closeout\lfile}}
\def\writestoppt{}\def\writedef#1{}

\def\seclab#1\par{\DefWarn#1%
\xdef #1{\noexpand\hyperref{}{section}{\the\secno}{\the\secno}}%
\writedef{#1\leftbracket#1}\wrlabeL{#1=#1}\par%
\nobreak\medskip\nobreak\noindent\ignorespaces}
\def\subseclab#1\par{\DefWarn#1%
\xdef #1{\noexpand\hyperref{}{subsection}{\the\secno.\the\subsecno}%
{\the\secno.\the\subsecno}}\writedef{#1\leftbracket#1}\wrlabeL{#1=#1}\par%
\nobreak\medskip\nobreak\noindent\ignorespaces}
\def\subsubseclab#1\par{\DefWarn#1%
\xdef#1{\noexpand\hyperref{}{subsubsection}{\the\secno.\the\subsecno.\the\subsubsecno}%
{\the\secno.\the\subsecno.\the\subsubsecno}}\writedef{#1\leftbracket#1}\wrlabeL{#1=#1}\par%
\nobreak\medskip\nobreak\noindent\ignorespaces}
\def\applab#1\par{\DefWarn#1%
\xdef#1{\noexpand\hyperref{}{appendix}{\secn@m}{\secn@m}}%
\writedef{#1\leftbracket#1}\wrlabeL{#1=#1}%
\par\nobreak\medskip\nobreak\noindent\ignorespaces}
\def\appsublab#1{\DefWarn#1%
\xdef #1{\noexpand\hyperref{}{appendix}{\secn@m.\the\subsecno}{\secn@m.\the\subsecno}}%
\writedef{#1\leftbracket#1}\wrlabeL{#1=#1}}
\newwrite\tfile \def\writetoca#1{}
\def\leaderfill{\leaders\hbox to 1em{\hss.\hss}\hfill}
\def\writetoc{\immediate\openout\tfile=\jobname.toc
   \def\writetoca##1{{\edef\next{\write\tfile{\noindent ##1
   \string\leaderfill{
   \string\hyperref{}{page}{\noexpand\number\pageno}%
   {\noexpand\number\pageno}} \par}}\next}}
}
\newread\ch@ckfile
\def\listtoc{\immediate\closeout\tfile\immediate\openin\ch@ckfile=\jobname.toc
\ifeof\ch@ckfile\message{no file \jobname.toc, no table of contents this pass}%
\else\closein\ch@ckfile\centerline{\bf Contents}\nobreak\medskip%
{\baselineskip=15.5pt\footnotefont\parskip=0pt\catcode`\@=11\input\jobname.toc
\catcode`\@=12\bigbreak\bigskip}\fi}
\catcode`\@=12 
\def\tenpoint{\def\rm{\fam0\tenrm}
\textfont0=\tenrm \scriptfont0=\sevenrm \scriptscriptfont0=\fiverm
\textfont1=\teni  \scriptfont1=\seveni  \scriptscriptfont1=\fivei
\textfont2=\tensy \scriptfont2=\sevensy \scriptscriptfont2=\fivesy
\textfont\itfam=\tenit \def\it{\fam\itfam\tenit}\def\footnotefont{\ninepoint}%
\textfont\bffam=\tenbf \def\bf{\fam\bffam\tenbf}\def\sl{\fam\slfam\tensl}\rm}
\font\ninerm=cmr9 \font\sixrm=cmr6 \font\ninei=cmmi9 \font\sixi=cmmi6
\font\ninesy=cmsy9 \font\sixsy=cmsy6 \font\ninebf=cmbx9
\font\nineit=cmti9 \font\ninesl=cmsl9 \skewchar\ninei='177
\skewchar\sixi='177 \skewchar\ninesy='60 \skewchar\sixsy='60
\def\ninepoint{\def\rm{\fam0\ninerm}
\textfont0=\ninerm \scriptfont0=\sixrm \scriptscriptfont0=\fiverm
\textfont1=\ninei \scriptfont1=\sixi \scriptscriptfont1=\fivei
\textfont2=\ninesy \scriptfont2=\sixsy \scriptscriptfont2=\fivesy
\textfont\itfam=\ninei \def\it{\fam\itfam\nineit}\def\sl{\fam\slfam\ninesl}%
\textfont\bffam=\ninebf \def\bf{\fam\bffam\ninebf}\rm}
%
\hyphenation{anom-aly anom-alies coun-ter-term coun-ter-terms}

\def\tikzcaption#1#2{\DefWarn#1\xdef#1{Fig.~\the\figno}
\writedef{#1\leftbracket Fig.\noexpand~\the\figno}%
{
\smallskip
\leftskip=20pt \rightskip=20pt \baselineskip12pt\noindent
{{\bf Fig.~\the\figno}\ \ninepoint #2}
\bigskip
\global\advance\figno by1 \par}}

\def\ntoalpha#1{%
\ifcase#1%
@%
\or A\or B\or C\or D\or E\or F\or G\or H\or I\or J\or K\or L\or M%
\fi
}

\global\newcount\appno \global\appno=1
\def\applab#1{\xdef #1{\ntoalpha{\appno}}\writedef{#1\leftbracket#1}\wrlabeL{#1=#1}
\global\advance\appno by1}

\def\preprint#1 #2\par{\rightline{\vbox{\baselineskip12pt\hbox{#1}\hbox{#2}}}\vskip2cm}
%
\def\title#1\par{\centerline{\bf #1}\nopagenumbers\pageno=0}
\def\author#1\par{\bigskip\bigskip\centerline{#1}}

\newcount\addressno

\def\email#1#2{
\footnote{\null}{\kern-\parindent \llap{$^#1$\hskip1pt}email: #2}}

\def\startcenter{%
  \par
  \begingroup
  \leftskip=0pt plus 1fil
  \rightskip=\leftskip
  \parindent=0pt
  \parfillskip=0pt
}
\def\stopcenter{\endgroup}

\def\address{\bigskip%
  \ifnum\the\addressno=0\else\stopcenter\endgroup\fi
  \advance\addressno by 1%
  \begingroup
  \startcenter
  \it
  \obeylines
  \addressAux
}
\def\addressAux#1{#1}

\def\abstract{\stopcenter\endgroup\bigskip\bigskip\noindent}

\def\Dsl{\,\raise.15ex\hbox{/}\mkern-13.5mu D} 
\def\dsl{\raise.15ex\hbox{/}\kern-.57em\partial}
 
\def\boxeqn#1{\vcenter{\vbox{\hrule\hbox{\vrule\kern3pt\vbox{\kern3pt
	\hbox{${\displaystyle #1}$}\kern3pt}\kern3pt\vrule}\hrule}}}


\def\a{\alpha}
\def\b{{\beta}}

\def\d{{\delta}}
\def\e{{\epsilon}}
\def\l{\lambda}

\def\t{{\theta}}
\def\om{{\omega}}

\def\half{{1\over 2}}
\def\p{{\partial}}

\def\bar{\overline}
\def\({\left(}
\def\){\right)}

\def\cA{{\cal A}}
\def\cF{{\cal F}}
\def\cJ{{\cal J}}
\def\cK{{\cal K}}
\def\cI{{\cal I}}
\def\cV{{\cal V}}
\def\cW{{\cal W}}
\def\cY{{\cal Y}}
\def\cZ{{\cal Z}}



\def\len#1{{%
\def\Dlen{\left|\mkern-1mu #1\mkern -0.5mu\right|}%
\def\Sslen{\left|\mkern-1.3mu #1\mkern -1.3mu\right|}%
\def\SSlen{\left|\mkern-2.8mu #1\mkern-1.3mu\right|}%
\mathchoice{\Dlen}{\Dlen}{\Sslen}{\SSlen}}}

\def\Im{\mathop{{\rm Im}}} 
\def\sfrac#1/#2{\kern.1em\raise.5ex\hbox{\the\scriptfont0 #1}%
\kern-.1em/\kern-.15em\lower.25ex\hbox{\the\scriptfont0 #2}}

\font\tenshuffle=shuffle10 \font\sevenshuffle=shuffle7 \font\fiveshuffle=shuffle7 at 5pt
\def\shuffle{{%
\def\Dshuffle{\mathbin{\hbox{\tenshuffle\char'001}}}%
\def\Sshuffle{\mathbin{\hbox{\sevenshuffle\char'001}}}%
\def\SSshuffle{\mathbin{\hbox{\fiveshuffle\char'001}}}%
\mathchoice{\Dshuffle}{\Dshuffle}{\Sshuffle}{\SSshuffle}}}


\def\qed{\hbox{\hskip 3pt
\vbox{\hrule\hbox to 7pt{\vrule height 7pt\hfill\vrule}
\hrule}}\hskip3pt}

\overfullrule=0pt\relax

\frenchspacing

\def\checkdef#1#2{
\ifx\UndeFined#1%
	\def#1{#2}
\else
	\immediate\write16{*** BUG ***: the label \string#1 is already defined ***}
\fi
}
\newread\instream

\openin\instream= label.defs
\ifeof\instream\message{No labels in advance yet. Wait till next pass.}
\else\closein\instream \input label.defs
\fi
\writedefs

\def\arXiv:#1].{\hepthStrip#1 \nil}
\def\hepthStrip#1 #2\nil{\href{http://arxiv.org/abs/#1}{arXiv:#1 #2\unskip}].}
